\documentclass[fleqn]{article}
\usepackage{fleqn}
\usepackage{graphics}
\usepackage{epsfig}
\usepackage{amsmath}
\usepackage{amssymb}
\usepackage{colortbl}
\usepackage{calc}
\newcommand{\beq}{\begin{equation}}
\newcommand{\eeq}{\end{equation}}
\newcommand{\bea}{\begin{eqnarray}}
\newcommand{\eea}{\end{eqnarray}}
\newcommand{\cg}{\cellcolor[gray]{0.8}}
\newcommand{\cgr}{\cellcolor[rgb]{0,0.8,0}}
\newcommand{\cpn}{\cellcolor[rgb]{1,0,1}}
\newcommand{\clb}{\cellcolor[rgb]{0,1,1}}
\newcommand{\crd}{\cellcolor[rgb]{1,0,0}}
\newcommand{\cyl}{\cellcolor[rgb]{1,1,0}}
\newcommand{\ctl}{\cellcolor[rgb]{0,0.5,0.5}}
\newcommand{\col}{\cellcolor[rgb]{0.5,0.5,0}}
\newcommand{\csl}{\cellcolor[rgb]{0.5,0,0}}
\begin{document}
\Large
\begin{center}
{\bf  A finite geometric toy model of space-time as an error correcting code}
\end{center}
\large
\vspace*{-.1cm}
\begin{center}
P\'eter L\'evay$^{1}$ , Fr\'ed\'eric Holweck$^{2}$
\end{center}
\vspace*{-.4cm} \normalsize
\begin{center}

$^{1}$Department of Theoretical Physics, Institute of Physics, Budapest University of\\
Technology, H-1521 Budapest, Hungary

$^2$
Laboratoire Interdisciplinaire Carnot de Bourgogne, ICB/UTBM, UMR 6303 CNRS,
Universit\'e Bourgogne Franche-Comt\'e, 90010 Belfort Cedex, France
\vspace*{.0cm}

\vspace*{.2cm} (17 December 2018)
 \end{center}
\vspace*{.1cm} \noindent {\bf Abstract:}
A finite geometric model of space-time (which we call the bulk) is shown to emerge as a set of error correcting codes. The bulk is encoding a set of messages located
in a blow up  of the Gibbons-Hoffman-Wootters (GHW) discrete phase space for $n$-qubits (which we call the boundary).
Our error correcting code is a geometric subspace code known from network coding, and the correspondence map is the finite geometric analogue of the Pl\"ucker map well-known form twistor theory.
The $n=2$ case of the bulk-boundary correspondence is precisely the twistor correspondence where the boundary is playing the role of the twistor space and the bulk is a finite geometric version of compactified Minkowski space-time. For $n\geq 3$  the bulk is identified with the finite geometric version of the Brody-Hughston quantum space-time. For special regions on both sides of the correspondence we associate certain collections of qubit observables. On the boundary side this association gives rise to the well-known GHW quantum net structure.
In this picture the messages are complete sets of commuting observables associated to Lagrangian subspaces giving a partition of the boundary. Incomplete subsets of observables corresponding to subspaces of the Lagrangian ones are regarded as corrupted messages. Such a partition of the boundary is represented on the bulk side as a special collection of space-time points. 
For a particular message residing in the boundary, the set of possible errors is described by the fine details of the light-cone structure of its representative space-time point in the bulk. 
The geometric arrangement of representative space-time points, playing the role of the variety of codewords, encapsulates an algebraic algorithm for recovery from errors on the boundary side.

\vspace*{.3cm}
  \noindent

   {\bf PACS:} 04.60Pp, 02.40.Dr, 03.65.Ud, 03.65.Ta, 03.65Aa, 03.67Pp, 04.20Gz \\
   
   {\bf Keywords:} Quantum Geometry, Stabilizer Codes, Quantum Entanglement, Twistors
    \\ \hspace*{1.95cm}
     \vspace*{-.2cm} \noindent \hrulefill

\section{Introduction}

Since the advent of holography\cite{Holo1,Holo2} and the AdS/CFT correspondence\cite{AdS} physicists have realized that in order to understand the properties of a physical system sometimes one can find clues for achieving this goal within the realm of another one of a wildly different kind.
In mathematics, an idea of a similar kind had already been followed by classical geometers of the 19th century.
The work of Pl\"ucker, Klein,
Grassmann and others has shown  us that it is rewarding to reformulate problems pertaining to geometrical structures of a space  in terms of geometrical structures of another space of very different kind.
The simplest instance of such a geometrical type of rephrasing called the Klein correspondence\cite{Hodge} revealed that the lines of the three dimensional (projective) space can be parametrized by the points of a four dimensional one.
Had the physicists of that time been mesmerized by philosophical ideas on unification of space and time as a four dimensional entity, they probably would have regarded this mathematical correspondence as a promising link between the geometry of space-time and a space of one dimension less.
Few decades later the idea of unification of space, time, matter and gravity emerged in the form of a geometric theory of a four dimensional continuum: Einstein's General Theory of Relativity. However, the notion of relating this unified space-time structure to a space of one dimension less occurred much later in the 1960s when Roger Penrose created twistor theory\cite{Penrose1}.

Twistor theory has given us a correspondence between spaces of wildly different geometries. Combining the ideas of geometers of the 19th century in twistor theory the concept of complexification linked with space-time structure has become a key ingredient.
Indeed, compactified and complexified Minkowski space-time turned out to be the right object to consider and real Minkowski space-time was arising merely as a real slice of this complex space. In the twistor correspondence\cite{Penrose1,WW,Hurd} physical data of four (complexified) dimensional space-time is to be expressed in terms of data of a three (complex) dimensional space: twistor space.
Then for example the causal structure of space-time manifests itself on the twistor side of the correspondence as data encoded in complex geometric structures.

Following the insights of string theory dualities and the spectacular succes of the AdS/CFT correspondence (for an introduction to these topics see \cite{Becker}), 
the trick of relating spaces of different dimensions and geometries has become a daily routine for string theorists. In the meantime some new ideas have been formulated within a different and rapidly evolving new research field: Quantum Information Theory (QIT)\cite{Nielsen}.
After the discoveries of Deutsch\cite{Deu},  Shor\cite{Shor}, Grover\cite{Grover} and others it has become clear that quantum algorithms in principle can be implemented on quantum computers, objects that are capable of outperforming their classical cousins for certain computational tasks.
Parallel to these developments it also turned out that QIT serves as an effective new language capable of articulating basic notions underlying quantum theory in a compelling way. 

It is then not surprising that QIT, as a new technique of elaboration, has been added to the plethora of stringy dualities in 2006 when some correspondences between QIT, String theory and holography have been reported\cite{RT1,RT2,KL,D,Ly}.
The result of Ryu and Takayanagi\cite{RT1,RT2} turned out to be a breakthrough.
The Ryu-Takayanagi proposal based on the notion of holographic entanglement entropy established a new method for exploring the meaning of the so-called bulk-boundary correspondence.
According to this idea, a gravitational theory in the bulk is encoding quantum information associated with degrees of freedom
residing in the boundary. 
As a consequence of this, the recurring theme of regarding space-time as an emergent entity showed up in a new fashion. 
In line with this proposal for the nuts and bolts of space-time, boundary entanglement manifesting itself in the bulk serves as a glue\cite{Raamsdonk,Swingle,Bartek}.
As far as the method of how this encoding supports the emerging space-time structure, the language of error correcting codes\footnote{For an early bird observation on the surprising relevance of the Hamming code in understanding the mathematical structure of the $E_7$ symmetric black hole entropy formula, see Ref.\cite{LevaySAM}.} has been invoked\cite{Verlinde,Preskill,Dong1,Dong2}.

The central idea underlying such error correcting schemes is bulk reconstruction using boundary data\cite{HKLL}.
This idea sheds some light on the important issue called "subregion-subregion duality". This notion suggests that a region of the boundary should contain complete information about a certain subregion of the bulk.
More precisely, according to this picture bulk operators in AdS can be reconstructed as CFT operators in a subregion provided they lie in 
a special region in the bulk\cite{Hubeny,Gravdual}.
For instance, in their error correcting scheme the authors of Ref.\cite{Dong1} choose a finite set of local bulk operators realized in the CFT via the global representation of\cite{HKLL}.
Then acting with these operators on some fixed state and forming the linear span produces a code subspace.
Different choices for this finite set of operators then yield different code subspaces.
Hence what we get by this procedure is not a single code but rather a collection of error correcting codes. Hence in the light of this concept bulk space-time is a collection of error correcting codes of a very special kind. In particular, studying the gauge-like degree of freedom manifesting itself in different choices of codewords should be connected in an intricate manner to the geometry of certain bulk domains.
Finally in this approach bulk effective field theory operators emerge as a set of logical operators on various encoded subspaces which are 
protected against local errors in the boundary CFT.

In  order to study some aspects of a concept similar to subregion-subregion duality we would like to propose an interesting finite geometric toy model in this paper. The model connects two spaces,
which we call "boundary" and "bulk", via a classical error correcting code. Moreover, we reveal that this classical code is associated with a "quantum net structure" in a natural manner. We find that a version of this structure on the boundary side is already well-known in the literature as the quantum net associated to the Gibbons-Hoffman-Wootters (GHW) discrete phase space\cite{w1,W2}. 
On the other hand, on the bulk side the structure we find is a finite geometric version of the Brody-Hughston "quantum space-time"\cite{Brody1,Brody2}. 

In our paper we will settle with some basic exposition of our ideas with many details left to be explored for future work. 
First of all we emphasize that at the present time we are not attempting to tie the duality in question to any discretization of the AdS/CFT correspondence, i.e to efforts showing up in studies like Refs.\cite{Swingle,Preskill}.
In fact our model at this stage contains merely two levels of elaboration. At one level it is a finite geometric illustration of how geometric data of one type of a space (boundary) determines the geometric data of the other (bulk) 
via an error correcting code
based on a simple and explicit map.
At the other level of progression, to special subregions of either space we associate certain collections of elementary quantum observables in a manner which is controlled (though in an ambiguous manner) by the mathematical form of our map.

The idea that a map of a simple kind is capable of grasping certain characteristic features of complex physical phenomena is well-known.
Let us just refer to the fact that some of the key aspects of the phenomenon of chaos in dynamical systems can be grasped
by chaotic maps like the Arnold cat map.
The map we elevate here to the status of an object having some new physical meaning is the Pl\"ucker map of Eq.(\ref{Pluckermap}).
A special instance of this map is the one responsible for the Klein correspondence (see Figure 2.). 
This is the map we already referred to in the beginning of this introduction by emphasizing its special role in connection with twistor theory.
The basic hint of our paper is coming from an observation that this map can be used to associate $2^{n-1}$-qubit Pauli operators to $n$-qubit ones\cite{LPS,HSL}. The new ingredient amenable to a new physical interpretation of this map is the fact that it also has an intimate connection with geometric subspace codes\cite{KK,Silva,Stokes,Gorla}.

Finally we would like to make some clarifying comments on our terminology in this paper.
Our reason for calling the bulk part of the correspondence as some sort of space-time is as follows.
First of all, we will see that the use of our map is just a generalization of a finite geometric instance of the twistor correspondence of Penrose which is hence featuring a finite geometric version of compactified Minkowski space-time on one side of the correspondence. Second, we will show that the incidence structure of subspaces on the boundary side manifests itself on the bulk side in a structure reminiscent of the usual causal structure of ordinary space-time.
Naturally, having a causal-like structure by itself is not justifying our bulk to be a representative of some sort of space-time.
It could be for instance a finite geometric analog of "kinematic space" as defined in Ref.\cite{Bartek}, which also has a causal structure related to partial ordering of different boundary regions. 
However, apart from some speculations presented in our conlusions, we refrain from making a distinction between these possibilities.
We must also stress that since we are working within a finite geometric context, our use of the words "bulk" and "boundary" is of course lacking the usual meaning as customary in AdS/CFT. These words are mainly used here as useful abbreviations for the geometrical objects we are intending to relate.
We must add however, that in spite of this, the results of Section 5.5 culminating in the appearance of our suggestive Figure 9 gives some support for our
nomenclature.

The organization of this paper is as follows.
In Section 2 we present the necessary finite geometric background. Observables of $n$-qubits, their associated projective spaces, Grassmannians and subspace codes are defined here. We are aware that these somewhat abstract concepts are probably not belonging to the conventional wisdom of most of the readers. Hence we opted to include extensive and detailed background material in five clarifying Appendices.
Moreover, we devoted Section 3 to a detailed elaboration of the simplest (two qubit case).

In Section 3, the reader is introduced to the basic concepts of our paper through a set of very simple calculations that are organized into a collection of subsections. In this case all of the relevant finite geometric structures also have a pictorial representation. 
The two-qubit case discussed here is in complete analogy with Penrose's twistor correspondence. 
The bulk is the finite geometric analog of compactified, complexified Minkowski space-time.
It also turns out that the boundary, which is the finite geometric version of twistor space, is related to the GHW discrete phase space for two qubits. For the messages residing in the boundary, we choose elements of a fibration partitioning the boundary into a set of lines. 
To the points of these message lines, one can associate a maximal set of commuting observables. A fibration to messages of that kind is reminiscent of a particular slicing up of the boundary to Cauchy slices known from the continuous case. 
The errors in this picture are either points contained in the message lines, or planes containing them.
Then we study how the set of possible errors is represented in the bulk. We regard the representatives of the messages in the bulk as codewords. Then we show that the possible set of errors in the boundary is represented in the bulk by the light cone structure of the codewords. This observation facilitates an algebraic description of an error correction process with the recovery from errors having an interesting physical interpretation. 

In Section 3.6 we realize that one really should consider not just a single copy of messages and their associated codewords, but instead a whole {\it collection} of them intertwined in a special manner. This possible set of codes corresponds to all possible slicings of the boundary into lines carrying maximal commuting sets of observables. We characterize this set of codes algebraically and discuss their physical meaning in the twistor language. 
Next we notice that by assigning a fixed state to a particular message line we can associate states even to the remaining message lines in a unique manner.
This construction is then identified with the rotationally covariant association of a quantum net to the lines of the GHW phase space\cite{w3}.
Then we initiate a study for understanding our classical error correction code in terms of the quantum states of the quantum net. We point out that the observables associated with plane errors have a clear-cut interpretation within the formalism of stabilizer codes. 
We close this section with a study on how one can relate the different codes and their associated states within
the total set of possible codes with the help of unitaries decomposed in terms of elementary quantum gates. These unitaries are encoding the entanglement properties of the stabilizer states associated to the lines.
We observe that when changing our code to another one the uniqueness of our association of quantum states to boundary lines is lost.

Section 4 is devoted to generalizations for an arbitrary number of $n$ qubits. Here when necessary the $n=3$ case is invoked as an illustration, with details found in Appendix D. First in Section 4.1 we discuss how the boundary can be fibered in terms of messages. Then in Section 4.2 we relate the boundary to the GHW phase space\cite{w1,W2} of $n$-qubits. It turns out that our boundary is just the blow up of the projectivization of this phase space. This term refers to the fact that the GHW phase space is a space with coordinates taken from the finite field $GF(2^n)$ regarded as an extension of $GF(2)\simeq {\mathbb Z}_2$. Then the blowing up process is effected by using the field reduction map for taking the vector space of the GHW phase space to the vector space underlying our boundary. 

We show in Section 4.3 that our bulk encoding  $n$-qubit messages is the Grassmannian image of the boundary under the Pl\"ucker map. We identify the bulk as the finite geometric version of an object called the Brody-Hughston "quantum space-time"\cite{Brody1,Brody2}. The bulk is equipped with a "chronometric form"\cite{Fink,Brody1,Brody2} encoding the intersection properties of the subspaces of different dimension of the boundary in the causal structure of the bulk. 
Since our bulk is regarded as the "hypertwistor" analog of the Klein quadric, which is related to a complexification of Minkowski space-time with the ordinary space-time being just a real slice, we then turn to a characterization of ithe "real" slice of it.
In Section 4.4. we identify such a real slice of the bulk as the image of the Lagrangian Grassmannian under the Pl\"ucker map.
We show that in our finite geometric version, just like in the $n=2$ case, this real slice of the bulk for $n\geq 3$ can be embedded into a hyperbolic quadric residing in a projective space ${\mathbb P}(V)$. However, this time the underlying vector space $V$ is a one of dimension $2^n$ taken over the finite field $GF(2)$, equipped with a natural symplectic structure.

Since the boundary was amenable to a nice reinterpretation in terms of a space (namely the GHW one) taken over the field extension $GF(2^n)$ in Section 4.5, we present an interesting reinterpretation along this line also valid for the bulk.
In order to cope with the $2^n$ dimensional vector space stucture of the previous section reminiscent of the space of $n$-qubits (taken however, over finite fields), we introduce a new term.
Namely we abbreviate the term "a quantum bit with amplitudes taken from a finite field" in short by the one - a {\it fibit}.
The basic idea we initiate in this section is the one of glueing together the bulk from $n$-fibits.
In Section 4.6 we find that there is a neat physical interpretation of the space of codewords in the bulk representing our messages of the boundary
in terms of $n$-fibits over $GF(2^n)$ correlated in a special manner via a twisted tensor product structure.
According to this idea, there is a minimalist representation of the bulk as an object encoding boundary data summarized in Figure 9. In this representation the GHW phase space can be regarded as a circle discretized into $n$-points. The $n$-fibits are then represented by $n$ such circles glued together in a special manner via an application of the Frobenius automorphism of our finite field.
Section 4.7 is containing an adaptation of the main ideas of Ref.\cite{Stokes}, on the generalization of the error correction scheme
we presented in Sections 3.4-3.6. This section contains merely the basic ideas, with the details needing further technical elaboration, this remains a challenge we are looking to take up in future work.
Our conclusions and some speculations are left to Section 5. Our paper is supplemented with five Appendices containing technical details and illustrations. They are included to help the reader to navigate in the field of finite geometric concepts.

\section{Observables, projective spaces and subspace codes}

Let  us consider the $2n$-dimensional vector space $V\equiv V(2n,2)$ over the field $GF(2)\equiv \mathbb{Z}_2=\{0,1\}$.
In the following we will refer to the dimension of a vector space as its {\it rank}, hence $V$ is of rank $2n$.
In the canonical basis $e_{\mu}, \mu=0,1,2,\dots 2n-1$ 
we arrange the components of a vector $v\in V$ in the form
\begin{equation}
v\leftrightarrow (q_0,q_1,\dots,q_{n-1},p_0,p_1,\dots,p_{n-1}).
\label{elsokonv}
\end{equation}
\noindent
We represent $n$-qubit observables by vectors of $V$  in the following manner.
Use the mapping
\beq
(00)\leftrightarrow I, \quad (01)\leftrightarrow X,
\quad (11)\leftrightarrow Y,\quad (10)\leftrightarrow Z
\label{alap}
\eeq
\noindent
where $(X,Y,Z)\equiv (\sigma_x,\sigma_y,\sigma_z)$ i.e they are the usual Pauli spin matrices.
Then for example in the $n=3$ case the array
$(q_0,q_1,q_2,p_0,p_1,p_2)$ of six numbers taken from $\mathbb{Z}_2$ encodes a three-qubit operator {\it up to sign}.
For example
\begin{equation}
(100110)\leftrightarrow \pm YXI\leftrightarrow \pm Y\otimes X\otimes I.
\label{pelda3}
\end{equation}
\noindent
The leftmost qubit operator is the "zeroth"-one ($Y$) with labels $(q_0,p_0)$, the middle one is the "first" ($X$) with labels 
$(q_1,p_1)$ and the rightmost one is the "second" ($I$) with labels $(q_2,p_2)$.
As a result of this procedure we have a map
\beq
v\mapsto \pm \mathcal{O}_v
\label{op}
\eeq
\noindent
between vectors of $V$ and $n$-qubit observables $\mathcal{O}$ up to sign. Note that since under multiplication the operators also pick up multiplicative factors of $\pm i$, the space of observables is {\it not} forming an algebra. In order to properly incorporate the multiplicative structure the right object to consider is the Pauli group\cite{Nielsen} which is the set of operators of the form $\{\pm \mathcal{O}_v,\pm i\mathcal{O}_v\}$.  
Then one can show that the center of the Pauli group is the group $\{\pm 1,\pm i\}$, and its central quotient is just the vector space $V$. Under this isomorphism vector addition in $V$ corresponds to multiplication of Pauli group elements up to $\pm 1$ and $\pm i$ times the identity operator.

The vector space $V$ is also equipped with a symplectic form $\langle\cdot,\cdot\rangle$ encoding the commutation properties of the corresponding observables. 
Namely, for two vectors $v,v^{\prime}\in V$ with components in the canonical basis 
\begin{equation}
v\leftrightarrow (q_0,q_1,\dots,q_{n-1},p_0,p_1,\dots,p_{n-1}),\qquad
v^{\prime}\leftrightarrow (q^{\prime}_0,q^{\prime}_1,\dots,q^{\prime}_{n-1},p^{\prime}_0,p^{\prime}_1,\dots,p^{\prime}_{n-1})
\label{familiar}
\end{equation}
\noindent
\begin{equation}
\langle v,v^{\prime}\rangle = 
\sum_{i=0}^{n-1}\left(q_ip^{\prime}_i+q^{\prime}_ip_i\right)\in \mathbb{Z}_2.
\label{symp}
\end{equation}
\noindent
In the symplectic vector space $(V,\langle\cdot,\cdot\rangle )$ we have $ \langle v,v^{\prime}\rangle=0$ or $1$, referring to the cases when the corresponding $n$-qubit observables are commuting or anticommuting: $[\mathcal{O}_v,\mathcal{O}_{v^{\prime}}]=0$ or
$\{\mathcal{O}_v,\mathcal{O}_{v^{\prime}}\}=0$.
Notice that over the two element field $GF(2)$ an alternating form like $\langle\cdot,\cdot\rangle$ is symmetric. 

Since $V$ is even dimensional and the symplectic form is
nondegenerate, the invariance group of the symplectic form is the
symplectic group $Sp(2n,{\mathbb Z}_2)\equiv Sp(2n,2)$. This group is acting on the row
vectors of $V$ via $2n\times 2n$ matrices $M\in Sp(2n,2)$ from the
right, leaving the matrix $J=\langle e_{\mu},e_{\nu}\rangle$ of the symplectic form invariant \beq
v\mapsto vM,\qquad MJM^t=J. \label{transz} \eeq \noindent It is
known that $Sp(2n,2)$ is generated by transvections\cite{Cherchiai,Shaw}
$T_w\in Sp(2n,2), w\in V$ of the form \beq T_w:V\to V,\qquad
v\mapsto T_wv=v+\langle v,w\rangle w \label{transvections} \eeq
\noindent and they are indeed symplectic, i.e. \beq \langle
T_wv,T_wv^{\prime}\rangle=\langle v,v^{\prime}\rangle. \label{szimpltulajd} \eeq

Given the symplectic form one can define a quadratic form $Q:V\to \mathbb{Z}_2$ by the formula
\beq
Q(v)\equiv \sum_{i=0}^{n-1}q_ip_i
\label{kvad}
\eeq
\noindent
which is related to the symplectic form via 
\beq
\langle v,v^{\prime}\rangle=Q(v+v^{\prime})+Q(v)+Q(v^{\prime}).
\label{kapcsolat}
\eeq
\noindent
Generally quadratic forms which, by a convenient choice of basis, can be given the (\ref{kvad}) canonical form are called {\it hyperbolic}\cite{Cherchiai,Shaw}.
In our special case the meaning of the (\ref{kvad}) quadratic form is clear: for $n$-qubit observables $\mathcal{O}_v$ containing an even (odd) number of tensor product factors of $Y$, the value of $Q$ is zero (one). Hence the observables which are symmetric under transposition are having $Q(v)=0$ and ones that are antisymmmetric under transposition are having $Q(v)=1$. 

In the following we will refer to the set of subspaces of rank $k=1,2,3,\dots ,2n-1$ of $V$ as the Grassmannians: $Gr(k,2n)$.
For $k=1,2,3,\dots,2n-1$ these spaces are arising by considering the spans of one, two, three, etc., $2n-1$  linearly independent vectors $v_1,v_2,v_3,\dots ,v_{2n-1}\in V$. These Grassmannians are just sets of {\it lines}, {\it planes}, {\it spaces}, etc., and {\it hyperplanes} through the origin. Since we are over $GF(2)$ these are sets of the form $\{au\}$, $\{au+bv\}$, $\{au+bv+cw\}$, etc. with $a,b,c\in \mathbb{Z}_2$, consisting of one, three, seven, etc., $2^{2n-1}-1$ nontrivial vectors.
A line $\{au\}$ through the origin (zero vector) is a {\it ray}. Over $GF(2)$ the number of rays equals the number of nonzero vectors of $V$. 

Regarding the set of rays as the set of points of a new space of one dimension less, gives rise to the projective space ${\mathbb P}(V)\equiv PG(2n-1,2)$. In this projective context lines, planes, spaces etc. of $V$ give rise to points, lines, planes etc. of ${\mathbb P}(V)$. Hence the collection of the projectivization of the Grassmannians $G(k,2n)$ 
denoted by $\mathcal{G}(k-1,2n-1)$
forms the projective geometry of $PG(2n-1,2)$. 
In the following the word {\it dimension} will be used for the dimension of a projective subspace ${\mathbb P}(S)$, and {\it rank} will refer to the dimension of the corresponding vector subspace $S$.
Notice that according to  Eq.(\ref{op}), the sets of projective subspaces (points, lines, planes, etc. hyperplanes) of $PG(2n-1,2)$, i.e. the Grassmannians $\mathcal{G}(0,2n-1), \mathcal{G}(1,2n-1), \mathcal{G}(2,2n-1),\dots \mathcal{G}(2n-2,2n-1)$, up to a sign correspond to the set of {\it nonzero} observables, certain triples, seven-tuples , etc., $2^{2n-1}-1$-tuples of them.

 For our quadratic form of Eq.(\ref{kvad}) the points satisfying the equation $Q(v)=0$ form a {\it hyperbolic quadric} in $PG(2n-1,2)$, denoted by $Q^+(2n-1,2)$.
Hence symmetric $n$-qubit observables are represented by points {\it on}, and antisymmetric ones {\it off} this hyperbolic quadric in $PG(2n-1,2)$.

Since we have the (\ref{symp}) symplectic form at our disposal one can specify further our projective subspaces and their corresponding subsets of observables. 
A subspace of $V$ is called {\it isotropic} if there is a vector in it which is orthogonal to the whole subspace. 
A subspace $\mathcal{I}$ is called {\it totally isotropic} if for all points $v$
and $u$ of $\mathcal{I}$ we have $\langle v,u\rangle=0$, i.e. a totally isotropic subspace is orthogonal to itself. 
Notice that in the case of rank one and two suspaces of $V$, i.e. dimension zero and one subspaces (points and lines) of $P(V)$ the notions isotropic and totally isotropic coincide.
Due to Eq. (\ref{op}) it is clear that a totally isotropic subspace is represented by a set of {\it  mutually commuting  observables}. 
Notice that the dimension of {\it maximally} totally isotropic subspaces of $PG(2n-1,2)$ is $n-1$. These are called Lagrangian subspaces. These are arising from totally isotropic subspaces of rank $n$ of the rank $2n$ vector 
space $(V,\langle\cdot,\cdot\rangle)$. The corresponding set of operators gives rise to a maximal set of $2^n-1$-tuples of commuting observables. 
The incidence structure of the set of totally isotropic subspaces of $PG(2n-1,2)$ defines the {\it symplectic polar space}
$\mathcal{W}(2n-1,2)$.
A projective subspace of maximal dimension of $\mathcal{W}(2n-1,2)$ is called its {\it generator}. The projective dimension of a maximal subspace is called the rank of the symplectic polar space. Hence in our case the rank of $\mathcal{W}(2n-1,2)$ is $n$.
$\mathcal{W}(2n-1,2)$ made its debut to physics in connection with observables of $n$-qubits in Ref.\cite{SP}.

A {\it subspace code}\cite{KK,Silva,Stokes} is a set of subspaces in $V$. Alternatively one can regard it as a set of projective subspaces in ${\mathbb P}(V)$. Let us suppose that all of the subspaces have the same dimension $k-1$. Then the code is contained in the Grassmannian $G(k,2n)$, or projectively in $\mathcal{G}(k-1,2n-1)$. These codes are called Grassmannian codes.
If two subspaces in $V$ intersect only in the zero vector then the corresponding subspaces in the projective space are non intersecting.
Then we will also call two subspaces $\mathcal{S}_1$ and $\mathcal{S}_2$ of $V$ nonintersecting as long as they only intersect in the zero vector.
This idea motivates the introduction of 
a natural distance between two subspaces $\mathcal{S}_1$ and $\mathcal{S}_2$  given by the formula\cite{KK}
\beq
d(\mathcal{S}_1,\mathcal{S}_2)={\rm dim}({\mathcal{S}}_1+{\mathcal{S}}_2)-{\rm dim}({\mathcal{S}}_1\cap{\mathcal{S}}_2)=
{\rm dim}({\mathcal{S}}_1)+{\rm dim}({\mathcal{S}}_2)-2{\rm dim}({\mathcal{S}}_1\cap{\mathcal{S}}_2)
\label{dist}
\eeq
\noindent
It can be shown\cite{KK} that $d(\cdot,\cdot)$ satisfies the axioms of a metric on ${\mathbb P}(V)$.
According to this metric two subspaces are close to each other when their dimension of intersection is large.
For example in the case of $G(2,4)$ the maximal possible distance is $4$.
It is realized by pairwise nonintersecting planes in $V(4,2)$, or alternatively for pairwise nonintersecting lines in $PG(3,2)$.
Another example is provided by Figure 8. where we see a series of planes with their distance decreasing as their dimension of intersection increasing.

In the error correcting process a codeword as a subspace is sent through a noisy channel. Then this message is corrupted, hence instead of the code subspace another subspace is received.
The basic idea of code construction is then to find a set of codewords, realized as subspaces in $G(k,2n)$, in such a way to be "well separated" with respect to $d(\cdot,\cdot)$, so that unique decoding is possible.
Moreover, it is also desirable to come up with efficient decoding algorithms.

In particular one would like to construct a code with maximum possible distance and a maximum number of elements. In order to do this one has to restrict the values of $k$ and $n$ conveniently.
This leads us to consider a set of subspaces $\mathcal{S}$, such that it partitions $V$. It means that there is no vector in $V$ such that it is not lying in an element of $\mathcal{S}$. As an extra constraint to be used, we demand that\cite{Gorla} any two elements of $\mathcal{S}$
are non intersecting if and only if $k$ divides $2n$. Such sets are called {\it spreads}\cite{Demb}, and in this paper we will be merely considering the special case of $k=n$.
For example a {\it spread of lines} in $PG(3,2)$ ($k=2$, and $n=4$) is a partition of the $15$ element point set of $PG(3,2)$ into a set of disjoint lines. Since over $GF(2)$ each line is featuring $3$ points, in this way we obtain 
$5$ lines to be used as codewords of a subspace code.
In the general case we have $n-1$ dimensional subspaces of $PG(2n-1,2)$ forming a spread. In this case the number of points of $PG(2n-1,2)$ is $2^{2n}-1$. Since an $n-1$-subspace is containing $2^n-1$ points with a choice of spread one can partition the point set of this projective space into $2^n+1$ such subspaces.

Let us now suppose that we have chosen a set of $n-1$ dimensional subspaces of $PG(2n-1,2)$ forming a spread\cite{Gorla} for building up a subspace code.
Then it is said that this spread defines a $[2n,n,2^n+1,2n]_2$ code\footnote{Generally we have $[r,k,M,D]_q$ codes where $V$ is a vector space of rank $r$ over the finite field $GF(q)$, and $M=\vert\mathcal{C}\vert$ is the number of codewords.}. 
Here the third entry is referring to the number $\vert\mathcal C\vert=2^n+1$ of codewords, and the last to the distance $D$ of the code defined to be the minimum of $d(\mathcal{C}_a,\mathcal{C}_b)$ where $\mathcal{C}_a$ and $\mathcal{C}_b$ are any two different codewords.
In our case according to Eq.(\ref{dist}) $D=2n$.

If we have a symplectic structure on $V$ one can specify further our constant dimension subspace codes.
A particularly interesting case of this kind is arising if we ask for the existence of spread codes with their codewords being isotropic $n-1$ subspaces. The reason for why these codes are interesting is as follows.
According to Eq.(\ref{op}) we have a mapping between subspaces of $V$ and certain subsets of $n$-qubit observables.
It then follows that the codewords, as special subspaces, will correspond to special sets of operators with physical meaning. In particular isotropic spreads, for $k=n$, will correpond to $2^n+1$ maximal sets of mutually commuting $2^n-1$ tuples of observables partitioning the the total set of $2^{2n}-1$ nontrivial observables.  
It is well-known that such sets of observables give rise to mutually unbiased bases systems used in quantum state tomography and for the definition of Wigner functions over a discrete phase space\cite{w1,W2,Fields}.

\section{The Klein Correspondence as a toy model of space time as an error correcting code}

In this section we work out the simplest case of encoding two qubit observables ({\it boundary observables}) into a special set of
three qubit ones ({\it bulk observables}) via error correction.
Being very simple this case serves as an excellent playing ground for showing our basic finite geometric structures at work. Moreover, since the number of qubits is merely two we are in a position of also having a pictorial representation for
many of the abstract concepts introduced in the previous section.
These detailed considerations aim at helping the reader to visualize our basic ideas, to be generalized later for an arbitrary number of qubits. 

\subsection{The Klein Correspondence and two-qubit observables}

For the two qubit case we use the three dimensional projective space ${\mathbb P}(V)=PG(3,2)$ with the underlying vector space of rank four: $V\equiv V(4,2)$. As we emphasized $V$ is equipped with a symplectic form, see Eq.(\ref{symp}) with $n=2$. The $15$ points of $PG(3,2)$ correspond to the $15$ nontrivial two-qubit observables.
We have altogether $35$ lines of $PG(3,2)$ corresponding to triples of observables with their product equals the identity ($I\otimes I\equiv II$) modulo factors of $\pm 1,\pm i$. From the $35$ lines $15$ ones are isotropic and $20$ ones are non-isotropic. 
At the level of observables an example for the former is $\{XY,YZ, ZX\}$,  and for the latter is $\{IX,IY,IZ\}$. The former defines a mutually commuting set of three observables.
The structure of the point-line incidence structure of isotropic and non isotropic lines is shown in Figure 1.
For the isotropic case the incidence structure gives rise to $\mathcal{W}(3,2)$. This structure happens to coincide with the one of a generalized quadrangle: $GQ(2,2)$ also called the "doily"\footnote{For the definition of generalized quadrangles, their extensions and their use in understanding the finite geometric aspects of form theories of gravity see Ref.\cite{LHSmagic}.}. 

\begin{figure}[pth!]
\centerline{\includegraphics[width=10truecm,clip=]{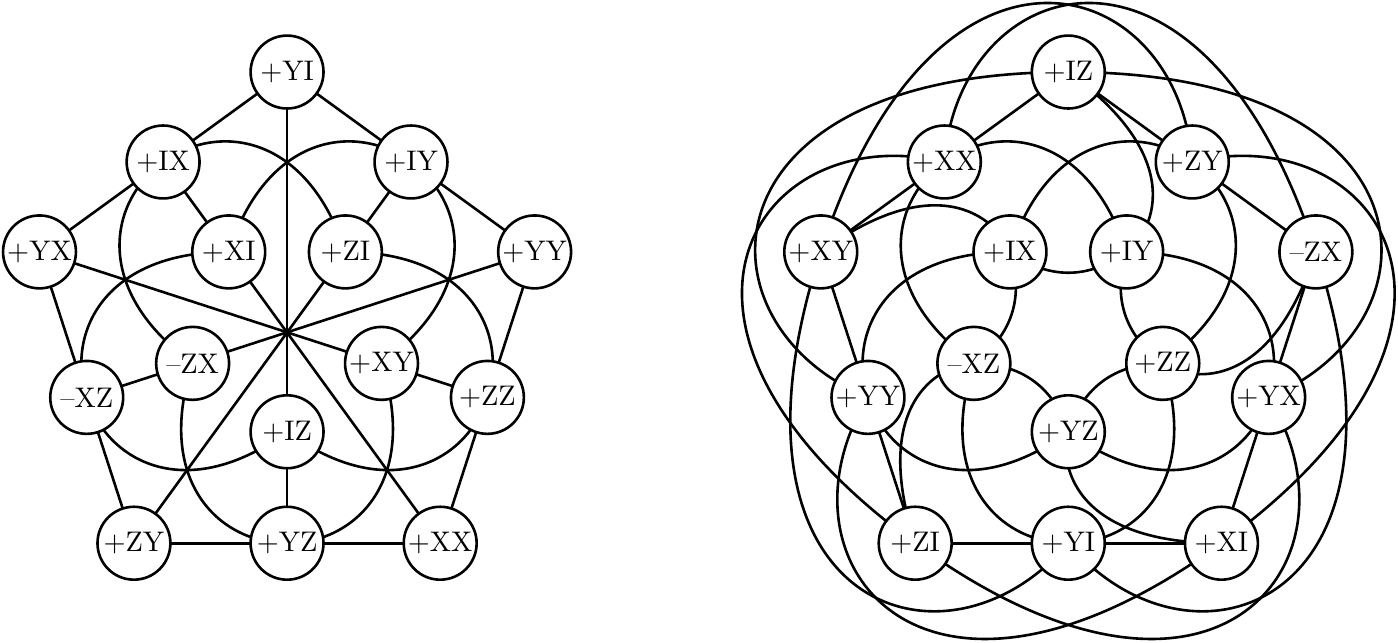}}
\caption{The doily ($\mathcal{W}(3,2)$) labelled by the nontrivial two-qubit observables (left). The commuting triples of observables correspond to its $15$ isotropic lines. The structure of nontrivial two-qubit observables giving rise to $20$ non-isotropic lines of $PG(3,2)$ (right) is encapsulated in the Cayley-Salmon ($[15_4,20_3]$) configuration. The isotropic and non-isotropic lines give altogether the $35$ lines of $PG(3,2)$. Notice that in $PG(3,2)$ there are seven lines incident with a point: three of them isotropic, and four of them non-isotropic.}
\end{figure}
\begin{figure}[pth!]
\centerline{\includegraphics[width=11truecm,clip=]{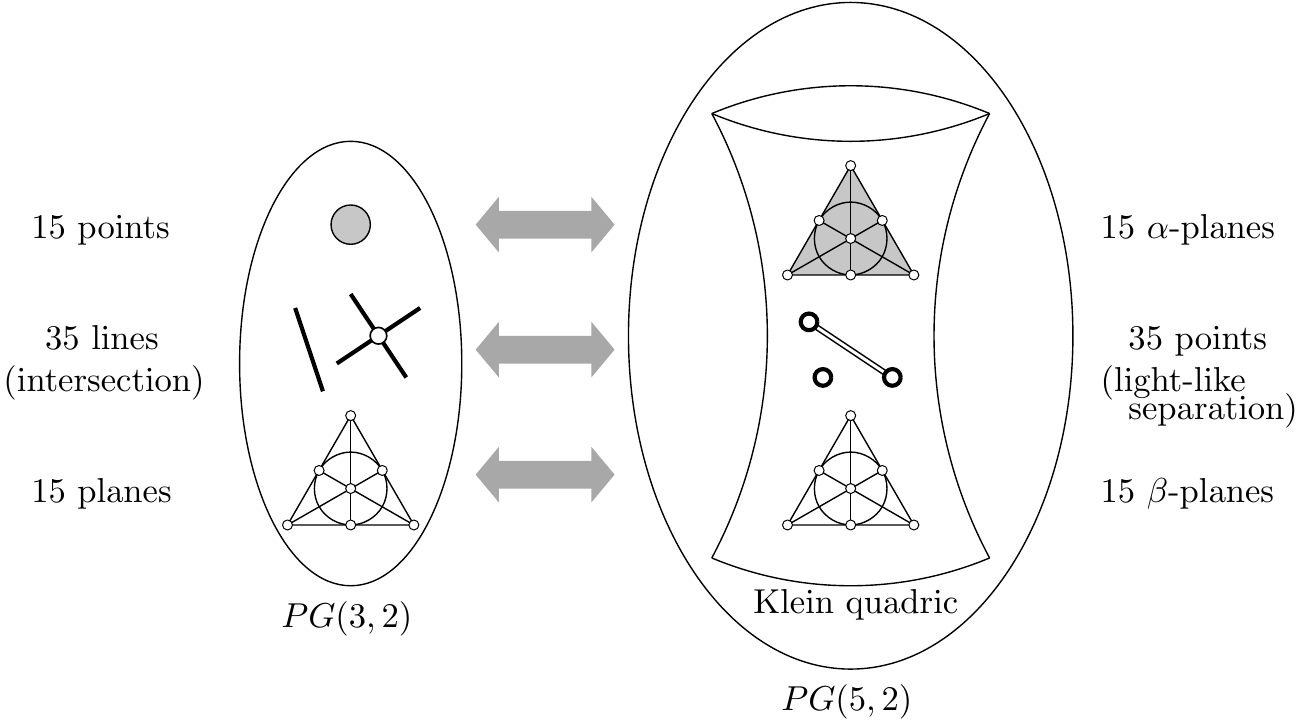}}
\caption{A pictorial representation of the Klein Correspondence.}
\end{figure}

In $PG(3,2)$ we also have $15$ planes containing $7$ points and $7$ lines. 
The incidence structure of these $7$ points and $7$ lines defines a Fano plane (see the diagram representing a plane in Figure 2.).
From the $7$ lines $3$ are isotropic and $4$ are non-isotropic. 
Representing a plane by a set of observables we obtain seven-tuples like:
$\{XX,IX,XI, ZY,YZ, YY, ZZ\}$. As we see one of the observables ($XX$) of this plane enjoys a special status.
Indeed, $XX$ is commuting with all of the remaining six ones. At the level of $V$ the corresponding vector $(0011)$ is orthogonal to the six vectors corresponding to the remaining six observables. Hence the orthogonal complement of this vector defines a subspace of $V$ of rank three. Projectively this corresponds to a hyperplane, i.e. projective subspace of dimension two, which is precisely our plane. 
This plane is an isotropic, but not a totally isotropic one.
As an illustration we note that the isotropic lines of our plane are represented by: the $3$ triples $\{XX,IX,XI\}$, $\{XX,ZY,YZ\}$, $\{XX,YY,ZZ\}$. On the other hand the non-isotropic ones
are represented by the $4$ triples: $\{XI,ZY,YY\}$, $\{IX,ZY,ZZ\}$, $\{XI,YZ,ZZ\}$, $\{IX,YY,YZ\}$.
For the physical meaning of these seven-tuples of observables we mention that these define generalized X-states\cite{Rau}.
For the convenience of the reader a complete list of observables representing points, lines and planes of $PG(3,2)$ can be found in Appendix A. 
The collection of these objects defines the projective geometry of our projective space $PG(3,2)$.

Now we establish a correspondence between two qubit observables and a special subset of three qubit ones.
It will then be used to establish an error correcting code, 
featuring certain subsets of operators on both sides of the correspondence.
The mathematical basis of our correspondence is the Klein Correspondence which we now discuss.

The Klein Correspondence (see Figure 2 and Appendix A) establishes a mapping between the $15$ points, $35$ lines and $15$ planes  of the projective space $PG(3,2)$ ({\it boundary}), and the $15$ $\alpha$-planes, $35$ lines and $15$ $\beta$-planes of a hyperbolic Klein quadric ({\it bulk}) embedded in $PG(5,2)$.
Since according to Eq.(\ref{op}) we have a mapping between subspaces (defined by vectors) of projective spaces and certain subsets of observables, this mathematical trick connects sets of observables of very different kind on both sides of the correspondence. Moreover, due to the very nature of the Klein Correspondence this correspondence between observables will be inherently {\it nonlocal}. 
We emphasize that at this stage our terminology of calling the spaces featuring this correspondence as "boundary" and "bulk" is dictated by mere convenience however, as we will see later, a suggestive one.

As a first step recall that a {\it line} ${\ell}$ in ${\mathbb P}(V)=PG(3,2)$ is described by two linearly independent vectors $v,v^{\prime}\in V$.
Indeed, these vectors span a subspace of rank two (a plane) in $V$, projectively a subspace of dimension one (a line) in ${\mathbb P}(V)$.
A representative of ${\ell}$ can be obtained by an arrangement of the four components of $v$ and $v^{\prime}$ as two row vectors as follows
\beq
\begin{pmatrix}Q\vert P\end{pmatrix}\equiv \begin{pmatrix}q_0&q_1&p_0&p_1\\
q^{\prime}_0&q^{\prime}_1&p^{\prime}_0&p^{\prime}_1\end{pmatrix},\qquad P=\begin{pmatrix}p_0&p_1\\p^{\prime}_0&p^{\prime}_1\end{pmatrix},\quad
Q=\begin{pmatrix}q_0&q_1\\q^{\prime}_0&q^{\prime}_1\end{pmatrix}.
\label{arrange}
\eeq
\noindent
Clearly left multiplication of $(Q\vert P)$ by any invertible $2\times 2$ matrix with elements taken from $\mathbb{Z}_2$, i.e. $GL(2,\mathbb{Z}_2)=SL(2,\mathbb{Z}_2)$ is not changing the line, it only changes the linearly independent vectors representing it.
The set of lines in $PG(3,2)$ forms the Grassmannian $\mathcal{G}(1,3)$.

Now we introduce coordinates for our lines: the Pl\"ucker coordinates. 
These are just the determinants of the six possible $2\times 2$ minors
one can form from the (\ref{arrange}) arrangement. 
We will regard the Pl\"ucker coordinates as the six components of a vector ${\mathcal P}$ taken in the canonical basis of a vector space $V(6,2)$ of rank six:
\beq
({\mathcal P}_{01},{\mathcal P}_{02},{\mathcal P}_{03},{\mathcal P}_{23},{\mathcal P}_{13},{\mathcal P}_{12}),\qquad \mathcal{P}\in V(6,2).
\label{pluck}
\eeq
\noindent
Clearly: ${\mathcal P}_{01}={\rm Det}Q$, and ${\mathcal P}_{23}={\rm Det}P$.
Moreover, since we are over the two element field we have for example
${\mathcal P}_{02}=q_0p^{\prime}_0+p_0q^{\prime}_0$.
The Pl\"ucker coordinates are called line coordinates since they are invariant under the left action of $SL(2,\mathbb{Z}_2)$ on $(P\vert Q)$, hence they are not depending on how we choose $v$ and $v^{\prime}$ representing the line.

Notice that ${\mathcal P}$ defines a {\it point} in $PG(5,2)$ with a special property.
Namely, one can check that
\beq
{\mathcal P}_{01}{\mathcal P}_{23}+{\mathcal P}_{02}{\mathcal P}_{13}+{\mathcal P}_{03}{\mathcal P}_{12}=0.
\label{hquad}
\eeq
\noindent
The left hand side is a quadratic combination of the (\ref{kvad}) form with $n=3$. Hence Eq.(\ref{hquad}) defines a {\it hyperbolic quadric}. This quadric lying inside $PG(5,2)$ will be denoted by $Q^{+}(5,2)$. Hence our point ${\mathcal P}$ is lying on a hyperbolic quadric in $PG(5,2)$. In the following we refer to this quadric as the Klein Quadric. (See Figure 2.) 

The six-dimensional vector space $V(6,2)$ can also be identified with $\wedge^2 V$ where $V\equiv V(4,2)$. In this representation $\mathcal{P}$ can be regarded as an element of
$\wedge^2 V$ which can be written as
\beq
\mathcal{P}=\mathcal{P}_{01}e_0\wedge e_1+\dots+ \mathcal{P}_{23}e_2\wedge e_3=v\wedge v^{\prime}.
\label{irottpe}
\eeq
\noindent
Recall in this respect that $\mathcal{P}$ is satisfying the (\ref{hquad}) Pl\"ucker relations if and only if $\mathcal{P}=u\wedge v$ 
for some $u,v\in V$, i.e. iff $\mathcal{P}$ is a separable bivector\cite{Hodge}.

Since the left hand side of Eq.(\ref{hquad}) is precisely of the (\ref{kvad}) form with $n=3$, 
then to the corresponding quadratic form one can associate the usual symplectic form of Eq.(\ref{symp}).
Explicitely we have
\beq
\langle\mathcal{P},\mathcal{P}^{\prime}\rangle=
{\mathcal P}_{01}{\mathcal P}_{23}^{\prime}+
{\mathcal P}_{23}{\mathcal P}_{01}^{\prime}+
{\mathcal P}_{02}{\mathcal P}_{13}^{\prime}+
{\mathcal P}_{13}{\mathcal P}_{02}^{\prime}+
{\mathcal P}_{03}{\mathcal P}_{12}^{\prime}+
{\mathcal P}_{12}{\mathcal P}_{03}^{\prime}
\label{szimpla2}
\eeq
\noindent
Then using the (\ref{op}) correspondence the symplectic vector space $(V(6,2),\langle\cdot,\cdot\rangle)$ can be used\footnote{For simplicity by an abuse of notation we denote the symplectic forms on {\it both spaces} $V(4,2)$ and $V(6,2)$ by the same symbol $\langle\cdot,\cdot\rangle$.} as a one representing a {\it special class} of nontrivial three-qubit observables as special points of $PG(5,2)$. 
Indeed, by virtue of (\ref{pelda3}), (\ref{pluck}) and (\ref{hquad}) the points lying on the Klein Quadric correspond to observables which are represented by {\it symmetric $8\times 8$} matrices.
The upshot of these considerations is that we have a bijective correspondece between lines ${\ell}$ of $PG(3,2)$ and points ${\mathcal P}$ lying on $Q^+(5,2)\subset PG(5,2)$. Moreover, at the level of observables this implies that we have a correspondence between $35$ triples of two-qubit observables and
the $35$ nontrivial symmetric three-qubit ones.

For example a simple calculation featuring $v\leftrightarrow (0111)$ and $v^{\prime}\leftrightarrow (1110)$, 
$\ell_1\leftrightarrow{\rm span}\{v,v^{\prime}\}$, $\mathcal{P}_1=v\wedge v^{\prime}$  shows that 
\beq
(\mathcal{O}_v,\mathcal{O}_{v^{\prime}},\mathcal{O}_{v+v^{\prime}})=(XY,YZ,ZX)\leftrightarrow \mathcal{O}_{v\wedge v^{\prime}}= YYZ
\nonumber
\eeq
\noindent
or with $u\leftrightarrow (0001)$ and $u^{\prime}\leftrightarrow (0101)$, $\ell_2\leftrightarrow{\rm span}\{ u,u^{\prime}\}$, $\mathcal{P}_2=u\wedge u^{\prime}$
\beq
(\mathcal{O}_u,\mathcal{O}_{u^{\prime}},\mathcal{O}_{u+u^{\prime}})=(IX,IY,IZ)\leftrightarrow \mathcal{O}_{u\wedge u^{\prime}}= IXI
\nonumber
\eeq
\noindent
where $\ell_1$ is an isotropic line and $\ell_2$ is a non-isotropic one. The corresponding observables $\mathcal{O}_{\mathcal{P}_1}$
and $\mathcal{O}_{\mathcal{P}_2}$ are symmetric.
The detailed dictionary can be found in Appendix A.

The lines represented in the $(Q\vert P)$ form can be partitioned into two classes depending on whether ${\rm Det}Q=1$ or $0$.
An equivalent representative for lines of the first class can be given the form: $(I\vert A)=(I\vert Q^{-1}P)$. Lines of the second class will be called {\it lines at infinity}. The simplest example of a line at infinity (called the {\it distinguished line at infinity}) is the isotropic one with representative $(0\vert I)$, which corresponds to the mutually commuting triple of observables $(XI,IX,XX)$.
One can check that a line is at infinity precisely when it has nonzero intersection with this distinguished one.

Using Eq.(\ref{szimpla2}) one can check that for two lines in the first class, i.e. ones of the form $(I\vert A)$ with Pl\"ucker coordinates $(1,a_{21},a_{22}, {\rm Det}A, a_{12},a_{11})$ and $(I\vert A^{\prime})$
with Pl\"ucker coordinates $(1,a^{\prime}_{21},a^{\prime}_{22}, {\rm Det}A^{\prime}, a^{\prime}_{12},a^{\prime}_{11})$ we have
\beq
\langle\mathcal{P},\mathcal{P}^{\prime}\rangle ={\rm Det}(A-A^{\prime}).
\label{mink}
\eeq
\noindent
One can also show that
\beq
\mathcal{P}\wedge\mathcal{P}^{\prime}=\langle\mathcal{P},\mathcal{P}^{\prime}\rangle e_0\wedge e_1\wedge e_2\wedge e_3.
\label{visszateres}
\eeq
\noindent
Note that being an element of the Klein Quadric, we have $\mathcal{P}=u\wedge v$ for some linearly independent vectors $u,v\in V$ with $\ell\leftrightarrow{\rm span}\{ u,v\}$.
As a result of this $\mathcal{P}\wedge\mathcal{P}^{\prime}=0$ iff the corresponding lines $\ell$ and $\ell^{\prime}$ are identical or intersecting in a point. (See Figure 2.) For $\mathcal{P}\wedge\mathcal{P}^{\prime}\neq 0$ iff the lines are concurrent.

Notice now that for $2\times 2$ matrices with {\it complex elements} satisfying the {\it reality constraint} $A=A^\dagger$ the 
right hand side of Eq.(\ref{mink}) gives rise to the Minkowski separation of space-time events represented by two four-vectors with coordinates $(a_0,a_1,a_2,a_3)$ and $(a^{\prime}_0,a^{\prime}_1,a^{\prime}_2,a^{\prime}_3)$, where
$a_{11}=a_0+a_3$, $a_{12}=a_1-ia_2$, $a_{21}=a_1+ia_2$, and $a_{22}=a_0-a_3$.

By analogy we call the points $\mathcal{P}$ and $\mathcal{P}^{\prime}$ on the Klein Quadric {\it light-like separated} if the left hand side of (\ref{mink}) equals zero, and {\it not light like separated} when it equals one.
We adopt this terminology for {\it all} $35$ points of the Klein Quadric which are representatives of the $35$ lines of $PG(3,2)$.
Then according to Eq.(\ref{mink}) for intersecting lines $\ell$, $\ell^{\prime}$ the corresponding points $\mathcal{P}$, $\mathcal{P}^{\prime}$ are light-like separated. Concurrent lines give rise to non-light like separation (see Figure 2.). 
These observations hint at an analogy with a finite geometric version of twistor theory. Indeed, this is the analogy which will enable us to regard the Klein quadric as a finite geometric model of space-time.

\subsection{An analogy with twistor theory}

In twistor theory\cite{Penrose1} one works over the field of complex numbers.  The lines of the form $(I\vert A)$ also satisfying
the reality constraint $A=A^\dagger$, via the Klein Correspondence, give rise to points of {\it real Minkowski space-time}.
By analogy when working over the field $GF(2)$,  lines of the form $(I\vert A)$ also satisfying
the special constraint $A=A^T$ under the Klein Correspondence give rise to the points of an object that will be called as a  "quantum space-time structure" over $GF(2)$. 
This structure of course has nothing to do with a discretized version of physical space-time.
This is just the $GF(2)$ version of a structure which has already appeared in the literature precisely under this name\cite{Brody1}.
However, reversing the philosophy of twistor theory, in the following we will regard this object as an emerging (bulk) "space-time" structure.

The meaning of the $A=A^T$ constraint is easy to clarify. One can check that isotropic lines of the form $(P\vert Q)$ are precisely the ones satisfying the constraint $PQ^T=QP^T$, hence for lines belonging to the first class (having the form $(I\vert A)$) we have $A=A^T$. 
Since the 
$PQ^T=QP^T$
constraint also works for lines at infinity, it is worth adopting this as a finite geometric analogue of the generalized reality condition\footnote{Unlike in our finite geometric setting where we use a symplectic polarity based on the symplectic form $\langle\cdot,\cdot\rangle$, in the complex setting of twistor theory a Hermitian polarity is used with the corresponding form having signature $(2,2)$. Hence unlike our constraint $PQ^T=QP^T$, in twistor theory its Hermitian analogue (also featuring complex conjugation) is used\cite{Penrose1}.}  used in twistor theory\cite{Penrose1}. Since in twistor theory under the Klein Correspondence inclusion of lines at infinity corresponds to taking the {\it conformal compactification} of Minkowski space-time, by analogy we arrive at the interpretation: the isotropic lines of $PG(3,2)$ correspond to points of a $GF(2)$ analogue of conformally compactified Minkowski space-time, living as a subset inside the Klein Quadric (see Figure 3).    

\begin{figure}[pth!]
\centerline{\includegraphics[width=12truecm,clip=]{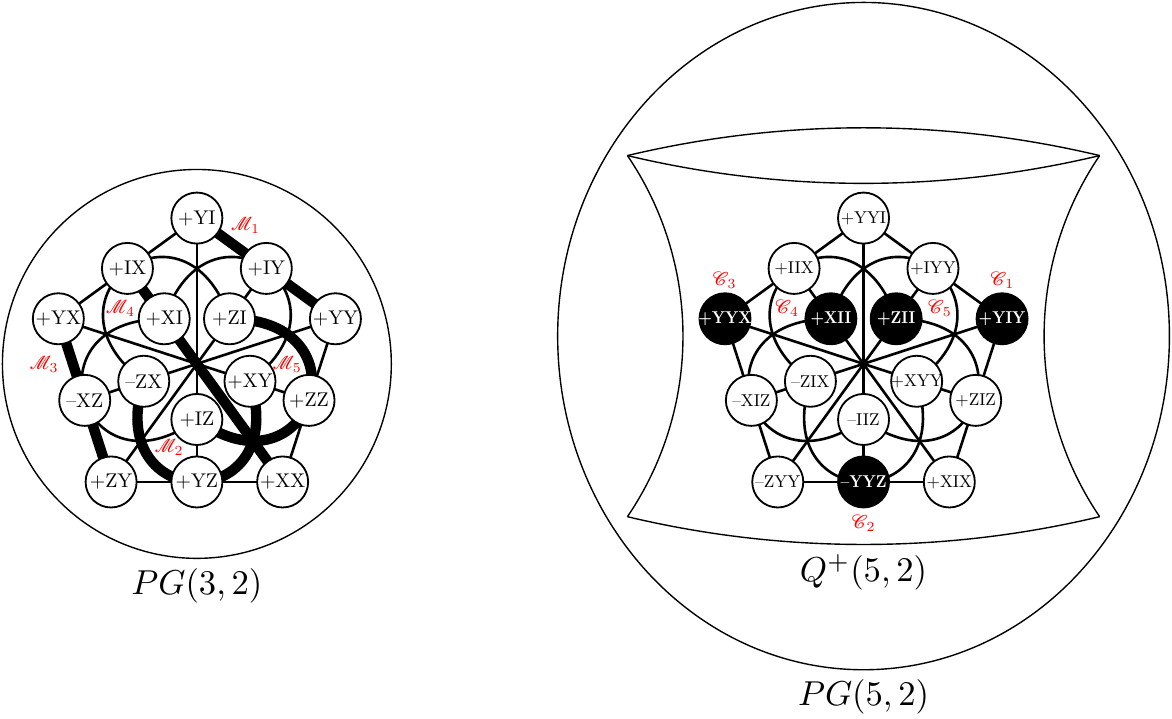}}
\caption{Codewords and the Klein Correspondence.
Under this correspondence the $15$ isotropic lines of $PG(3,2)$ are mapped bijectively to $15$ points, forming a subset, of the Klein Quadric: $Q^+(5,2)$. The collection of these points is defined by the hyperplane section given by Eq.(\ref{laggrass}), and has the interpretation as
 the $GF(2)$ analogue of conformally compactified Minkowski space-time known from twistor theory\cite{Penrose1}.
 The totality of $35$ points of the Klein Quadric is the $GF(2)$ version of {\it complexified} conformally compactified Minkowski space-time, an object we refer to as the "bulk".
On the other hand $PG(3,2)$ is the $GF(2)$ analogue of projective twistor space. It has the physical interpretation as the blow up of the projectivization of the Gibbons-Hoffman-Wootters discrete phase space\cite{W2} for two qubits, and will be referred to as the "boundary".
The $5$ message words $\mathcal{M}_a, a=1,2,3,4,5$  are forming a special subset of $5$ isotropic lines in the boundary: an {\it isotropic spread}. These correspond to $5$ special points associated with the $5$ codewords $\mathcal{C}_a, a=1,2,3,4,5$ encoding the messages of the boundary. They are forming an {\it ovoid} in the bulk.
The signs showing up in the figure are generated by Eqs.(\ref{atteres}) and (\ref{kakukk}).}
\end{figure}

At the level of Pl\"ucker coordinates the meaning of this constraint is as follows.  
Let us consider the (\ref{arrange}) arrangement taken together with the constraint $PQ^T=QP^T$.
Then a calculation of the Pl\"ucker coordinates shows that for isotropic lines the relation
\beq
\mathcal{P}_{02}=\mathcal{P}_{13}
\label{laggrass}
\eeq
\noindent
holds.
Using (\ref{pluck}) we see that triples of commuting operators on the boundary correspond to symmetric operators of the form
$\cdot Y\cdot$ or $\cdot I\cdot$ in the bulk, i.e. three-qubit ones for which the  middle slot is either $Y$ or $I$. 
The $15$ isotropic lines in $PG(3,2)$ form a special subset of the Grassmannian of lines $\mathcal{G}(1,3)$: the {\it Lagrangian Grassmannian} $\mathcal{LG}(1,3)$. Under the Klein Correpondence the lines of $\mathcal{LG}(1,3)$
are represented by those points of $Q^+(5,2)$ which are also lying on the (\ref{laggrass}) hyperplane in $PG(5,2)$.  

In the following we will refer to $PG(3,2)$ as the "boundary" and $Q^+(5,2)$ aka Klein Quadric as the "bulk".
In twistor theory language $PG(3,2)$ is the $GF(2)$-version of projective twistor space\cite{Penrose1,Brody1}.
In the next section we also relate the "boundary" to the projectivization of the Gibbons-Hoffman-Wotters's (GHW) discretized phase space for two-qubits\cite{W2}.
The {\it lines} comprising the Lagrangian Grassmannian living in the boundary, correspond to 
the {\it points} of a $GF(2)$-analogue of conformally compactified Minkowski space-time living in the bulk.
This substructure living inside the bulk is just a new copy of the doily
 (See Figure 3.). Moreover, precisely as in twistor theory: intersecting lines in the "boundary", correspond to light-like separated points in the "bulk".
However, unlike in twistor theory here (thanks to the rule of Eq.(\ref{op})) there is  also a correspondence between two-qubit observables in the "boundary" and certain three-qubit ones in the "bulk".
Notice, that the (\ref{op}) rule works merely {\it up to sign}. We will have something important to say in connection with this later.

However, our bulk is more then  a $GF(2)$-analogue of conformally compactified Minkowski space-time.
In the following we argue that the bulk is a $GF(2)$-analogue of conformally compactified 
{\it complexified} Minkowski space-time of twistor theory.
 
As a first step seeing this note that in twistor theory\cite{Penrose1} one can regard the fully complexified version $M^{\ast}$ of conformally compactified Minkowski spacetime $M$ as the Klein representation of lines in $PG(3,\mathbb{C})$. 
Moreover, the complexified null lines of $M$ correspond to the points of $PG(3,\mathbb{C})$.
Geometrically the complexified null lines are pairs of complex planes, one $\alpha$-plane and one $\beta$-plane\cite{Penrose1}.
The operation of complex conjugation in $M^{\ast}$,
which leaves the real space $M$ invariant, in the $PG(3,\mathbb{C})$ picture corresponds to the action of a Hermitian
polarity (See footnote 2.) It is also known that this polarity corresponds to\cite{Penrose1} a point$\leftrightarrow$ plane association in $PG(3,\mathbb{C})$.  

One can easily demonstrate how the corresponding structures show up in our $GF(2)$ case. First of all, the Klein Quadric $Q^+(5,2)$ indeed serves as the Klein representation of lines in $PG(3,2)$.
A null line residing inside the doily in the bulk is lying in the intersection of an $\alpha$ and a $\beta$-plane.
For example the null line $\{ZIZ,ZII,IIZ\}$ is lying at the intersection of the planes $\{ZZZ,IZZ,IZI,ZZI,ZIZ,ZII,IIZ\}$
and$\{ZXZ,IXZ,IXI,ZXI,ZIZ,ZII,IIZ\}$. (See Appendix A.)
However, instead of the Hermitian polarity in our case we have the symplectic polarity. Then the point$\leftrightarrow$ plane association is the one that works at the level of observables by associating to an observable (point) the set of observables commuting with it (plane). For example, under the operation of "conjugation" the observable $XX$ is associated to the set of observables $\{XX,IX,XI,ZY,YZ,YY,ZZ\}$. 
At the level of the bulk, this conjugation gives rise to the exchange between the observables associated with $\alpha$-planes and $\beta$-planes.
Since these planes share an isotropic line (for an illustration of this see Figure 4.), under conjugation this line
is left invariant. This invariant set of isotropic lines is precisely the Lagrangian Grassmannian $\mathcal{LG}(1,3)$.
\begin{figure}[pth!]
\centerline{\includegraphics[width=8truecm,clip=]{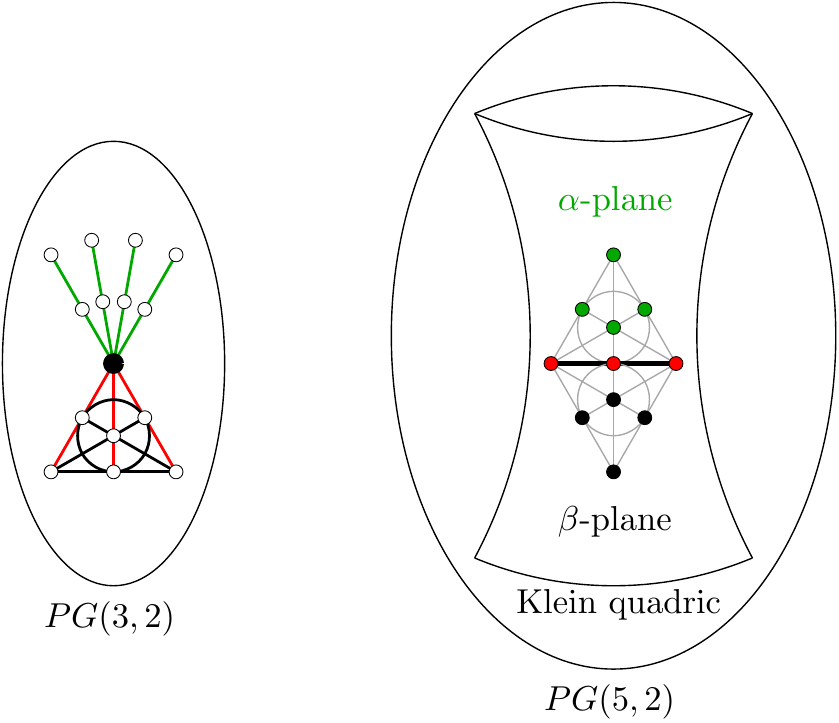}}
\caption{The intersection of $\alpha$ and $\beta$ planes defines isotropic lines in the bulk.}
\end{figure}
Since being invariant under conjugation means "real" in this context, for the finite geometric analogue of $M$ what we get is the set of $15$ isotropic lines on $15$ points, i.e the incidence structure of the doily (Figure 3.). 
We remark that on the $35$ symmetric three-qubit observables one can explicitely construct the action of an unitary operator, which is implementing this operation of conjugation. For the details see Appendix A.

In summary, we have shown that one can reinterpret the $GF(2)$-version of the Klein Correspondence as a one relating
two qubit observables on one side with an underlying finite geometric structure (boundary), and a special set of three-qubit ones with an underlying finite geometric structure (bulk). The boundary is just the $GF(2)$ version of twistor space, and the bulk is the $GF(2)$ analogue of complexified conformally compactified Minkowski space-time.
In twistor theory physical data concerning space-time are reformulated in terms of data of twistor space. 
In our model this philosophy is reversed: from the  data of the $GF(2)$ version of twistor space emerges the $GF(2)$ version of space-time data. In the next section we show that the boundary data on observables is encoded into the bulk data by
an error correcting code: a geometric subspace code.

\subsection{Encoding the boundary into the bulk}

Our mapping of boundary observables into bulk ones can be used to interpret the bulk as an emerging object implementing naturally an error correcting code, with message data residing at the boundary.
In our very special case the message "word" to be "sent" is an element of an initially fixed {\it spread of observables}  partitioning the $15$ nontrivial observables located at the boundary. 
By a {\it spread of observables} we mean a set of mutually commuting triples of observables such that every observable belongs exactly to one of such triples.
This notion is coming from the finite geometric one\cite{Demb} of the {\it spread of isotropic lines}, which is a collection of isotropic lines in $PG(3,2)$
such that every point belongs exactly to one of such lines.
In our case a spread of observables is consisting of $5$ triples. A particular spread to be used below, highlighted by shaded lines inside the doily, is depicted on the left hand side of Figure 3.
Spread codes have been used as subspace codes for network coding\cite{Gorla}.
In the following we reinterpret them as a method for encoding boundary observables into bulk ones.

Our spread of isotropic lines defines a set of constant dimension subspaces in $V$, hence defines a subspace code.
The spread is consisiting of $5$ lines, these are the $5$ possible words containing the message.  
According to Eq.(\ref{dist}) this spread of lines realizes the maximal possible distance (i.e. $4$) for the  words.
Now in terms of finite geometry: the sent data is a line, the recieved data is a point (one error down in dimension), or a plane (one error up in dimension). 
In terms of observables: the sent data is a triple of commuting observables, the received data is either just a single observable, or a seven-tuple of observables commuting with a fixed particular one (forming the building blocks of a generalized X-state\cite{Rau}).

The boundary message (a particular line of the spread)  is encoded into the bulk in the form of a bulk space-time
point as shown in Table 1 and Figure 3. Notice that to a boundary message a quantum state can be associated in a unique manner. 
These states are stabilized by the correponding message observables.
Using the notation
\beq
\vert\overline{0}\rangle=\frac{1}{\sqrt{2}}(\vert 0\rangle +\vert 1\rangle),\qquad
\vert\overline{1}\rangle=\frac{1}{\sqrt{2}}(\vert 0\rangle -\vert 1\rangle),\qquad
\label{underline}
\eeq
\noindent
\beq
\vert\tilde{0}\rangle=\frac{1}{\sqrt{2}}(\vert 0\rangle +i\vert 1\rangle),\qquad
\vert\tilde{1}\rangle=\frac{1}{\sqrt{2}}(\vert 0\rangle -i\vert 1\rangle).\qquad
\label{tilde}
\eeq
\noindent
these stabilized states can be written in a simple manner as can be seen in Table 1.

For the $5$ lines of the spread one can bijectively associate $5$
points (the ones with red labels of Figure 3).
From these labels one can immediately verify that the $5$ bulk three-qubit observables, corresponding to the $5$ elements of the  boundary spread of observables, are pairwise anticommuting, i.e. they form a five dimensional Clifford algebra. In twistor geomerical terms: the $5$ space-time points representing the boundary message words are pairwise non light-like separated. 

Notice also that in the bulk all the representatives of the errors, namely points and planes in the boundary, are planes (the $\alpha$ and $\beta$ planes) that are isotropic with respect to the bulk symplectic form. This can be checked by inspection in Appendix A by observing that all such planes are labelled by seven-tuples of mutually commuting observables. Moreover, they are maximal totally isotropic subspaces of the embedding space $PG(5,2)$ equipped with this simplectic form.
More importantly consulting Appendix A, one can also verify that every maximal totally isotropic subspace lying in the bulk {\it contains precisely one point} of the $5$ special space-time points encoding the boundary message.
In finite geometric terms our $5$ bulk points form an {\it ovoid}\cite{Coss1}. 
This ovoid property makes it possible to use the bulk spacetime as an error correcting code in the following manner. 

\begin{table}[h!]
\centering
\begin{tabular}{|c|c|c|c|c|}
\hline Name& Boundary Message & Stabilized state & Name & Bulk Code
\\ \hline \hline
$\mathcal{M}_1$ & $IY$,$YI$,$YY$ & $\vert\psi_1\rangle=\vert\tilde{0}\tilde{0}\rangle$ & $\mathcal{C}_1$ & $YIY$\\
$\mathcal{M}_2$ & $YZ,XY,-ZX$ & $\vert\psi_2\rangle=
\frac{1}{\sqrt{2}}\left(\vert \tilde{0}0\rangle -\vert \tilde{1}1\rangle\right)$ & $\mathcal{C}_2$ & $-YYZ$\\
$\mathcal{M}_3$ &$YX,ZY,-XZ$& $\vert\psi_3\rangle=\frac{1}{\sqrt{2}}\left(\vert 0\tilde{0}\rangle -\vert 1\tilde{1}\rangle\right)$ & $\mathcal{C}_3$ & $YYX$\\
$\mathcal{M}_4$ &$XI,XX,IX$& $
\vert\psi_4\rangle=\vert\overline{0}\overline{0}\rangle$ & $\mathcal{C}_4$ & $XII$\\
$\mathcal{M}_5$ &$IZ,ZZ,ZI$& $
\vert\psi_5\rangle =\vert 00\rangle$ & $\mathcal{C}_5$ & $ZII$\\
\hline
\end{tabular}
\caption{Boundary-Bulk encoding for two qubits. For an explanation for our choices of signs see Sec. 3.7. and Eq.(\ref{kakukk}).
For a pictorial representation of an alternative, rotationally covariant encoding see Figure 7.}
\label{tab:1}
\end{table}

\subsection{Error correction}

Suppose that $5$ message words of the boundary are encoded into $5$ codewords of the bulk by this method. Hence we know in advance that the bulk codewords correspond to the observables: $YYX,YYZ,YIY,ZII,XII$.
Suppose now that one of the message words $\{ZI,IZ,ZZ\}$ is sent, but due to error what we get is $ZZ$. There are three possible isotropic lines containing $ZZ$: which one was the message?
As a first step we send the corrupted information (a point corresponding to $ZZ$) of the boundary to the 
bulk via the Pl\"ucker map. What we get is the $\alpha$-plane corresponding to $\{ZXZ,ZZX,IZX,IXZ,IYY,ZYY,ZII\}$.
By the ovoid property of the codewords we know that precisely one of them should show up in this $\alpha$-plane. As a second step we identify it: it is $ZII$. Knowing that the method of coding is the Pl\"ucker map as the third step one identifies the inverse image of this bulk point, 
namely the boundary triple $\{ZI,IZ,ZZ\}$, which was the original message.  

Dually, let us suppose that the same message $\{ZI,IZ,ZZ\}$ have been sent but due to an error what we get is: $\{XX,YY,XY,YX,IZ,ZI,ZZ\}$. Note that this is the plane dual to the point $ZZ$. 
The image of this error is the $\beta$-plane $\{ZZZ,ZXX,IXX,IZZ,IYY,ZYY,ZII\}$.
This is the conjugate plane of the $\alpha$ plane of the previous paragraph.
Again: our $\beta$-plane contains precisely one codeword: it is again $ZII$, which identifies the same message. 

It is useful to illustrate the geometrical meaning of this error correction process.
Consider first the case when the errors are points. In our example starting with the message $\{IZ,ZI,ZZ\}$, there are three different possible errors, corresponding to the three points $P_1,P_2,P_3\leftrightarrow IZ,ZI,ZZ$ on the boundary.
These points are represented by three $\alpha$-planes of the bulk.  According to the results of the previous subsection these planes represent complexified light rays meeting in a point which is precisely our codeword $\mathcal{C}_5$.
In the boundary through each "error point" there are seven lines, three of them are isotropic ones. One of them is just the line corresponding to the message. Since in the boundary they are intersecting in the same point, in the bulk they constitute three points of a light ray going through $\mathcal{C}_5$. 
For the three possible errors there are three such light rays all of them going through $\mathcal{C}_5$. They are lying inside the corresponding complexified light rays. Hence the possible point errors of a message located in the boundary are represented by the
{\it ligh cone} of the corresponding codeword in the bulk (Figure 5.).

\begin{figure}[pth!]

\centerline{\includegraphics[width=12truecm,clip=]{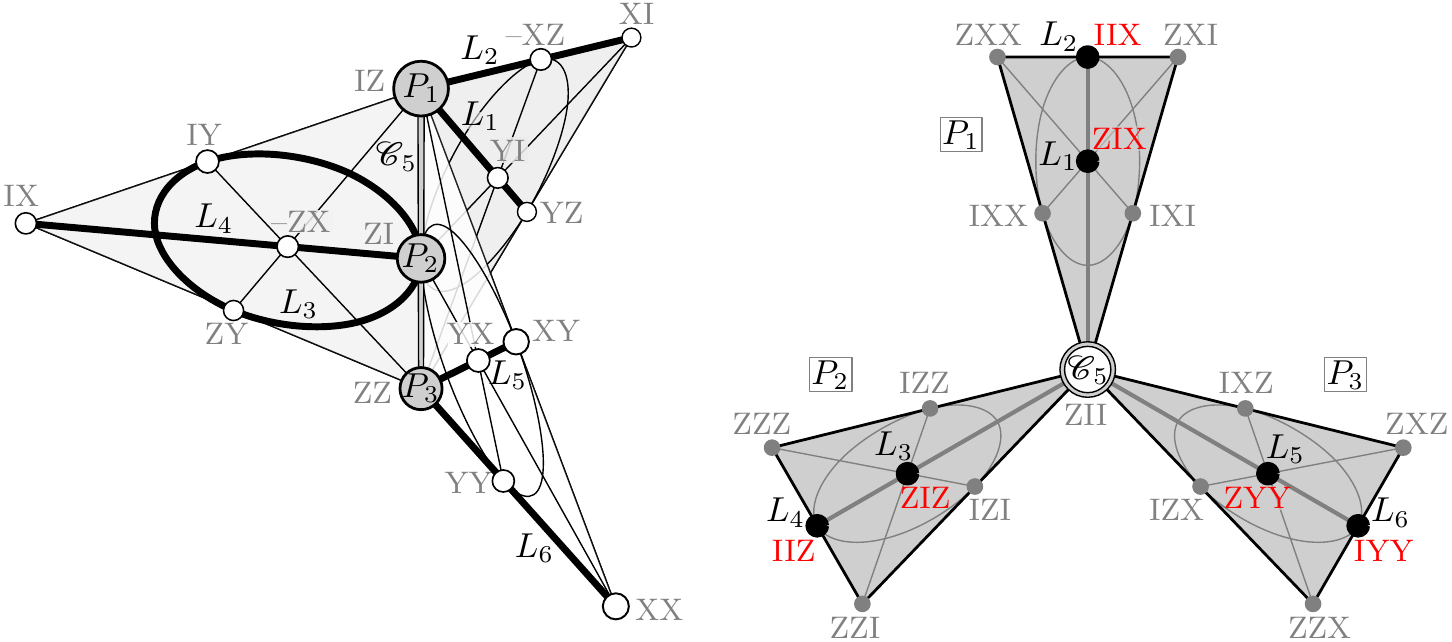}}
\caption{Point errors of a message located in the boundary represented by the ligh cone structure of the corresponding codeword in the bulk. For illustrative purposes by an abuse of notation the objects corresponding to each other in the boundary (left) and in the bulk (right) are denoted by the same letters. For example both the boundary message (left) and its bulk codeword (right) is denoted by the same symbol: $\mathcal{C}_5$. Similarly the three error points (left), and their bulk representatives as planes are denoted by the same symbols: $P_1,P_2,P_3$. The isotropic lines $L_j, j=1,2,\dots 6$ meeting in the error points (left), are corresponding to the points $L_j$ constituting the light cone of $\mathcal{C}_5$ (right). The slight distortion of planes on the right, represents the fact that they are really complexified light rays, containing the ordinary ones.}
\end{figure}

In the second case the errors are planes. Then we have three possibilities for these planes intersecting in the message line $\{IZ,ZI,ZZ\}$.
The error planes contain $3$ isotropic and $4$ non-isotropic lines (see the left hand side of Figure 6).
We emphasize that the light cone structure corresponding to this situation in the bulk is the same as it was in the case of point errors. However, the {\it complexified} light cone structures are different. Indeed, for point errors we obtain three $\alpha$-planes $P_1,P_2,P_3$ (see the right of Figure 5) and for plane errors three $\beta$-planes
$\Pi_1,\Pi_2,\Pi_3$ (see the right of Figure 6). As in Figure 5 the planes $\Pi_1,\Pi_2,\Pi_3$ of Figure 6 (right) are intersecting in the {\it same point}. Since the $\alpha$ planes are related to the $\beta$ ones via conjugation their labels are related by an
$X\leftrightarrow Z$ flip in the middle qubit.
Recall, that in the bulk, the intersections $P_i\cap\Pi_i$ of the  $\alpha$-planes $P_i$ and the $\beta$-planes $\Pi_i$ 
for $i=1,2,3$ are the light rays through the codeword $\mathcal{C}_5$, see also Figure 4.

\begin{figure}[pth!]
\centerline{\includegraphics[width=12truecm,clip=]{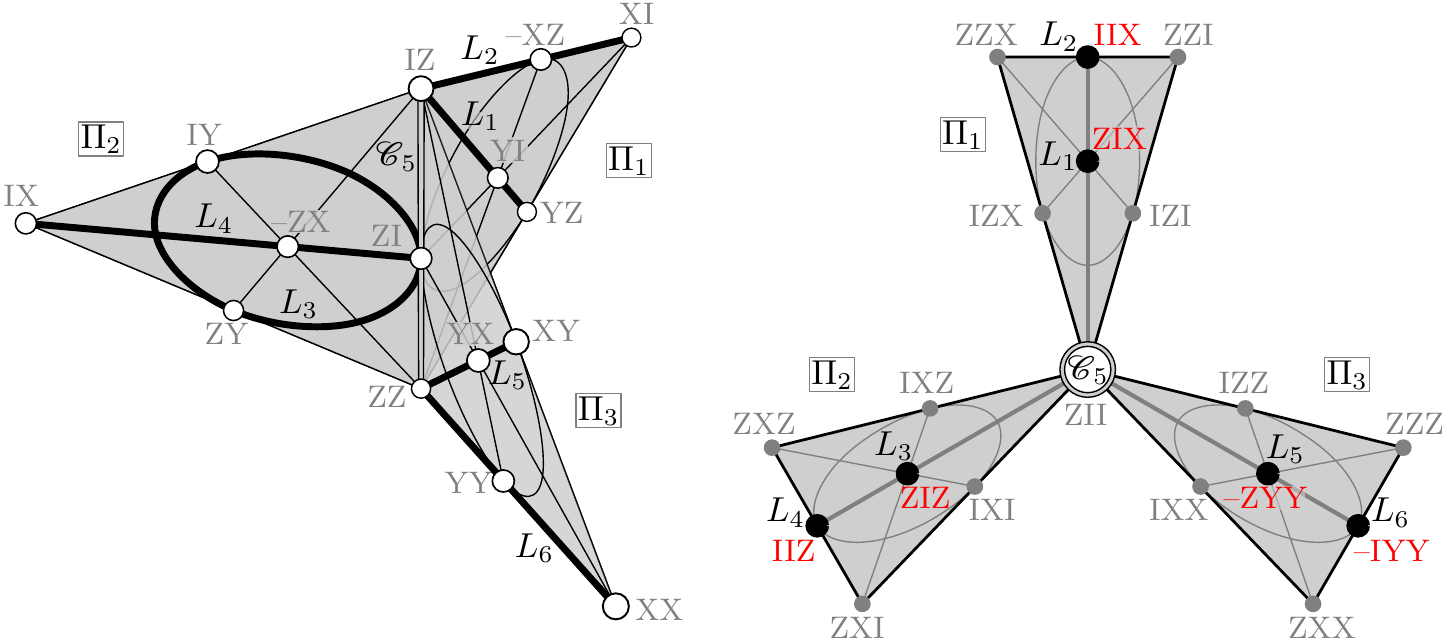}}
\caption{Plane errors of a message located in the boundary represented by the ligh cone structure of the corresponding codeword in the bulk. 
These plane errors $\Pi_1,\Pi_2,\Pi_3$ (left) are intersecting in the message line. Each of them contains $3$ isotropic lines (shaded in black) and $4$ non-isotropic ones. They are mapped to the seven points of the corresponding planes (right). 
We emphasize that the light cone structure (right) is the same as the one in Figure 5.
However, the {\it complexified} light cones,  the $\alpha$ planes $P_1,P_2,P_3$ (right) of Figure 5., and the $\beta$-planes $\Pi_1,\Pi_2,\Pi_3$ (right) of Figure 6. are different. The difference manifests itself in different labels for the remaining points of these planes. Notice that for the totally isotropic $\beta$-planes in the bulk we have chosen positive ones (see Section 3.7). This accounts for the negative signs showing up in $\Pi_3$. However, as will be shown later, no consistent choices of signs on the right hand side rendering all $\beta$-planes corresponding to all of our codewords is possible.}
\end{figure}

\subsection{Algebraic description of the error correction process}

In order to be able to generalize our considerations for an arbitrary number of qubits, we also need an algebraic characterization of our error correction method.
In Appendix A by calculating Pl\"ucker coordinates, we have presented a detailed dictionary of observables on both sides of the correpondence.
This very explicit structure was useful for illuminating the basic geometric structures involved, but it is of limited value for generalization. 
Luckily there is also a nice algebraic characterization of the decoding process\cite{Stokes} without passing to Pl\"ucker coordinates.
In the following we reformulate and develop the results of Ref.\cite{Stokes} convenient for our purposes.

First recall that isotropic lines in the boundary, corresponding to message words, are encoded into codewords in the bulk via the (\ref{laggrass}) constraint. This means that the {\it symmetric} bulk observables encoding boundary messages are commuting with the special {\it antisymmetric} observable $IYI$.
This reduces the number of $35$ bulk observables to $15$ ones. In order to reduce this number to $5$, picking out our codewords $\{YIY,YYZ,YYX,XII,ZII\}$, we need further restrictions.

One can notice that if we choose either of the antisymmetric observables $IZY, IXY$ as an extra one, these ones will commute merely with our codewords. One can also notice that the set $\{IYI,IZY,IXY\}$ corresponds to a non-isotropic line $\ell_{\ast}$ of $PG(5,2)$ {\it off} the bulk.
Hence we have a line $\ell_{\ast}$ which does not intersect the bulk. Let us now consider the subspace $\ell_{\ast}^{\perp}$, where the orthogonal complement is meant with respect to the symplectic form $\langle\cdot,\cdot\rangle$ of $PG(5,2)$.
Physically this is the set of {\it all} three-qubit observables, modelled by $PG(5,2)$,  commuting with the triple $\{IYI,IZY,IXY\}$. Since from the triple only two observables are independent, this means that we have two constraints on the six component vectors taken from $V(6,2)$, representing three-qubit observables. Hence $\ell_{\ast}^{\perp}$ is a rank four subspace, projectively a subspace of dimension three, i.e. $\Delta\equiv\ell_{\ast}^{\perp}$ is a copy of $PG(3,2)$. 
This projective subspace of dimension three is intersecting our bulk quadric precisely in our five codewords:
\beq
 \{\mathcal{C}_1,\mathcal{C}_2,\mathcal{C}_3,\mathcal{C}_4,\mathcal{C}_5\}=\Delta\cap Q^+(5,2). 
\label{intersect}
\eeq
\noindent

Let us formalize this in terms of $PG(5,2)$ data.
The codewords are a collection of special points in $PG(5,2)$. Let us denote this collection of points collectively by the homogeneous coordinates arranged in a {\it column vector} 
\beq
\mathcal{C}^t=(\xi_0,\xi_1,\xi_2,\eta_0,\eta_1,\eta_2).
\eeq
\noindent
Let us also introduce two more vectors
\beq
\Gamma^t=(0,1,0,0,1,0),\qquad \mathcal{R}^t=(r_0,r_1,r_2,s_0,s_1,s_2).
\label{recov}
\eeq
\noindent
The first of these vectors corresponds to the three-qubit observable $IYI$, and the second vector will be called the {\it recovery} vector.
In our special case of the codewords of Table 1 $\mathcal{R}^T=(0,1,1,0,0,1))$ corresponds to  $IZY$.
However, since later we would like to describe a {\it collection of codes} we regard $\mathcal{R}$ as a collection of vectors (to be specified in the next subsection) rather than a particular vector.
Clearly for each of our special codewords showing up in Table 1 we have
\beq
\langle\Gamma,\mathcal{C}\rangle=
\langle\mathcal{R},\mathcal{C}\rangle=0,\qquad \langle \Gamma,\mathcal{R}\rangle =1,\qquad \ell_{\ast}=\{\Gamma,\mathcal{R},\Gamma+\mathcal{R}\},\qquad\mathcal{Q}(\mathcal{C})=0
\label{important}
\eeq
\noindent
summarizing the fact that $\ell_{\ast}$ is a non-isotropic line of $PG(5,2)$ and $\mathcal{C}\in\ell_{\ast}^{\perp}\cap Q^+(5,2)$.

Now after these bulk related considerations consider the boundary.
Suppose we want to send the {\it message} $\mathcal{M}$, for example a one of Table 1.
Suppose further that this message is corrupted by a point error, 
hence what is "transmitted" is the fixed {\it error} vector $\mathcal{E}^T=(q_0,q_1,p_0,p_1)$.
This point can be on any of three possible isotropic lines. In order to find the message line we have to find at least one extra point on this line
\beq
\chi^T=(x_0,x_1,y_0,y_1).
\label{extra}
\eeq
The column vector $\chi$ refers to a collection of possible points collinear with $\mathcal{E}$.
By calculating the Pl\"ucker coordinates\footnote{For example for the line ${\rm span}\{\mathcal{E},\chi\}$ we have $\mathcal{P}^{(\mathcal{E},\chi)}_{02}=q_0y_0+p_0x_0$, etc.}
of the set of possible lines $\{\mathcal{E},\chi,\mathcal{E}+\chi\}$ then 
one gets a collection of points in the bulk: $\mathcal{P}^{(\mathcal{E},\chi)}$.
Now our geometric method of bulk encoding of boundary messages says that from the set of possible bulk points, the ones representing messages are satisfying Eqs.(\ref{important}).

The first constraint to be met in the bulk is Eq.(\ref{laggrass}), or equivalently $\langle\Gamma,\mathcal{P}^{(\mathcal{E},\chi)}\rangle=0$, which in boundary terms is just
$q_0y_0+q_1y_1+p_0x_0+p_1x_1=0$ i.e. $\langle\mathcal{E},{\chi}\rangle =0$, the condition of isotropy in the boundary.
The second one will be called the constraint of recovery. It is of the form:
$\langle\mathcal{R},\mathcal{P}^{(\mathcal{E},\chi)}\rangle =0$. 
Arranging the six components of $\mathcal{R}$ into a $4\times 4$ "antisymmetric" matrix $R$ over $GF(2)$,
and also introducing the matrix $J$ of the symplectic form
\beq
R\equiv \begin{pmatrix}0&s_0&s_1&s_2\\s_0&0&r_2&r_1\\s_1&r_2&0&r_0\\s_2&r_1&r_0&0\end{pmatrix},\qquad
J\equiv \begin{pmatrix}0&0&1&0\\0&0&0&1\\1&0&0&0\\0&1&0&0\end{pmatrix}
\label{antika}
\eeq
\noindent
these constraints can be written as
\beq
\mathcal{E}^TJ\chi=0,\qquad
\mathcal{E}^TR\chi=0.
\label{elegans}
\eeq
\noindent
These two equations describe two distinct planes intersecting in a line: precisely our message line.
In projective geometry the vectors
\beq
\mathcal{E}^{\ast}=J\mathcal{E},\qquad \mathcal{F}^{\ast}\equiv R\mathcal{E}
\label{dualsik}
\eeq
\noindent
are the coordinates of the intersecting planes. 
Notice that in these equations no Pl\"ucker coordinates show up. The only bulk related quantity is the recovery matrix $R$, on the other hand $\mathcal{E}$ and $\chi$ are boundary related. 

For plane errors we can use projective duality between points and planes. Note that a plane error is fixed by the seven points of the plane, described by the vectors $\chi$, satisfying $\langle\mathcal{E},\chi\rangle =0$.
This plane is determined by the fixed vector $\mathcal{E}$. This equation is of course just the first one of Eq.(\ref{elegans}), which after introducing the dual vector $\mathcal{E}^{\ast}$ gives back the usual description of a plane in projective geometry. 
Now in the case of point errors after recovery we obtained the message line as a one characterized by the two vectors: $\mathcal{E}^{\ast}$ and $\mathcal{F}^{\ast}$, hence using duality in our new case of plane errors we obtain the message line as the one characterized by the two vectors $\mathcal{E}$ and $\mathcal{F}$. Hence the message line is
\beq
\mathcal{M}=\{\mathcal{E},\mathcal{F},\mathcal{E}+\mathcal{F}\},\qquad \mathcal{F}=JR\mathcal{E}.
\label{nice}
\eeq
\noindent
Notice that since the matrices $R$ and $RJR$ are symmetric, and we are over $GF(2)$, the choice $\chi\equiv\mathcal{F}=JR\mathcal{E}$ explicitely solves Eqs.(\ref{elegans}), characterizing the recovery process for point errors. Hence the explicit (\ref{nice})
method for the recovery from plane errors is universal. It can also be used as a very simple algorithm for the recovery from errors of {\it both type}. 
Indeed, in our setting, one merely has to take care of recovery from one type of error. Recovery from the other type is automatically taken into account by projective duality and isotropicity of the message lines.

There is an ambiguity in this recovery process. According to the third formula of Eqs.(\ref{important}) we can also use a new matrix $R^{\prime}$ for recovery. Indeed, one can define $R^{\prime}$ as the one with the same entries as $R$ except for $r_1^{\prime}=r_1+1$ and $s_1^{\prime}=s_1+1$.
Then we have $JR+JR^{\prime}={\bf 1}$ ($4\times 4$ identity matrix), hence $\mathcal{F}^{\prime}\equiv JR^{\prime}\mathcal{E}=\mathcal{E}+\mathcal{F}$. Hence this ambiguity merely effects which of the two points on the message line of Eq.(\ref{nice}) we obtain. 

Appreciate the elegance of the mathematical representation of the (\ref{nice}) recovery process. The corrupted {\it boundary} data $\mathcal{E}$ is linearly transformed into the message data, via the calculation of the additional boundary data $\mathcal{F}=JR\mathcal{E}$. The recovery is due to the {\it bulk} related matrix $R$. However, the relationship between the structure of $R$ and our codewords residing in the bulk needs further elaboration. Furthermore, it would be also desirable to clarify the physical meaning of the recovery matrix. 

\subsection{The meaning of the recovery matrix}

In order to learn more about the role played by the recovery matrix $R$ of Eq.({\ref{antika}) in our story, it is worth considering instead of a {\it single} error correcting code the set of {\it all possible codes}. 
Since our code of Table 1 was based on a special isotropic spread of $PG(3,2)$, interpreted as a system of message words built from boundary observables, the set of all possible codes is just the set of all possible isotropic spreads of the boundary. In our case it is known that we have merely {\it six} isotropic spreads.

In Table 2 we listed all such spreads, not paying any attention this time to the signs of the corresponding observables.
The reader can recognize that the distinguished spread of Table 1, is the spread $\mathcal{S}_6$.
A pictorial representation for five of these spreads can be obtained by successive rotations by $72$ degrees of the pattern of the spread $\mathcal{S}_6$ contained in the doily of the left hand side of Figure 3. The remaining spread ($\mathcal{S}_2$) is just the one coming from the five "diagonal lines" of the doily.
The reader can convince themselves that this is indeed the full set of spreads, giving rise to our {\it system of message words}
$\mathcal{S}_j, j=1,2,\dots 6$. 

\begin{table}[h!]
\centering
\begin{tabular}{|c|c|c|c|c|c|}
\hline Spread& $\mathcal{M}_{j0}$ & $\mathcal{M}_{j1}$ & $\mathcal{M}_{j2}$ & $\mathcal{M}_{j3}$ & $\mathcal{M}_{j4}$
\\ \hline \hline
$\mathcal{S}_1$ & $IY,YI,YY$  & $ZY,XX,YZ$ & $ZX,ZI,IX$ & $XY,YX,ZZ$ & $XI,IZ,XZ$\\
$\mathcal{S}_2$ & $XY,YX,ZZ$  & $YI,IZ,YZ$ & $XI,IX,XX$ & $YY,XZ,ZX$ & $IY,ZY,ZI$\\
$\mathcal{S}_3$ & $YY,XZ,ZX$  & $IZ,ZZ,ZI$ & $ZY,XX,YZ$ & $IY,XY,XI$ &  $YX,YI,IX$\\
$\mathcal{S}_4$ & $XY,ZX,YZ$  & $XX,YY,ZZ$ & $YX,YI,IX$ & $IY,ZY,ZI$ & $XI,IZ,XZ$\\
$\mathcal{S}_5$ & $XZ,YX,ZY$ & $XX,YY,ZZ$ & $YI,IZ,YZ$ & $IX,ZI,ZX$ & $XI,IY,XY$\\
$\mathcal{S}_6$ & $IY,YI,YY$  & $YZ,XY,ZX$ & $YX,ZY,XZ$ & $XI,XX,IX$ & $IZ,ZZ,ZI$\\
\hline
\end{tabular}
\caption{The six system of message words $\mathcal{S}_j, j=1,2,\dots 6$ of the boundary. Finite geometrically they are isotropic spreads of lines of $PG(3,2)$.
Each spread is containing $5$ message words $\mathcal{M}_{ja}, a=1,\dots 5$.
A pictorial representation for five of these spreads can be obtained by successive rotations by $72$ degrees of the pattern of the spread $\mathcal{S}_6$ contained in the doily of the left hand side of Figure 3. The remaining spread ($\mathcal{S}_2$) is just the one coming from the five diagonal lines of the doily.
}
\label{tab:2}
\end{table}

The system of message words is mapped to the {\it system of code words} by the Pl\"ucker map. As we discussed, this mapping is a one relating spreads of the boundary to ovoids in the bulk. This "Grassmannian image" of the boundary spreads in the bulk can elegantly be described by the lines $\ell_{\ast}$ of Eqs.(\ref{important}) off the bulk.
In Table 3. one can find the system of code words, with the explicit structure of the associated lines.
One can realize that the set of recovery vectors $\mathcal{R}$ in the fourth column forms a sixi-dimensional Clifford algebra ${\rm Cliff}(6)$, explicitely we have
\beq
(\Gamma_1,\Gamma_2,\Gamma_3,\Gamma_4,\Gamma_5,\Gamma_6)\equiv (YXI,YZX,YZZ,XXY,ZXY,IZY),\qquad \{\Gamma_j,\Gamma_k\}=2\delta_{jk}{\bf 1}.
\label{cliff6}
\eeq
\noindent
In fact one can also take into account the fact that $\Gamma\equiv \Gamma_7$ anticommutes with all of these observables hence we have a seven-dimensional Clifford algebra ${\rm Cliff}(7)$.
Similarly the set of observables giving rise to the recovery vectors $\mathcal{R}^{\prime}$ in the fourth column forms another copy of a ${\rm Cliff}(6)$
\beq
(i\Gamma_{17},i\Gamma_{27},i\Gamma_{37},i\Gamma_{47},i\Gamma_{57},i\Gamma_{67})\equiv (-YZI,-YXX,YXZ,-XZY,-ZZY,IXY),
\label{cliff6v}
\eeq
\noindent
where $\Gamma_{I_1I_2I_3\dots}\equiv\Gamma_{I_1}\Gamma_{I_2}\Gamma_{I_3}\cdots, 1\leq I_1<I_3<I_3<\dots\leq 7$.
This configuration of two copies of six-dimensional Clifford algebras interchanged by $\Gamma$ in finite geometry is called a "double six".

\begin{table}[h!]
\centering
\begin{tabular}{|c|c|c|c|c|}
\hline Spread&Bulk code words $\mathcal{C}_{ja}$ & $\Gamma$ & $\mathcal{R}$ & $\mathcal{R}+\Gamma$
\\ \hline \hline
$\mathcal{S}_1$ & $YIY,XYY,IIZ,ZYY,IIX$  & $IYI$ & $YXI$ & $YZI$\\
$\mathcal{S}_2$ & $ZYY,ZIX,XII,YYI,ZIZ$  & $IYI$ & $XXY$ & $XZY$\\
$\mathcal{S}_3$ & $YYI,ZII,XYY,XIX,XIZ$  & $IYI$ & $ZXY$ & $ZZY$\\
$\mathcal{S}_4$ & $YYZ,IYY,XIZ,ZIZ,IIX$  & $IYI$ & $YZX$ & $YXX$\\
$\mathcal{S}_5$ & $YYX,IYY,ZIX,IIZ,XIX$  & $IYI$ & $YZZ$ & $YXZ$\\
$\mathcal{S}_6$ & $YIY,YYZ,YYX,XII,ZII$  & $IYI$ & $IZY$ & $IXY$\\
\hline
\end{tabular}
\caption{The system of code words related to the system of message words of Table 2.
Each spread is containing the $5$ message words $\mathcal{M}_{ja}, a=1,\dots 5$ of Table 2. The sixth spread, with its message and code words, coincides with the one of Table 1. The last three columns are featuring the points of the line $\ell_{\ast}$, off the bulk, defining the code via Eq.(\ref{intersect}). They are featuring the recovery vectors $\mathcal{R}$ and $\mathcal{R}^{\prime}\equiv\mathcal{R}+\Gamma$. Finite geometrically the two sets of recovery vectors $\mathcal{R}_j$, $\mathcal{R}^{\prime}_j$ form a "double-six". They are two copies of six-dimensional Clifford algebras $Cliff(6)$ interchanged by $\Gamma$.}
\label{tab:3}
\end{table}

Now one can observe that the set of codewords corresponding to a particular line $\ell_{\ast}$ forms again a Clifford algebra. This time a five-dimensional one. For example the codewords corresponding to $\mathcal{S}_6$ can be described as
\beq
\{\mathcal{C}_1,\mathcal{C}_2,\mathcal{C}_3,\mathcal{C}_4,\mathcal{C}_5,\}=
\{i\Gamma_{167},i\Gamma_{267},i\Gamma_{367},i\Gamma_{467},i\Gamma_{567}\}.
\eeq
\noindent  
One can easily check that in terms of the set of possible recovery vectors 
\beq
\mathcal{R}\leftrightarrow
\Gamma_j,\qquad j=1,2,\dots 6
\label{recliff}
\eeq
\noindent 
the codewords of the $j$-th row of Table 3 are given (up to a sign) by the formula
\beq
\mathcal{C}_{kj}=i\Gamma_{kj7},\qquad j=1,2,\dots 6,\qquad k\neq j.
\label{codewordscliff}
\eeq
\noindent
In this formula fixing $j$ amounts to fixing the row of Table 3, corresponding to choosing the $j$-th code. On the other hand for a fixed row (code), $k$ is running through all values from $1$ to $6$ except the fixed value of $j$, producing the set of codewords within the particular code. Clearly for a fixed $j$ the observables as codewords are commuting with the observables comprising the corresponding line $\ell_{\ast}$.

Eq.(\ref{codewordscliff}) gives an elegant characterization of codewords of the bulk in terms of the possible set of (\ref{recliff}) recovery vectors.
Eqs.(\ref{nice}), (\ref{cliff6}), (\ref{recliff}) and (\ref{codewordscliff}) then neatly summarize the boundary-bulk error correcting picture.

Amusingly this error correcting picture is related to a correspondence between {\it commuting sets} of two-qubit boundary observables and {\it anticommuting sets} of three-qubit bulk ones, based on the Klein Correspondence.
We also emphasize that our formalism describes the relationship between the boundary and the bulk as {\it collection of error correcting codes}. Hence our finite geometric model fits into the philosophy  coming from the AdS/CFT correspondence of regarding an asymptotically AdS space-time as an error correcting code\cite{Verlinde,Preskill,Dong1,Dong2}.

Finally let us try to clarify the physical meaning of the set of recovery vectors showing up in (\ref{recliff}).
As we have shown, the bulk is the $GF(2)$ analogue of compactified complexified Minkowski space-time embedded in $PG(5,2)$. We have also seen that although our recovery vectors are living {\it off the bulk},
their special (\ref{intersect}),(\ref{important}) relationship
to the bulk makes it possible to define the codewords in a natural manner.

Notice that in conventional twistor theory where instead of $PG(5,2)$ we have $PG(5,\mathbb{C})$ similar structures are used to define {\it conformally flat spacetimes}. For example in the simplest non-flat example of a complex de-Sitter space the analogue of the recovery vector $\mathcal{R}$ is a vector $\mathcal{I}\in \mathbb{C}^6$ {\it off the Klein Quadric}. 
Moreover, Penrose\cite{PR} even characterized the conformal factor of Robertson-Walker type cosmological models in terms of a {\it pair} of such vectors $\mathcal{I}$ and $\tilde{\mathcal{I}}$. Depending on the properties of this pair, namely whether they are real or complex or lie on or off the quadric, we get different types of models. 
In the case of complex de-Sitter space quantities like $I(X)\equiv\langle\mathcal{ I},X\rangle$ describe the conformal factor, where $X$ is a point on and $\mathcal{ I}$ is a one off the Klein Quadric.
In conformally flat space-times, characterized by such "fields" $I(X)$, the zeroes and singularities of $I(X)$ correspond to notions like "infinity" and "singular points"\cite{Hurd,Brody2,PR}.
Clearly in our finite geometric context the analogue of the pair $(\mathcal{I},\tilde{\mathcal{I}})$ is $(\Gamma,\mathcal{R}$), with the corresponding $GF(2)$-valued "fields" are quantities like $\Gamma(X)\equiv\langle\Gamma,X\rangle$ and $\mathcal{R}(X)\equiv\langle\mathcal{R},X\rangle$. We already know that the zeroes of $\Gamma(X)$ define the real section of the bulk. Now the important point we would like to make is the one that according to (\ref{important}) the common zeroes of $\Gamma(X)$ and $\mathcal{R}(X)$ are precisely our code words.
Moreover, in twistor theory linear combinations like $\mu\mathcal{I}+\nu\tilde{\mathcal{I}}$  give rise to the hypersurfaces of homogenity of the space-time. Over $GF(2)$ such linear combinations are precisely the ones defining our lines $\ell_{\ast}$ whose orthogonal complement defines the codewords. Hence the six sets of codewords are just the $GF(2)$ analogues of such hypersurfaces. 

We note in closing that one need not have to choose $\mathcal{I}$ off the bulk in order to get structures of physical relevance via imposing the constraint $I(X)=0$. Indeed, if we choose $\mathcal{I}$ corresponding to the infinity twistor, then what we get is conformal infinity\cite{Hurd}. In our $GF(2)$ case the analogue of $\mathcal{I}$ is the vector $(000100)$ answering the bulk code word $\mathcal{C}_4=XII$ (see Table 1). Under the Klein Correspondence $\mathcal{C}_4=\mathcal{I}$ corresponds to the isotropic line
$(0\vert I)$ answering the boundary message word\footnote{See the considerations
 before Eq.(\ref{mink}).}: $\mathcal{M}_4=\{IX,XI,XX\}$. Since at the boundary lines at infinity are precisely the ones having nonzero intersection with this distinguished line, this means that at the bulk these correspond to points lying on the light cone of $\mathcal{I}$. Since this constraint is precisely the condition of $I(X)=0$, this means that the zero locus of this $GF(2)$ field is just {\it the light cone at infinity}. 
A pictorial representation of this structure coincides with the one of Figure 5. (with labels suitably adjusted) where $\mathcal{M}_4$ is used in the left and $\mathcal{C}_4$ on the right hand side. This is precisely the structure we used for our representation of the error correction process.

\subsection{States and signs}

According to the theory of stabilizer codes certain sets of mutually commuting observables uniquely determine states\cite{Nielsen,Gottesman}. Such states are stabilized by these observables.
This observation makes it possible to recast our error correction picture in terms of states rather than observables.
More precisely, in order to define stabilizer states one has to leave the realm of observables (objects of the form $\pm\mathcal{O}$) in favour of elements of the $n$-qubit Pauli group. The latter is containing objects of the form $G_n\equiv\{\pm\mathcal{O},\pm i\mathcal{O}\}$. 

If $S$ is a subgroup of $G_n$ and $\mathcal{H}_{S}$ is a subspace of the $n$-qubit Hilbert space $\mathcal{H}$ such that every element of $\mathcal{H}_{S}$ is fixed by the action of elements taken from $S$ then $\mathcal{H}_{S}$ is called the {\it vector space stabilized} by $S$. 
$S$ is called the {\it stabilizer} of $\mathcal{H}_{S}$.
The sufficient and necessary conditions to be satisfied by $S$ in order to stabilize a nontrivial $\mathcal{H}_{S}$ are as follows\cite{Nielsen,Gottesman}. 1. the elements of $S$ should commute and 2. $-{\bf 1}\notin S$, where ${\bf 1}=I\otimes I\otimes\cdots\otimes I$ is the $n$-qubit identity operator.
Notice that the second condition implies that $\pm i{\bf 1}\notin S$, hence the elements of $S$ are taken from our set
 $\pm\mathcal{O}$ of observables. 

In the following we suppose that $S$ is a {\it stabilizer subgroup} of $G_n$. We write a presentation\footnote{A presentation is given in terms of {\it independent generators} of the group. In terms of our $V(2n,2)$ representation of observables this means that the corresponding vectors are linearly independent.} of $S$ in terms of its {\it commuting} generators in the following form: $S=\langle\mathcal{O}_1,\mathcal{O}_2,\dots,\mathcal{O}_{n-k}\rangle$. 
Then we have the following basic result\cite{Nielsen}: if $S$ is given by a presentation as above then $\mathcal{H}_S$ is a $2^k$ dimensional vector subspace of the $2^n$ dimensional Hilbert space.

Now in the case of $k=0$ sets of $n$ commuting observables generate a stabilizer group $S=\langle\mathcal{O}_1,\mathcal{O}_2,\dots,\mathcal{O}_{n}\rangle$.  This group determines a vector $\vert\psi_S\rangle$ up to a phase, i.e. a state.
The cardinality of $S$ is $2^n$ which is the number of vectors in a rank $n$ subspace $W$ of $V(2n,2)$. Projectively this means that $P(W)$ is a subspace of projective projective dimension $n-1$, with the number of its points being $2^n-1$.
Moreover, since $PG(2n-1,2)$ comes equipped with a symplectic form, $2^n-1$-tuples of commuting observables give rise to the set of maximally totally isotropic subspaces, i.e. the set of totally isotropic (Lagrangian) $n-1$-planes. They are comprising the Lagrangian Grassmannian $\mathcal{LG}(n-1,2n-1)$. 
Hence for $k=0$ a particular group $S$ can be used as a representative of an isotropic $n-1$ plane. Clearly by, playing with signs, for a particular isotropic $n-1$ plane one can associate $2^n$ possible representatives $S$, hence $2^n$ representative states $\vert\psi_{S}\rangle$.

For example for $n=2$ choosing $S=\langle\mathcal{O}_1,\mathcal{O}_2\rangle$ with $\mathcal{O}_1=YZ$ and $\mathcal{O}_2=XY$ the group $S$ is containing the $4$ elements:
$\{II,XY,YZ,-ZX\}$. The three nontrivial observables $\{XY,YZ,-ZX\}$ represent an isotropic line in $PG(3,2)$.
The state 
$\vert\psi_{S}\rangle$ 
that one can uniquely associate to this triple of observables is the state $\vert\psi_2\rangle$ of Table 1, which is fixed by all elements of $S$, e.g. for $\mathcal{O}_3\equiv\mathcal{O}_1\mathcal{O}_2=-ZX$ we have $\mathcal{O}_3\vert\psi_2\rangle =\vert\psi_2\rangle$ etc.
However, one could have tried another representative of this isotropic line as $S^{\prime}=\langle XY,ZX\rangle$ which is containing the $4$ elements:
$\{II,XY,-YZ,ZX\}$. In this case the vector $\vert\psi_{S}\rangle =\vert\psi_2\rangle$ is changing to $\vert\psi_{S^{\prime}}\rangle=\frac{1}{\sqrt{2}}(\vert\tilde{0}1\rangle +\vert\tilde{1}0\rangle)$. 

In the $n$-qubit case totally isotropic $n-1$-spreads of $PG(2n-1,2)$ will be used to give rise to message words, consisting of certain strings of observables. According to our philosophy $PG(2n-1,2)$ then will be called the {\it boundary}. 
As in the $n=2$ case of Table 1 an isotropic spread of the boundary is a partition of the $(2^n-1)(2^n+1)$ points of $PG(2n-1,2)$ to $2^n+1$ totally isotropic $n-1$ planes. Each totally isotropic $n-1$ plane is containing $2^n-1$ points. A particular isotropic plane can be represented by $2^n$ different strings of observables, differring only in their distribution of signs. The different representatives of a plane give rise to $2^n$ different states fixed by the corresponding strings of observables. These states form the $2^n$ basis states of the $n$-qubit Hilbert space $\mathcal{H}$. Hence to a totally isotropic $n-1$ plane one can associate in a unique manner a basis of $\mathcal{H}$. It can be shown that for an isotropic $n-1$-spread
the collection of $2^n+1$ basis systems can be choosen such that each element of one basis is an equal magnitude superposition of any of the other bases. Such basis sets are said to satisfy the MUB property, i.e.they are mutually unbiased\cite{W2,Thasmub}. 
Since a MUB can be used effectively for determining an unknown mixed $n$-qubit state via quantum state tomography\cite{w1}, our choice of message words of the boundary is intimately connected to a choice of possible measurements one should perform during the protocol of effective state determination.
For example the states showing up in Table 1 are members of the well-known MUB set for two qubits\cite{W2}. They clearly satisfy the MUB property for $n=2$ namely: $\vert\langle\psi_a\vert\psi_b\rangle\vert=1/\sqrt{2^n}$ with $a\neq b$ and $a,b=1,\dots 5$.

In Table 1 and the left hand side of Figure 3 we have choosen a particular distribution of signs which makes all of the isotropic {\it message} lines of the spread $S_6$ {\it positive lines}. This notion means that if we multiply the commuting observables along the line then we get the identity with a {\it positive sign}, hence the observables of the message lines form a stabilizer $S$. To these lines one can associate {\it states} in a unique manner.
However, in order to achieve the same goal for the other spreads $\mathcal{S}_j, j=1,2,\dots 5$ of Table 2. giving rise to other message sets $\mathcal{M}_{ja}$, 
some {\it other distribution of signs} is needed. In other words there is no distribution of signs for the full set of observables which is compatible with the set of all possible codes of Table 2 and 3.
In finite geometric terms: although it is possible to associate states to the lines of an isotropic spread of lines, there is no way of doing this consistently for all of the isotropic lines of the boundary, $PG(3,2)$. 

Recall that the set of isotropic lines on the $15$ points of $PG(3,2)$ is the doily.
Now the reason for the incompability of signs is a theorem which states that the number of negative lines of the doily is always an odd number\cite{LHSmagic}. Hence one will always encounter at least one negative line. Since for negative lines $-{\bf 1}\in S$ one cannot elevate an $S$, corresponding to the observables of such a line, to the status of a stabilizer, hence there is no possibility for associating stabilizer states to all of the lines of the doily.
Interestingly the reason for this theorem to hold is the fact that the doily is containing grids as geometric hyperplanes\cite{PS}.
A grid is a collection of $9$ points in a rectangular arrangement having $6$ lines, each line containing $3$ points.
This arrangement labelled by qubit observables is known to physicists as a {\it Mermin square}. 
A Mermin square\cite{Mermin1,Mermin2} can be used for proving (without the use of probabilities) that there are no noncontextual local hidden variable theories compatible with the predictions of quantum theory. The proof is based on the simple observation that for a grid the number of negative lines is always odd. Since the doily is containing grids this proof can be carried through even for the doily. For a study of the interplay between elementary contextual configurations and finite geometry see Ref.\cite{HS}.

In summary we have learnt that for a particular code one can associate states to message lines. However, one cannot do this to the set of all possible codes.
Moreover, the reason for this incompatibility of codes is the same as the one responsible for the incompatibility of noncontextual local hidden variable theories and quantum theory.

\subsection{Error correction and states}

Let us now have a look at our error correction process from the perspective of stabilizer states.
For a fixed code one can associate states to the spread of lines, see Table 1.
We have two types of errors connected by projective duality: point errors and plane ones. For point errors instead of an isotropic line (message) we get a point (corrupted message). Notice now that in the boundary to a message line a state, and to a corrupted message which is a point, a two dimensional subspace of states is associated\footnote{Recall that if the stabilizer is given by the presentation  $S=\langle\mathcal{O}_1,\mathcal{O}_2,\dots,\mathcal{O}_{n-k}\rangle$ then $\mathcal{H}_S$ is a $2^k$ dimensional subspace. In our special case of a point error: $n=2$, $k=1$.}.
Hence in this stabilizer picture the transition form a message to a corrupted message corresponds to a transition from a state to a subspace containing this state. 

For example to the message line $\mathcal{M}_5\equiv\{IZ,ZI,ZZ\}$ the ray of $\vert 00\rangle$ denoted by $\mathcal{H}_{\mathcal{M}_5}$ is associated. Let us suppose that we have a point error with the corrupted message being the point $P\equiv\{ZZ\}$. Since $\vert 00\rangle$ and $\vert 11\rangle$ are eigenvectors of $ZZ$ with eigenvalue $+1$, the corrupted subspace is just the span of these two vectors  $\mathcal{H}_{P}={\rm span}\{\vert 00\rangle,\vert 11\rangle\}$.
Clearly  $\mathcal{H}_{\mathcal{M}_4}\subset\mathcal{H}_{P}$.

In the bulk we have the codeword $\mathcal{C}_5\equiv\{ZII\}$ encoding the message $\mathcal{M}_5$. In the stabilizer language this determines a $4$-dimensional subspace $\mathcal{H}_{\mathcal{C}_5}={\rm span}\{\vert 000\rangle,\vert 001\rangle,\vert 010\rangle,\vert 011\rangle\}$. Hence a message state $\mathcal{H}_{\mathcal{M}_5}$ on two-qubits is encoded into $\mathcal{H}_{\mathcal{C}_5}$ which is a subspace on three-qubits. Now to point errors in the boundary correspond the $\alpha$-planes. Since these are totally isotropic planes, represented by commuting seven-tuples of three-qubit observables, in the stabilizer formalism to these seven-tuples one can associate states.    
In our example\footnote{See Appendix A.} the corresponding $\alpha$-plane is $\Pi^{(\alpha)}=\{ZII,IXZ,ZXZ,IZX,ZZX,IYY,ZYY\}$ which gives rise to the stabilizer $\langle ZII,ZXZ,ZZX\rangle$ stabilizing the ray $\mathcal{H}_{\Pi^{(\alpha)}}\equiv\{
\frac{1}{\sqrt{2}}(\vert 0\overline{0}0\rangle +\vert 0\overline{1}1\rangle)\}\subset \mathcal{H}_{\mathcal{C}_5}$.
The state representing this ray is a biseparable three-qubit state containing bipartite entanglement in its last two qubits.
In summary we have
\beq
\mathcal{H}_{\mathcal{M}_5}\subset\mathcal{H}_{P},\quad({\rm boundary})\qquad\leftrightarrow\qquad
\mathcal{H}_{\Pi^{\alpha}}\subset \mathcal{H}_{\mathcal{C}_5}\quad({\rm bulk}).
\label{meaning}
\eeq
\noindent
Notice that under a transition from the boundary to the bulk, on the two sides of the $\subset$ symbol the roles of message and error subspaces are exchanged. Whilst in the boundary messages, in the bulk errors correspond to states.  
And dually: whilst in the boundary point errors, in the bulk (coded) messages correspond to subspaces.

Let us now turn to a similar elaboration for plane errors.
On the boundary for representatives of plane errors we have seven-tuples of observables. However, these $15$ seven-tuples of observables are not mutually commuting. Luckily these $15$ planes are dual to $15$ points. Indeed, a point labelled by an observable $\mathcal{O}$ is dual to a plane labelled by the same observable. In this case the seven observables of that plane are the ones commuting with this fixed $\mathcal{O}$. Moreover, we have already seen that the error correction process is the same, no matter whether we have point or plane errors. Hence one expects that to a plane labelled by $\mathcal{O}$ one should associate the same two dimensional {\it subspace} $\mathcal{H}_{\mathcal{O}}$, which we associated to the point $\mathcal{O}$.  

In order to prove this notice that for a fixed message line there are three possible plane errors. They correspond to the three possible planes containing the same message line with observables: $\mathcal{O},\mathcal{O}^{\prime},\mathcal{O}\mathcal{O}^{\prime}$, see Figure 6 for an example. The labels of the three possible error planes are just $\mathcal{O}$, $\mathcal{O}^{\prime}$, and $\mathcal{O}\mathcal{O}^{\prime}$. Let us suppose that our message line is corrupted, and the error plane arising is the one labelled by $\mathcal{O}$. 
In this case our message line is conveniently represented by the stabilizer $S=\langle\mathcal{O},\mathcal{O}^{\prime}\}$, where $\mathcal{O}$ is the special observable which is commuting with all observables of the error plane $\Pi_{\mathcal{O}}$. Clearly one can choose\footnote{For illustrations of this representation see Figure 6. or Appendix A.}  the seven observables of this plane as: $\{\mathcal{O},\mathcal{O}^{\prime},\mathcal{O}\mathcal{O}^{\prime}, A,B,C,D\}$ where 
$\mathcal{O}^{\prime}$ is {\it anticommuting} with each member of the set $\{A,B,C,D\}$.
To our fixed message line as usual we associate the state $\vert\psi_S\rangle$. 
Now using the stabilizer property $\mathcal{O}^{\prime}\vert\psi_S\rangle=\vert\psi_S\rangle$ and the anticommutation, one can immediately see that all of the states $A\vert\psi_S\rangle$,
$B\vert\psi_S\rangle$,
$C\vert\psi_S\rangle$,
$D\vert\psi_S\rangle$, are orthogonal to $\vert\psi_S\rangle$.
However, since $\mathcal{O}$ is commuting with everybody, all of these vectors are belonging to its eigensubspace with eigenvalue $+1$. Since this eigensubspace is two dimensional, for example the set $\{\vert\psi_S\rangle,A\vert\psi_S\rangle\}$ can be choosen as an orthonormal basis spanning this subspace. Hence $\mathcal{H}_{\Pi_{\mathcal{O}}}={\rm span}\{\vert\psi_S\rangle,A\vert\psi_S\rangle\}$. This is just the same subspace we associated to point errors.
This means that $\mathcal{H}_{\Pi_{\mathcal{O}}}=\mathcal{H}_{\mathcal{O}}$.
Hence point errors and plane ones are sharing the same subspaces containing the message state $\vert\psi_S\rangle$.

An important consequence of these considerations is the following.
Plane errors are the ones containing the message lines. The observables {\it off} the message line (i.e. $A,B,C,D$) are the ones that are anticommuting with {\it some} of the observables of the message (i.e. $\mathcal{O}^{\prime}$ and $\mathcal{O}\mathcal{O}^{\prime}$).
Hence these observables can be regarded as {\it error operators}. The action of these operators on the stabilizer state has the effect of moving this state to its orthogonal complement. This is just like in the usual stabilizer formalism of error correction\cite{Nielsen,Gottesman}.
Hence the physical interpretation of a plane error is very similar to the conventional interpretation of errors in the stabilizer formalism. Since plane errors are dual to point errors, and the recovery process is the same for these errors, we conclude that our observable based reinterpretation of geometric subspace codes\cite{Stokes} is very similar to the one of stabilizer codes\cite{Nielsen,Gottesman}.

There are however, some important differences to be noticed. Indeed, we have not examined the bulk representation of plane errors yet. A boundary plane error $\Pi_{\mathcal{O}}$ is represented by a totally isotropic $\beta$-plane 
$\Pi_{\mathcal{O}}^{(\beta)}$
in the bulk. For example the message $\mathcal{M}_5$ of the boundary is encoded into the bulk in the form of the codeword $\mathcal{C}_5$. According to Figure 6 to a plane error $\Pi$ of this message its bulk representative, namely
the isotropic plane $\Pi^{(\beta)}=\{ZII,IZZ,ZZZ,IXX,ZXX,-IYY,-ZYY\}$, is associated.
Then this plane represented by the stabilizer $\langle ZII,ZZZ,ZXX\rangle$ has
the ray $\mathcal{H}_{\Pi^{(\beta)}}\equiv\{
\frac{1}{\sqrt{2}}(\vert 000\rangle +\vert 011\rangle)\}\subset \mathcal{H}_{\mathcal{C}_5}$.

Hence although the process of error correction is the same for point errors and plane ones, the states which are representing the errors in the bulk are different. However, the rays $\mathcal{H}_{\Pi^{(\alpha)}}$ and $\mathcal{H}_{\Pi^{(\beta)}}$  are not independent. They are local unitary equivalent states belonging to the same entanglement class. Indeed, they are connected by a discrete Fourier transformation in the second qubit, i.e. a transformation of the form $I\otimes H\otimes I$.
Recall also in this respect that in twistor theory $\alpha$ and $\beta$ planes are conjugate planes. In our finite geometric setting it means that up to a sign their observables are related by an interchange of an $X$ with a $Z$ in the middle slot,
leaving invariant the $I$ and $Y$ (see Figures 5 and 6). If we would like to also respect the sign structure of these planes, i.e. we demand that these planes should be {\it positive ones} hence eligible to be interpreted as stabilizers, their observables should correspond to each other via a conjugation\footnote{As an example just compare the observables of the positive planes $\mathcal{H}_{\Pi^{\alpha}}$ and 
$\mathcal{H}_{\Pi^{\beta}}$.}
by $I\otimes H\otimes I$. This indeed effectively takes care of the signs since: $HYH=-Y$.

\subsection{Transformations}

We have seen that the bulk can be regarded as a collection of error correcting codes, encoding a collection of messages in the boundary. Now in closing this section we would like to see how these error correcting codes and their messages are related to each other. 

First of all recall that for message lines one can associate states, and these states belong to different entanglement classes. Under the SLOCC group for two qubits we have merely two entanglement classes: the separable one and the entangled one. Their SLOCC representatives are the state $\vert 00\rangle$ and any of the Bell states e.g. the one $\frac{1}{\sqrt{2}}(\vert 00\rangle+\vert 11\rangle)$.
For the message lines the associated states are shown in Table 1. We see that the states $\vert\psi_a\rangle$ are entangled for $a=2,3$ and separable in the remaining cases.

How can we relate these message words and their associated states?
Our message words are $\mathcal{M}_a$ where $a=1,2,3,4,5$, see Table 1. One can relate the 
$\mathcal{M}_k$ where $k=1,2,3$ as follows.
Define 
\beq
C_{12}\equiv {\rm CNOT}_{12},\qquad S_{12}={\rm SWAP}_{12}
\label{defigate}
\eeq
\noindent
the controlled-not and swap operations acting on two qubits. 
Let us define the unitary operator
\beq
\mathcal{D}\equiv C_{12}S_{12}=C_{21}C_{12}.
\label{replambda}
\eeq
\noindent
Now we have
\beq
\mathcal{M}_k=\mathcal{D}^{-k+1}\mathcal{M}_1\mathcal{D}^{k-1}, \qquad \vert\psi_k\rangle=\mathcal{D}^{-k+1}\vert\mathcal\psi_1\rangle,\qquad k=1,2,3.
\label{atteres}
\eeq
\noindent
The detailed action of this conjugation on the observables is
\beq
YY\mapsto -ZX\mapsto -XZ\mapsto YY,\quad
YI\mapsto XY\mapsto ZY\mapsto YI,\quad
IY\mapsto YZ\mapsto YX\mapsto IY.
\label{detail1}
\eeq
\noindent
In other words conjugation by $\mathcal{D}^{-1}=C_{12}C_{21}$ cyclically permutes the corresponding entries of $\mathcal{M}_k, k=1,2,3$ of the second column of Table 1.
The remaining messages 
$\mathcal{M}_4$ and $\mathcal{M}_5$ are left invariant since under conjugation by  $\mathcal{D}^{-1}$ we have
\beq
XI\mapsto XX \mapsto IX \mapsto XI,\qquad IZ\mapsto ZZ \mapsto ZI \mapsto IZ.
\label{detail2}
\eeq
\noindent
Notice also that due to the presence of the CNOT gate the unitary transformation of (\ref{replambda}) can change the entanglement class of the stabilized states. 

We can map the third order map $\mathcal{D}^{-1}=C_{12}C_{21}$ of Eq.(\ref{atteres}) acting by conjugation on message words in the boundary, to a third order map ${\bf D}^{-1}$ acting by conjugation on code words in the bulk.
A calculation carried out in Appendix B shows that this map is
\beq
{\bf D}^{-1}=I\otimes \mathcal{U}(M),\qquad \mathcal{U}(M)=i(H\otimes I)C_{12}(I\otimes H)C_{21}(X\otimes X).
\label{kakukk1}
\eeq
\noindent
Then we have
\beq
\mathcal{C}_k={\bf D}^{-k+1}\mathcal{C}_1{\bf D}^{k-1},\qquad k=1,2,3
\label{kakukk}
\eeq
where the codewords are given by Table 1 depicted by Figure 3. The codewords $\mathcal{C}_4$ and $\mathcal{C}_5$ are left invariant.

Let us finally see how we can obtain other codes starting from the special one of Table 1. As we have seen this special code corresponds to the spread $\mathcal{S}_6$ of Tables 2 and 3. We are in search of unitary operators acting by conjugation connecting the message words of $\mathcal{S}_6$ to the remaining ones of $\mathcal{S}_1,\dots, \mathcal{S}_5$. Based on our experience with Eqs.(\ref{atteres}) and (\ref{kakukk}) we are in search of unitary representations of certain transformations of the group $Sp(4,2)\simeq S_6$ generated by transvections. Moreover in forming our representations, we can use the trick of lifting transvections explained in Appendix B.

Using the duad labelling of the doily of Appendix B one can easily see that the cyclic permutation of spreads $(\mathcal{S}_6,\mathcal{S}_4,\mathcal{S}_3)$ is implemented by the permutation $(123)=(12)(23)$ generating a cyclic group of order three.
This generator is represented by the transvection $T_{12}T_{23}$. Now under the (\ref{d1})-(\ref{d3}) labelling $ZY$ and $XI$ in Figure 1 corresponds to the duads $12$ and $23$. Neglecting now the phase factors of Eq.(\ref{liftoper}) one can lift these transvections to the unitaries
\beq
\mathcal{U}(T_{12})=\frac{1}{\sqrt{2}}\left(I\otimes I+iZ\otimes Y\right),\qquad
\mathcal{U}(T_{23})=\frac{1}{\sqrt{2}}\left(I\otimes I+iX\otimes I\right).
\eeq
\noindent
Now $\mathcal{U}(T_{123})=\mathcal{U}(T_{12})\mathcal{U}(T_{23})$ is acting on the observables as
\beq
\mathcal{O}\mapsto \mathcal{U}(T_{123})\mathcal{O}\mathcal{U}^{\dagger}(T_{123}).
\eeq
\noindent
One can check that in terms of elementary quantum gates this unitary can be written as
\beq
\mathcal{U}(T_{123})=C_{12}(I\otimes HX)C_{12}(PHP\otimes I),\qquad P\equiv\begin{pmatrix}1&0\\0&i\end{pmatrix}.
\label{elementary}
\eeq
\noindent
Here we see that $\mathcal{U}(T_{123})$ can be written in terms of the CNOT ($C_{12}$), the Hadamard ($H$), the phase ($P$), and the bit-flip gate ($X$).
An explicit form of the action is $(\mathcal{S}_6,\mathcal{S}_4,\mathcal{S}_3)$
\beq
\{IY,YI,YY\}\mapsto\{IY,-ZI,-ZY\}
\eeq
\noindent
\beq
\{YZ,XY,-ZX\}\mapsto\{IX,-YI,-YX\}
\eeq
\noindent
\beq
\{YX,ZY,-XZ\}\mapsto\{-IZ,XI,-XZ\}
\eeq
\noindent
\beq
\{XI,IX,XX\}\mapsto\{-YY,ZZ,XX\}
\eeq
\noindent
\beq
\{IZ,ZI,ZZ\}\mapsto\{-ZX,XY,YZ\}
\label{naitt}
\eeq
\noindent
The action of this unitary has two orbits on the space of messages. They are of the form
\beq
(\mathcal{S}_6,\mathcal{S}_4,\mathcal{S}_3),\qquad  (\mathcal{S}_5,\mathcal{S}_1,\mathcal{S}_2)
\label{ciklus1}
\eeq
\noindent
where we have used the cycle notation, hence the entries are permuted cyclically.
Note that the transvections $T_{12}$ and $T_{23}$ are anticommuting hence according to Ref.\cite{Cherchiai} we have
$(T_{12}T_{23})^3=1$. Interestingly the unitary representative has the property
\beq
\mathcal{U}(T_{123})^3=-I\otimes I.
\label{kob}
\eeq
\noindent

In order to reach each spread from our distinguished one $\mathcal{S}_6$ we need an unitary relating the two orbits of (\ref{ciklus1}). This can be constructed for example from the lift of the permutation $(153)$. A similar construction then above yields
\beq
\mathcal{U}(T_{153})=\frac{1}{2}(I\otimes I-iZ\otimes I)(I\otimes I-iY\otimes Y).
\label{amasik}
\eeq
\noindent
Its explicit action is
\beq
\{IY,YI,YY\}\mapsto\{IY,-XI,-XY\}
\eeq
\noindent
\beq
\{YZ,XY,-ZX\}\mapsto\{IX,-ZI,-ZX\}
\eeq
\noindent
\beq
\{YX,ZY,-XZ\}\mapsto\{-IZ,YI,-YZ\}
\eeq
\noindent
\beq
\{XI,IX,XX\}\mapsto\{-ZY,XZ,YX\}
\eeq
\noindent
\beq
\{IZ,ZI,ZZ\}\mapsto\{-XX,YY,ZZ\}
\label{naitt1}
\eeq
\noindent
hence according to Table 2 $\mathcal{S}_6$ is connected to $\mathcal{S}_5$. 
We note that this unitary can be written in terms of elementary gates (generating the Clifford group $C_2$) in the following form
\beq
\mathcal{U}(T_{153})=(P\otimes I)\mathcal{V}(I\otimes PHP)\mathcal{V}^{\dagger},\qquad \mathcal{V}=C_{21}(I\otimes H)C_{12}(I\otimes H)C_{12}.
\eeq
\noindent

Looking at Eqs.(\ref{atteres}) and (\ref{kakukk}) one notices that the message words $\mathcal{M}_{4,5}$ and their corresponding codewords $\mathcal{C}_{4,5}$ seem to play a special role. Indeed, they are left invariant by the corresponding transformations $\mathcal{D}$ and ${\bf D}$, and the remaining message and code words are cyclically transformed into each other by the corresponding unitaries of order three.
Now the question arises can we find another unitary transformation $U$, this time of order {\it five}, which is treating all the five message words on an equal footing? An answer to this question is yes, and to cap all this an unitary (this time of order $2^n+1$) can be found for all values of $n$ \cite{w3}.
We will have something more to say on the role $U$ playing in our story later, here we are content with giving its explicit form in terms of elementary two-qubit gates.
Some calculational details for finding such an $U$ for $n=2$ are given at the end of Appendix B.

First let us define another unitary operator $W$ given by Eq.(\ref{w}).
Then in terms of this unitary we have
\beq
U=(P\otimes I)(X\otimes X)W(P\otimes I)(X\otimes X)S_{12}WS_{12},\qquad U^5={\bf 1}.
\label{unitary9}
\eeq
\noindent
where for some definitions see (\ref{defigate}).
Now the action of $U$ on observables is just the usual one by conjugation: $\mathcal{O}\mapsto U\mathcal{O}U^{\dagger}$, and one can check that starting this time with $\mathcal{M}_4$ under a sequence of conjugations we are completing a cycle:
$(\mathcal{M}_4,\mathcal{M}_0,\mathcal{M}_3,\mathcal{M}_1,\mathcal{M}_2)$. For the detailed form of this action see Eq.(\ref{orbitexplicit}) of Appendix B.

However, there is a subtlety here we have to point out.
We were unable to find a $U$ which is compatible with our choice of signs of Table 1.
Indeed, the best our $U$ can do is to reproduce the signs of each of our message words except for the one $\mathcal{M}_1$.
This means that under conjugation this message word is reproduced as $\mathcal{M}_1^{\prime}\equiv\{IY,-YI,-YY\}$ 
instead of the desired one $\mathcal{M}_1=\{IY,YI,YY\}$. However, since both of the lines being positive they are amenable to a stabilizer interpretation. This means that in Table 1 the difference will manifest itself via the appearance of a different stabilizer state. Namely instead of $\vert\psi_1\rangle$ of Table 1 one has to use the new state
\beq
\vert\psi_1^{\prime}\rangle\equiv\frac{1}{\sqrt{2}}(\vert 0\rangle +i\vert 1\rangle)\otimes \frac{1}{\sqrt{2}}
(\vert 0\rangle -i\vert 1\rangle)=\vert\tilde{0}\tilde{1}\rangle
\eeq
\noindent
where the (\ref{tilde}) definition has been used.
Hence our $U$ is cyclically permuting the states, i.e. we have in permutation notation
$(\psi_5,\psi_1^{\prime},\psi_4,\psi_2,\psi_3)$. Hence in this case by the repeated application of $U$ from the simple initial state $\vert\psi_5\rangle=\vert 00\rangle$ all the other stabilizer states can be generated in a unique manner.

\begin{figure}[pth!]
\centerline{\includegraphics[width=12truecm,clip=]{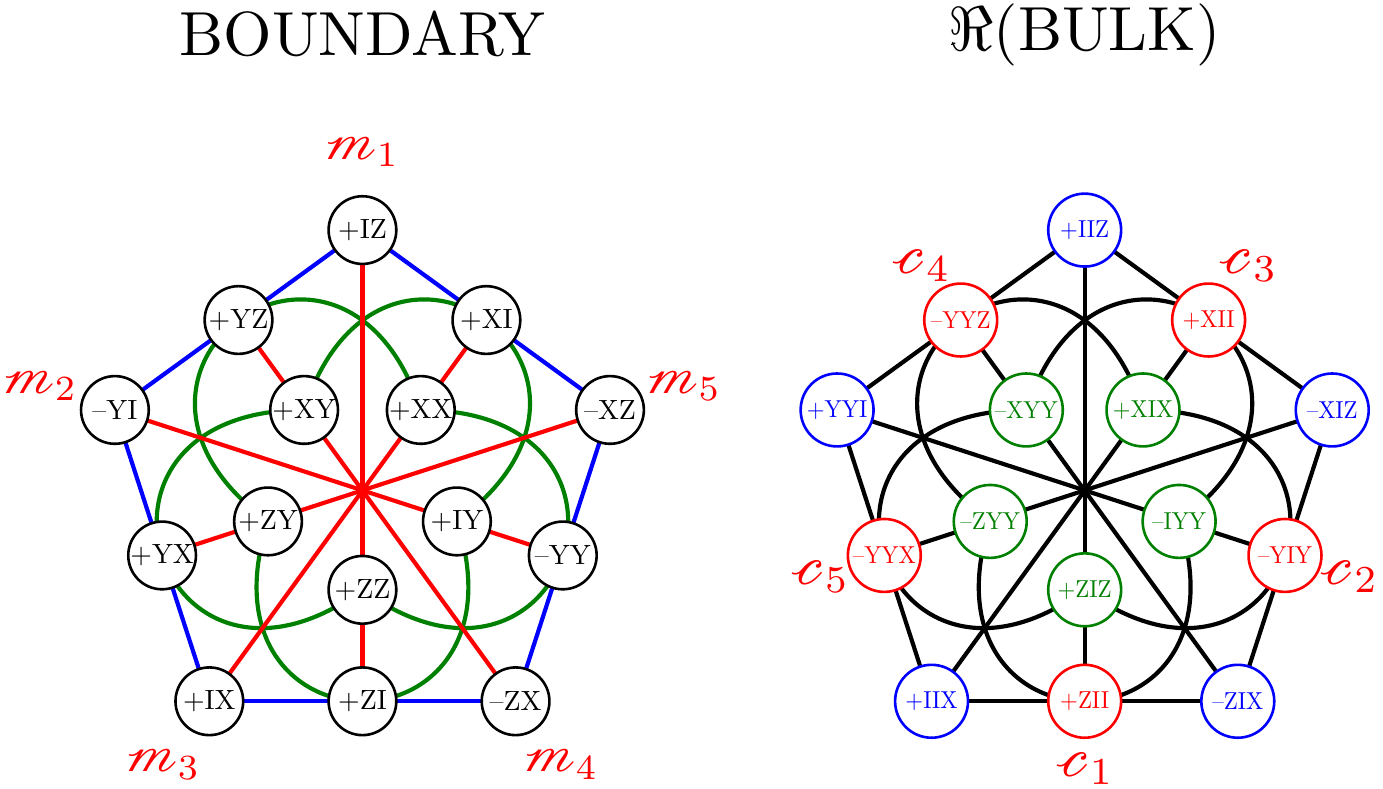}}
\caption{Left: A rotationally covariant set of messages in the boundary. The three different orbits of lines under $U$ of Eq.(\ref{unitary9}) are represented by colours. The five red (diagonal) message lines form a spread, with the corresponding triples of operators comprising a stabilizer.
Right: The real part of the bulk in a rotationally covariant labelling. The three different orbits under ${\bf U}$ of Eq.(\ref{5thorder}) are represented by differently coloured points. For a fixed set of messages in the boundary, the relative signs of different orbits in the bulk are ambiguous. In the figure one particular choice can be seen where ${m}_1$ and the vertical line going through ${c}_1$ were choosen positive.
}
\end{figure}

Hence we could have chosen either the message set 
$\{\mathcal{M}_1^{\prime},\mathcal{M}_2,\mathcal{M}_3,\mathcal{M}_4,\mathcal{M}_5\}$
compatible with our unitary $U$ of order five, or the usual message set $\{\mathcal{M}_1,\mathcal{M}_2,\mathcal{M}_3,\mathcal{M}_4,\mathcal{M}_5\}$
of Table 1 compatible with our unitary $\mathcal{D}$ of order three.
The reason for we opted for the latter one is the fact that the simple (\ref{replambda}) expression of $\mathcal{D}$ is straightforward 
to generalize\cite{W2} for arbitrary $n$, by adding extra $C_{ij}$s and $S_{ij}$s according to a simple algorithm.
On the other hand one can also argue in favour of $U$, that unlike $\mathcal{D}$ it provides a rotationally covariant\cite{w3} assignment of states to the lines of our boundary, starting from the state $\vert 00\rangle$.
Hence in this language our preferred choice of Table 1, though simple to handle, is not rotationally covariant.

For a rotationally covariant set one can define the new set $\{m_1,m_2,m_3,m_4,m_5\}=\{\mathcal{M}_5,\mathcal{M}_1^{\prime},\mathcal{M}_4,\mathcal{M}_2,\mathcal{M}_3\}$.
On this set the unitary of Eq.(\ref{marhajo}) is acting in cyclic manner by permuting the labels as $(12345)$ see Eq.(\ref{orbitexplicit}).
One can also define the lift of this unitary acting on the corresponding codewords 
$\{c_1,c_2,c_3,c_4,c_5\}=\{ZII,-YIY,XII,-YYZ,-YYX\}$
in a cyclic manner. This unitary ${\bf U}$ is of fifth order and of the form

\beq
{\bf U}=\frac{1}{2}\begin{pmatrix}iY&-Z&iY&Z\\Z&iY&-Z&iY\\I&-X&-I&-X\\X&I&X&-I\end{pmatrix},\qquad
c_a={\bf U}^{a-1}c_1{\bf U}^{1-a},\quad a=1,\dots 5.
\label{5thorder}
\eeq
\noindent
Since this $8\times 8$ matrix is real it is rather to be called an orthogonal transformation than a unitary one.
This is in accord with the fact that the observables featuring the Klein quadric are all real.
In Figure 7 one can find the details of this rotationally covariant encoding of boundary messages into codewords of the real part of the bulk. Notice that in this approach one can fix the boundary quantum net uniquely however, there is a sign ambiguity still left for bulk observables.
In the figure one particular consistent choice of signs is shown.
This can be regarded as an approach for also fixing the bulk quantum net provided we are already given a fixed one for the boundary.

\section{Generalization for $n$-qubits}

After this detailed excercise we can turn to the general case of $n$-qubits.
First we consider the boundary which is a higher dimensional $GF(2)$ generalization of projective twistor space: $PG(2n-1,2)$.
Then we show that the finite geometric representative of the set of possible messages is the set of totally isotropic $n-1$-spreads of $PG(2n-1,2)$.
Next we consider the Gibbons-Hoffman-Wotters (GHW) discrete phase space for $n$-qubits. It is mathematically $V(2,2^n)$ a two dimensional vector space over the field extension $GF(2^n)$ of $GF(2)$. 
This means that a point of this discrete phase space is labelled by the pairs $(q,p)$ where the coordinates and momenta are elements of $GF(2^n)$. We will use the projectivization of this space which is $PG(1,2^n)$.
Applying field reduction to $PG(1,2^n)$ we obtain our boundary $PG(2n-1,2)$.
In this manner we can relate our boundary to a space of physical importance. In particular it turns out that, in the terminology of Ref.\cite{W2}, our $n=2$ association of quantum states (or signed observables) to the isotropic lines of $PG(3,2)$ of Table 1. is just an instance of assigning to the GHW phase space a "quantum net".

As a next step we start an invesigation of the Grassmannian image of these structures in the bulk.
The points of our bulk are just the images of the $n-1$ dimensional subspaces of $PG(2n-1,2)$ under the Pl\"ucker map.
What we get by this process as the bulk is just the $GF(2)$ version of the Brody-Hughston "quantum space-time"\cite{Brody1}. 
In this picture the code words as points in the bulk, are arising as the images of message words as $n-1$ subspaces in the boundary. We will see that the set of $2^n+1$ code words forms an algebraic variety in the bulk which is the Grassmannian image of the $2^n+1$ message words of the boundary. 
As a generalization of Eq.(\ref{intersect}) this variety is arising as the complete intersection of the bulk with a suitable $2^n-1$ dimensional space $\Delta$, i.e. a $PG(2^n-1,2)$. 
The $2^n+1$ points residing in a real slice of the bulk can be embedded into a quadric playing a similar role in the general case than the Klein Quadric for the $n=2$ one.  The points corresponding to the codewords turn out to form a complete set of partial ovoids on this quadric.
Based on these results and the ones of Ref.\cite{Stokes} these structures then enable us an $n$-qubit generalization of the error correcting picture as we developed in the previous section in our $n=2$ special case.

\subsection{Spreads as messages}

Our boundary is the generalization of $(PG(3,2),\langle\cdot,\cdot\rangle)$ which is $(PG(2n-1,2),\langle\cdot,\cdot\rangle)$.
As message words we choose a spread $\mathcal{S}$ of totally isotropic $n-1$-subspaces living in $PG(2n-1,2)$.
We have $2^n+1$ such subspaces partitioning the $2^{2n}-1$ element point set of $PG(2n-1,2)$. Clearly each of such planes is containing $2^n-1$ points. They correspond to mutually commuting sets of $2^n-1$ observables. 
For a fixed choice of signs (which is always possible for spreads) one has a fixed set of stabilizer states associated to our messages.

As an illustration for these results consult Table 4. of Appendix D. where the $n=3$ case is elaborated. In this case we have a collection of $9$ totally isotropic planes consisting of $7$ points each. They are partitioning the $63$ nontrivial three-qubit observables i.e. the point set of $PG(5,2)$. For a choice of signs and their associated stabilizer states see Table 4.

A method for creating totally isotropic spreads is based on the method presented in Refs.\cite{Demb,Lunardon}.
Note that finding totally isotropic $n-1$-spreads in $PG(2n-1,2)$ is equivalent to finding MUBs for $n$-qubits, a problem whose solution is well-known\cite{Fields,Roetteler,Bandi}. Here we present a very brief finite
geometrically motivated reformulation.

An $n-1$-plane in $PG(2n-1,2)$ is arising from a rank $n$ subspace of $V\equiv V(2n,2)$. Such subspaces are generated by taking the linear span of $n$ linearly independent vectors: ${\rm span}\{v_1,v_2,\dots ,v_n\}\equiv\langle v_1,v_2,\dots,v_n\rangle$. Following the pattern of Eq.(\ref{arrange}) of the $n=2$ case one can write the $n$ linearly independent vectors as a $n\times 2n$ matrix in the form $(Q\vert P)$. In the coordinate patch where ${\rm Det}Q\neq 0$ one can alternatively write a representative as: $({\bf 1}\vert {\bf A})$ where ${\bf 1}$ is the $n\times n$ unit matrix.
Now for the canonical choice of basis in the general construction presented in Ref.\cite{Demb} one can present spreads in the form
\beq
({\bf 1}\vert {\bf A}_{k}),\quad ({\bf 0}\vert {\bf 1}),\quad ({\bf 1}\vert {\bf 0}), \quad k=0,1,2,\dots 2^n-2
\label{hurra}
\eeq
\noindent
where the $2^n-1$ $n\times n$ matrices ${\bf A}_{k}$ are symmetric, invertible and also satisfying the constraint
\beq
{\rm Det}({\bf A}_{k}-{\bf A}_{k^{\prime}})=1,\qquad k\neq k^{\prime}.
\label{jaj}
\eeq
\noindent
In order to understand these constraints recall that $({\bf 1}\vert {\bf A})$ for ${\bf A}$ symmetric represents a totally isotropic plane. See the discussion before Eq.(\ref{laggrass}). Moreover, the constraint of Eq.(\ref{jaj}) ensures that the corresponding planes
are disjoint. The constraint of invertibility is equivalent to the one that also the special plane $({\bf 1}\vert {\bf 0})$
is disjoint from the ones $({\bf 1}\vert {\bf A}_k)$. See also Eq.(\ref{mink}) in this respect.
For an explicit set of $3\times 3$ ${\bf A}_{k}$ matrices corresponding to plane spread in the three qubit case see Eqs.(\ref{elso1})-(\ref{masodik2}) in Appendix D.

According to our philosophy an $n-1$-subspace of a particular totally isotropic spread represents a message word. 
Corrupted messages of the first kind are just transitions to
$n-2,n-3,\dots $ dimensional subspaces, ordinary planes, and finally to points contained in our message $n-1$-subspace.
Note that all these subspaces are totally isotropic ones too. Hence after a particular choice of sign distribution one can associate stabilizers to their corresponding  observables.  
Then at the level of stabilizers the transition from messages to corrupted messages corresponds to transitions from states (rays) to a nested sequence of Hilbert subspaces of states of ever increasing dimension: $1,2,4\dots 2^{n-1}$. This process is the obvious generalization of point errors of the $n=2$ case.

Message corruption of the second kind can also happen by embedding our message planes into ever increasing higher dimensional error subspaces (the analogue of plane errors of the $n=2$ case).
However, thanks to projective duality we expect that one merely has to consider the errors of the first kind.
Errors of the second kind are closer to the spirit of stabilizer codes, since in this case higher dimensional subspaces containing our $n-1$-dimensional message subspaces will necessarily contain points mapped to observables anticommuting with the maximal set
characterizing our message plane. Hence in this case the anticommuting operators of the error subspace will be moving the corresponding stabilizer states to orthogonal subspaces of the Hilbert space. Hence the net result will be again the occurrence of a nested sequence of Hilbert subspaces of states of ever increasing dimension: $1,2,4\dots 2^{n-1}$.
Moreover, learning from the $n=2$ case we conjecture that our error recovery process will be the same for errors of both kind.
Then errors of the first type are geometrically more transparent, on the other hand errors of the second kind are easier to put into the context of stabilizer codes we are already familiar with.
Anyway, since our message subspaces are in the middle of the lattice of subspaces of $PG(2n-1,2)$ with distance smaller than $n$ from the possible error subspaces, our error correction picture dualizes nicely. Hence we can regard both pictures as equivalent descriptions.

\subsection{Relating the boundary to the GHW phase space}

The GHW discrete phase space\cite{W2} for $n$-qubits is ${\bf V}\equiv V(2,2^n)$. It is a two dimensional vector space over the field extension $GF(2^n)$ of $GF(2)$. (Some necessary background information on $GF(2^n)$ is summarized in Appendix C.) This means that a phase space point in the canonical basis with basis vectors $E$ and $F$ is
represented as
\beq
x=qE+pF\in{\bf V},\qquad x\leftrightarrow (q,p), \qquad q,p\in GF(2^n).   
\label{fibit}
\eeq
\noindent
We introduce a name for this two component quantity and call it a {\it fibit}. It is the finite field analogue of a qubit. Introducing it here will turn out to be of some value in Sec. 4.5.
Being a phase space ${\bf V}$ comes equipped with the symplectic form
\beq
\langle x,x^{\prime}\rangle_0=qp^{\prime}+pq^{\prime}\in GF(2^n)
\label{fibitsymp}
\eeq
\noindent
i.e. we have $\langle E,F\rangle_0=
\langle F,E\rangle_0=1$ and the remaining combinations are zero.

The projectivization of the GHW phase-space ${\mathbb P}({\bf V})$ will be denoted by $PG(1,2^n)$. This space is just the space of lines through the origin of our phase space
${\bf V}$. Alternatively one can regard this space as the space of states (rays) of fibits.
Using our (\ref{fibitsymp}) symplectic form turns $PG(1,2^n)$ to a symplectic polar space.
In order to reveal its physical meaning we would like to obtain the boundary from the GHW phase space.
Mathematically this means that we would like to obtain the symplectic polar space in
$(PG(2n-1,2),\langle\cdot,\cdot\rangle)$ from the one
 $(PG(1,2^n),\langle\cdot,\cdot\rangle_0)$. 
This means that apart from doing field reduction, we also have to relate the corresponding symplectic forms.

Since $GF(2^n)$ can be regarded as a vector space over $GF(2)$ using the trace map of Appendix C field reduction means that we can write
\beq
q=\sum_{k=0}^{n-1}q_k\tilde{e}_k,\qquad p=\sum_{m=0}^{n-1}p_me_m,\qquad {\rm Tr}(\tilde{e}_ke_m)=\delta_{km}. 
\label{choicebasis}
\eeq
Notice that a priori we need not have to choose the same {\it field basis} for our coordinates and momenta.
In (\ref{choicebasis}) we have chosen for the momenta a basis and for the coordinates the dual basis.
In this manner for an element $x\in {\bf V}$ with coordinates $(q,p)\in GF(2^n)\times GF(2^n)$ one can associate
an element $v\in V=V(2n,2)$ with $GF(2)$ coordinates arranged in the familiar (\ref{familiar}) form.
More importantly
using the conventions of (\ref{fibit}) and (\ref{choicebasis}) now we can relate the symplectic forms of our spaces as follows
\beq
\langle v,v^{\prime}\rangle={\rm Tr}\left(\langle x,x^{\prime}\rangle_0\right).
\label{valasztas}
\eeq
\noindent

Notice that this choice leads to a mathematically consistent way of relating the relevant symplectic forms\cite{W2,Gil,Lavrauw}.
However, this trick of relating the symplectic forms is not unique. Indeed for any nonzero $\lambda\in GF(2^n)$ one can introduce the map
\beq
L_{\lambda}: GF(2^n)\to GF(2): w\mapsto {\rm Tr}(\lambda w).
\eeq
\noindent
Composing this map with an arbitrary form (quadratic, symplectic) shows that any form on the GHW phase space is nondegenerate if and only if the corresponding form on our boundary is nondegenerate\cite{Gil,Lavrauw}.
Clearly including a $\lambda\neq 1$ in our formula under the trace of Eq.(\ref{valasztas}) amounts to changing e.g. our  dual basis $\tilde{e}_k$ to ${\lambda}\tilde{e}_k$. For a careful consideration of the physical meaning of this point see Ref.\cite{W2}. 

Let us now consider a fixed point in $PG(1,2^n)$. In the ${\bf V}$ perspective this correponds to the set of vectors $\{cx:
c\in GF(2^n)\}$ for a fixed vector $x\in {\bf V}$. After field reduction this set of vectors gives a rank $n$ subspace in $V$, which gives rise to an $n-1$ dimensional subspace of $PG(2n-1,2)$.
Thanks to our assignment of symplectic forms according to the rule 
(\ref{valasztas}) this $n-1$ dimensional subspace will be a totally isotropic one in $PG(2n-1,2)$.
This subspace will give rise to a maximal set of $2^n-1$ nontrivial mutually commuting observables.
There are $2^n$ possible choices of sign distributions for these observables yielding $2^n$ possible stabilizers.
The stabilized vectors of these stabilizers correspond to a particular member of a MUB for $n$-qubits.
There are $2^n+1$ points in $PG(1,2^n)$. They are mapped to the $n-1$ subspaces of a totally isotropic spread of $PG(2n-1,2)$.

For example for the $n=2$ case we have $GF(4)$ with its $4$ elements: $\{0,1,\omega,\omega^2\}$, where $\omega^2+\omega+1=0$ and $\omega^3=1$.
We expand 
\beq 
q=q_0\cdot 1+q_2\cdot \omega^2,\qquad p=p_0\cdot \omega+p_1\cdot 1
\eeq
\noindent
i.e. $(\tilde{e}_0,\tilde{e}_1)=(1,\omega^2)$ and
$(e_0,e_0)=(\omega,1)$, ${\rm Tr}(\lambda)\equiv\lambda+\lambda^2$.
Now for the pair $x=(q,p)=(1,\omega)$ one has $(q_0,q_1,p_0,p_1)=(1010)$. The set 
$\{cx:c\in GF(2^n)\}$ consists of the
 vectors: $(1,\omega),(\omega,\omega^2),(\omega^2,1)$. These correspond to the set of vectors: $(1010),(1111),(0101)$ in $V(4,2)$. Only two of them is linearly independent, hence they define a subspace of rank $2$, i.e. a $1$-dimensional subspace (a line) in $PG(3,2)$.
The mutually commuting set of nontrivial observables in this case is $(YI,YY,IY)$. The $4$ possible stabilizers in generator notation are: $\langle \pm YI,\pm IY\rangle$. In the notation of Eq.(\ref{tilde}) they give rise to the four stabilized states $\{\vert\tilde{0}\tilde{0}\rangle,
\vert\tilde{0}\tilde{1}\rangle,\vert\tilde{1}\tilde{0}\rangle,\vert\tilde{1}\tilde{1}\rangle\}$, forming a particular member of the well-known MUB for two-qubits\cite{W2}.
There are $5$ points in $PG(1,4)$. These points are mapped under the field reduction map to the $5$ lines of the spread of $PG(3,2)$ we used in Table 1.
 
In the (\ref{hurra}) representation of a spread two $n-1$ subspaces play a special role. They are the ones represented as 
$({\bf 1}\vert{\bf 0})$ and
$({\bf 0}\vert{\bf 1})$. The first plane corresponds to mutually commuting sets containing besides $I$ only $Z$ and the second one containing merely $X$ observables.
Having for all of them positive signs, their stabilizers single out the stabilized states $\vert 00\dots 0\rangle$ and
$\vert \overline{0}\overline{0}\dots \overline{0}\rangle$. 
Choosing a convenient distribution of signs, as in Table 1, to the remaining $2^n-1$ elements of the spread one can associate stabilizers and stabilized states in the following manner. 

Let us choose any of these planes. For example the plane $({\bf 1}\vert{\bf 1})$ will do. A convenient stabilizer correponding to a positive plane is generated as $\langle II\cdots Y,\dots ,IY\cdots I,YI\cdots I\rangle$.
The corresponding stabilized state is 
$\vert\tilde{0}\tilde{0}\cdots\tilde{0}\rangle$.
In order to find the remaining stabilizers and their stabilized states one can proceed as follows\cite{W2}. Take  
the order $2^n-1$ generators of $SL(2,2^n)$
of the form
\beq
\Lambda_{\omega}\equiv\begin{pmatrix}\omega&0\\0&\omega^{-1}\end{pmatrix}
\label{Lambda}
\eeq
\noindent 
where $\omega$ is a root of the primitive polynomial defining $GF(2^n)$.
This matrix is acting on a column vector $(q,p)^t$. (In the language of quantum information it is a SLOCC transformation\cite{Dur} on our fibit.)
According to our field reduction procedure this transformation boils down to an $Sp(2n,2)$ transformation
acting on the vector $(q_0,q_1,\dots,p_0,p_1,\dots p_{n-1})^t$. This transformation can be decomposed into transvections, and we can lift these transvections to unitary operators according to the method of Appendix B.
The result of this procedure is the unitary\cite{W2}
\beq
\mathcal{D}(\Lambda_{\omega})=\prod_{j=2}^n\left(C_{1j}^{a_{j-1}}\right)S_{1n}S_{1n-1}\cdots S_{12}
\label{Wootlift}
\eeq
\noindent
where the $a_j$ with $j=0,1,2,\dots n-1$ are $GF(2)$ elements showing up in the (\ref{primike}) primitive polynomial defining the field $GF(2^n)$.
Note that this transformation is the generalization of the $n=2$ one we used in Eq.(\ref{replambda}). 
The unitary of (\ref{Wootlift}) is of order $2^n-1$. Now as in Eq.(\ref{atteres}), starting with $\mathcal{M}_1$ and $\vert\psi_1\rangle$, we can use the action of $\mathcal{D}$ by conjugation to generate all of the remaining message words and stabilized states.
Via this procedure one can generate the message words $\{\mathcal{M}_1,\mathcal{M}_2,\dots,\mathcal{M}_{2^n-1}\}$, and
their corresponding stabilized states $\{\vert\psi_1\rangle,\vert\psi_2\rangle,\dots,\vert\psi_{2^n-1}\rangle\}$, the $n$-qubit analogues of the first three columns of Table 1.

However, based on the paper of Wootters and Sussman\cite{w3} we can elaborate on this idea even better. The idea is to generete cyclically {\it all} of our message words from the reference one $\mathcal{M}_{2^n+1}$ with stabilized state $\vert 00\cdots 0\rangle$. This idea is motivated by the desire to fix the GHW quantum net uniquely by exploiting rotational covariance.
In order to achieve this task one has to construct the $n$-qubit analogue of the unitary operator $U$ of order $2^n+1$
we defined for  $n=2$ in
Eq.(\ref{unitary9}). Unfortunately in the general $n$-qubit case we were unable to construct $U$  in a simple and instructuve manner. In any case employing such an $U$ for the construction of message words is related to a projective version of the Wootters Sussman construction\footnote{For an alternative method for constructing $U$ see also Ref.\cite{Chau}}. This amounts to choosing a special point of $PG(1,2^n)$ from the $2^n+1$ possible ones. Associate to this special point the state $\vert00\cdots 0\rangle$ and then use the $k$-th powers of $U$ for $k=1,2,\dots 2^n$ to assign states even to the remaining points of $PG(1,2^n)$. In this way after field reduction one ends up with our full set of message words in our boundary $PG(2n-1,2)$.  

The attractive feature of this technique is the cyclic generation of message words. See the left hand side of Figure 7. for an illustration. Notice also that since $PG(1,\mathbb{R})$ is the circle $S^1$ the cyclic visualization of the points of $PG(1,2^n)$ gives a geometric intuition of some sort of discretization of a circle based on the technique of using $GF(2^n)$ instead of $\mathbb{R}$. In this naive picture increasing the number of qubits $n$, represents making the discretization ever finer. Based on this observation in the following we picture $PG(1,2^n)$ as a collection of $2^n+1$ points arranged on the perimeter of a circle. See Figure 9 in this respect.
Of course we must bear in mind that what we are representing in this manner is a projective version of a phase space, not a configuration space. Hence though tempting to regard our circle as some sort of discretized analogue of a "boundary circle" analogous to the one showing up in asymptotically AdS spaces in $2\oplus 1$ dimensions, this analogy is misleading.
However, as we will see in the next section, this picture will turn out to be quite instructive for a visualization of our finite geometric space
$PG(2n-1,2)$ as some sort of "boundary" of a "bulk".

\subsection{The bulk as the Grassmannian image of the boundary.}

Having clarified the physical meaning of our boundary now we turn to a similar investigation for the bulk.
Our boundary is the twistor space $PG(2n-1,2)$. It can be regarded as a space "fibered" by $n-1$ dimensional totally isotropic subspaces.
This fibration corresponds to cutting $PG(2n-1,2)$ into $2^n+1$ disjoint slices (fibers). Each fiber is containing $2^n-1$ points 
of our twistor space giving the total number of points $2^{2n}-1$.
Clearly one particular fibration corresponds to a choice of an isotropic spread. Different fibrations of our twistor space correspond to different choices of such spreads. 
As far as error correction is concerned choosing a particular fibration corresponds to a choice of messages in the boundary. 
As an example for an $n=2$ fibration via the canonical spread see Figure 3.
According to this philosophy a mental representation for the totality of fibrations of our twistor space corresponds to conceiving our boundary as a collection of messages in all possible ways.

Alternatively one can think in terms of a familiar physical picture as follows. A slice of the fibered boundary comes naturally equipped with commuting sets of observables of cardinality $2^n-1$ differing only in their sign distribution. 
The $2^n$ possible sign distributions correspond to $2^n$ stabilizers of $2^n$ states forming a basis in the $n$-qubit Hilbert space.
Hence
to each slice of the fibration a basis of the $n$-qubit Hilbert space is associated.
In this picture a particular collection of slices corresponds to a choice of MUB.

An $n-1$-subspace of $PG(2n-1,2)$ is represented by a rank $n$-subspace of $V(2n,2)$ of the form $\langle v_1,v_2,\dots,v_n\rangle\equiv{\rm span}\{v_1,v_2,\dots,v_n\}$. 
Now our finite geometric bulk-boundary correspondence is simply the one provided by the Pl\"ucker map which is of the form
\beq
\pi:{\cal G}(n-1,2n-1)\to P(\wedge^n(V))\qquad {\rm with}\qquad \langle v_1,v_2,\dots,v_n\rangle\mapsto v_1\wedge v_2\wedge\dots\wedge v_n.
\label{Pluckermap}
\eeq
\noindent
i.e. $\pi$ is a map from the Grassmannian of $n-1$-planes in $PG(2n-1,2)$ to the projectivization of the space of $n$-vectors of $V=V(2n,2)$. 
Since the dimension of the vector space of $n$-vectors is $N={2n\choose n}$ this is a map from a Grassmannian to $PG(N-1,2)$.

Now an arbitrary element of $\wedge^nV$ can be written in the form
\beq
\mathcal{P}=\sum_{0\leq k_1<k_2<\dots k_n\leq 2n-1}\mathcal{P}_{k_1k_2\dots k_n}e_{k_1}\wedge e_{k_2}\wedge\cdots\wedge e_{k_n}.
\label{kifejtem}
\eeq
\noindent
However, the $n$-vectors we are interested in are merely the {\it separable} ones. These are the $\mathcal{P}\in\wedge^n V$ ones such that a representation of the form $\mathcal{P}=v_1\wedge v_2\wedge\dots\wedge v_n$ for some $v_1,v_2,\dots v_n$ exists.
$\mathcal{P}$ is separable if and only if the Pl\"ucker relations hold\cite{Hodge}. These are quadratic relations in terms of the components of $\mathcal{P}$, whose explicit form is not important for us.
Clearly over $GF(2)$ the image of the Grassmannian ${\cal G}(n-1,2n-1)$ under $\pi$ is the set of nonzero separable $n$-vectors.
This is an algebraic variety in  $PG(N-1,2)$. We will refer to this variety as "the Grassmannian image".

For the $n=2$ case of Section 2. the Grassmannian image was the Klein quadric. Indeed, in this special case the Pl\"ucker relations boil down to just a single quadratic relation coinciding with Eq.(\ref{hquad}), the defining equation of the Klein quadric. 
Since for our two qubit case we referred to this object as the {\it bulk}
in the following we extend this terminology for arbitrary $n$. Hence from now on we will use the terms "Grassmannian image" and "bulk" as synonyms. 
Hence in mathematical terms we have
\beq
{\rm BULK}\equiv\{v_1\wedge v_2\wedge\cdots\wedge v_n\in \mathbb{P}(\wedge^nV)\quad\vert\quad v_1,v_2,\dots v_n\in V(2n,2)\},
\label{bulkdef}
\eeq
\noindent
where $v_1,\dots ,v_n$ are linearly independent. Our bulk is just the $GF(2)$ version of the Brody-Hughston quantum space-time\cite{Brody1,Brody2}.

On the space $\wedge^n V$ there is a nondegenerate symplectic form $\langle\cdot,\cdot\rangle$ generalizing Eq.(\ref{szimpla2}). 
Namely, let 
\beq
\Omega=\{0,1,\dots n-1,n,n+1,\dots 2n-1\}=\{0,1,\dots,n-1,\overline{0},\overline{1},\dots ,\overline{n-1}\},
\label{Omegalab}
\eeq
\noindent
furthermore let us denote the $n$-subsets of $\Omega$ by $\Lambda,\Sigma$ etc.
Moreover, for any $n$-subset $\Lambda=\{i_1,i_2,\dots i_n\}\subset\Omega$  we introduce the notation
\beq
e_{\Lambda}\equiv e_{i_1}\wedge e_{i_2}\wedge\cdots\wedge e_{i_n}.
\eeq
\noindent
The complement of $\Lambda$ in $\Omega$ will be denoted by $\overline{\Lambda}$.
Hence for example for $n=3$, $\Lambda=\{0,2,\overline{2}\}$ we have
$\overline{\Lambda}=\{\overline{0},1,\overline{1}\}$. Note that since we are over $GF(2)$ when building up $e_{\Lambda}$ the order of the $e_{i_k}$ showing up
in $e_{\Lambda}$ is not important. 
Now our symplectic form is defined as
\beq
\langle e_{\Lambda},e_{\Sigma}\rangle =0,\quad {\rm if}\quad \Lambda\neq \overline{\Lambda},\qquad \langle e_{\Lambda},e_{\overline{\Lambda}}\rangle =1.
\label{simplasympl}
\eeq
\noindent
Clearly the analogue of Eq.(\ref{visszateres}) holds, i.e. we have
\beq
\mathcal{P}\wedge\mathcal{P}^{\prime}=\langle\mathcal{P},\mathcal{P}^{\prime}\rangle e_0\wedge e_1\wedge\cdots\wedge e_{2n-1}.
\eeq
\noindent
Now if we represent an $n-1$-plane $\mathcal{A}$ of the first class in the (\ref{arrange}) form $(\bf{1}\vert A)$ then one can show that the $n\times n$ matrix version of Eq.(\ref{mink}) also holds
i.e. we have
\beq
\langle\mathcal{P},\mathcal{P}^{\prime}\rangle ={\rm Det}(\bf{A}-{\bf A}^{\prime}).
\label{ujra}
\eeq
\noindent

The right hand side of this equation gives rise to the $GF(2)$ version of the chronometric form introduced by Brody and Hughston in Ref.\cite{Brody2}. As discussed in this paper\footnote{For the $n=3$ case see also the paper of Finkelstein\cite{Fink}.} if we work over the field of complex numbers and after employing the reality constraint ${\bf A}={\bf A}^{\dagger}$ the chronometric form can be regarded as the natural generalization of the one which gives rise to the usual Minkowski line element.
Then the authors show that the Grassmannian image of the set of $n-1$-planes in the "hypertwistor space" $\mathbb{C}P^{2n-1}$
forms a manifold of dimension $n^2$. This space is then identified as the complexified compactified version of an object which is called by them the "quantum space-time".
Clearly here we are faced with the $GF(2)$-version of this construction yielding our bulk.
Hence we see that our bulk as the Grassmannian image of the set of $n-1$-planes in $PG(2n-1,2)$ is just the finite geometric version of the Brody-Hughston quantum space-time. Moreover, after implementing the condition of isotropy on these
$n-1$-planes the Grassmannian image yields the $GF(2)$ analogue of  "real quantum space-time"\cite{Brody1,Brody2} living inside the "complex" one, an object that will be discussed in the next subsection. 

When two $n-1$-planes of our boundary intersect then the corresponding two points in the bulk have a degenerate separation (see Figure 2. for an illustration).
These two points are represented by the two separable $n$-vectors $\mathcal{P}$ and $\mathcal{P}^{\prime}$.
The degeneracy ($r$) of the separation, encapsulated in the structure of Eq.(\ref{ujra}), is just the rank of the separation matrix $\bf{A}-{\bf A}^{\prime}$.
If we denote by $\delta$ the dimensionality\footnote{The dimension is understood projectively. Hence if we have no intersection then $\delta=-1$.} of the intersection of the two $n-1$-planes, then we have $r=n-1-\delta$.
Hence for no intersection we have maximal rank $r=n$, and in this case ${\rm Det}({\bf A}-{\bf A}^{\prime})\neq 0$.
This is the case we have for two elements of a particular spread represented by $({\bf 1}\vert {\bf A})$ and $({\bf 1}\vert {\bf A}^{\prime})$.
On the other hand the different degrees of degeneracy are distinguished by the possible values $r=1,2,\dots n-1$.
In these cases ${\rm Det}({\bf A}-{\bf A}^{\prime})=0$, i.e. the separation matrix is degenerate.
Clearly, unlike in the $n=2$ Minkowski case where we have merely light-like and non-light like separation, in the general case we have an intricate structure of degenerate cases. 
This is illustrated for $n=3$ in Figure 8. 
\begin{figure}[pth!]
\centerline{\includegraphics[width=10truecm,clip=]{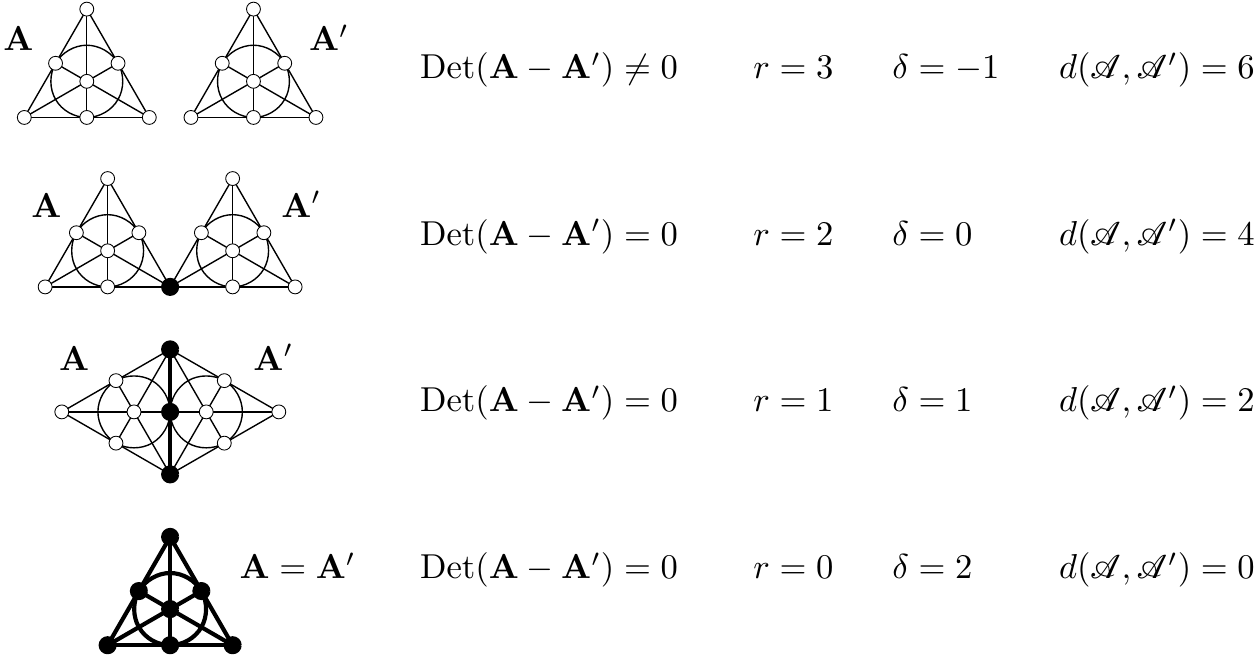}}
\caption{The intersection properties of $n-1$ planes in the boundary for $n=3$ interpreted as different levels of degeneracy for the separation of points in the bulk. We have $r$ standing for the rank, and $\delta$ standing for the projective dimension of the intersection. We have $r=n-1-\delta$. According to Eq.(\ref{dist}) for the corresponding vector subspaces $\mathcal{A}$ and $\mathcal{A}^{\prime}$ of rank three one can define a distance. This is given in the last column.}
\end{figure}

\subsection{The real slice of the bulk}

As we have seen the bulk is the Grassmannian image of the boundary, which we can also regard as the image of a space fibered by messages.
There are many different ways that the boundary can be fibered. In finite geometric terms there are many possible spreads one can choose for the twistor space $PG(2n-1,2)$.
However, our choice for the set of fibrations is special. Indeed, this set is the one of isotropic spreads.
The Grassmannian image of this set of messages comprises a special subset of points in the bulk.
Now we would like to present a geometric characterization of this subset of points.

According to our philosophy these points in the bulk are comprising the codewords encoding the messages.
For the $n=2$ case Figure 3 shows that the set of points corresponding to the codewords gives an ovoid
inside the doily. The set of all possible such codewords is the set of ovoids covering all the points of the doily.
We have also seen that the doily inside the bulk is just the finite geometric version of a real slice of complexified compactified Minkowski space-time. 
A particular set of codewords inside this real slice is geometrically described by Eq.(\ref{intersect}). In Section 3.6
we elucidated the physical meaning of taking this code slice. We have seen that taking this slice is the same technique by which in twistor theory conformally flat spacetimes are obtained.
In the following we would like to see how one can achieve a generalization of these ideas for arbitrary $n$.

First of all we define the real slice of the bulk as
\beq
\Re({BULK})
\equiv\{v_1\wedge v_2\wedge\cdots\wedge v_n\in \mathbb{P}(\wedge^nV)\quad\vert\quad \langle v_i,v_j\rangle=0\}.
\label{zaldivar}
\eeq
\noindent
This is the Grassmannian image of the space of maximal totally isotropic (Lagrangian) subspaces of $PG(2n-1,2)$, i.e. the image of the Lagrangian Grassmannian\cite{LagCode}.
Since Lagrangian subspaces are (locally) represented by the arrangement $({\bf 1}\vert {\bf A})$ where
 ${\bf A}$ is symmetric, our real slice is an algebraic subvariety of the bulk of dimension $n(n+1)/2$.
The number of real points in the bulk is\cite{LagCode} $\prod_{i=1}^{n}(1+2^i)$, i.e. for $n=2,3,4$ we have $15,135,2295$ points.

Let us now introduce the map\footnote{This map is the standard one used in Ref.\cite{Fulton} on page 260. However, over $GF(2)$
no factors of $(-1)^{i+j-1}$ are needed.}
 ${\partial}_k: \wedge
^kV\to\wedge^{k-2}V$ by the formula\cite{Gow}
\beq
{\partial}_k(v_1\wedge v_2\wedge\dots\wedge v_k)=\sum_{i<j}\langle v_i,v_j\rangle v_1\wedge\cdots\wedge\hat{v}_i\wedge\cdots\wedge\hat{v}_j\wedge\cdots\wedge v_k
\eeq
\noindent
where as usual $V=(V(2n,2),\langle\cdot,\cdot\rangle)$, and hats refer to omission of the corresponding factors.
We extend this action of $\partial$ linearly on elements of $\wedge^kV$.
As an example one can check that acting with ${\partial}_2$ on $\mathcal{P}$ of Eq.(\ref{irottpe}) the constraint ${\partial}_2\mathcal{P}=0$ gives rise to the one of Eq.(\ref{laggrass}).
This constraint is precisely the constraint of isotropy of a line of the boundary $PG(3,2)$, appearing at the level of Pl\"ucker coordinates. 

To see how this constraint works over $GF(2)$ for $n\geq 3$
let us define the $2$-subsets $B_1,\dots B_n$ of $\Omega$ by $B_i\equiv\{i\overline{i}\}$ where $0\leq i\leq n-1$.
Then if $\Lambda$ is an $n$-set containing $s\geq 1$ $2$-subsets $B_{i_1},\dots B_{i_s}$ we have 
\beq
{\partial}(v_{\Lambda})=\sum_{j=1}^sv_{\Lambda-B_{i_j}},
\eeq
\noindent
where in the following we omit the subscript of $\partial$ when from the context it is clear on which space it acts.
One can check\cite{Fulton} that since ${\partial}$ commutes with the action of $Sp(2n,2)$ the kernel of this map 
\beq
W\equiv
{\rm ker}\left({\partial}\cap\wedge^nV\right)=\{\mathcal{P}\in\wedge^nV\vert
{\partial}\mathcal{P}=0\}
\label{kernelke}
\eeq
\noindent
forms a representation of $Sp(2n,2)$.
One can also show\cite{LagCode} that
\beq
\Re({\rm BULK})={\rm BULK}\cap {\mathbb P}(W).
\label{rebulkdef}
\eeq
\noindent

 When the characteristic of the field is zero the module $W$ is known to be irreducible\cite{Fulton} however, over $GF(2)$ it is not.
Indeed, over $GF(2)$ we have $\partial^2=0$ and $\partial(\wedge^{n+2}V)\subset W$.
Moreover\cite{Gow}, $\partial(\wedge^{n+2}V)$ has codimension $2^n$ in $W$ and the quotient 
$W/\partial(\wedge^{n+2}V)$ forms an irreducible representation of $Sp(2n,2)$ of dimension $2^n$.
A detailed illustration of these results for $n=3$ can be found in Ref.\cite{LPS}.
Here we are content with a brief illustration of this case.

Let $\mathcal{S}\in\wedge^5V$, where $V=V(6,2)$ \beq
\mathcal{S}=ae_{01\overline{1}2\overline{2}}+be_{10\overline{0}2
\overline{2}}+ce_{20\overline{0}1\overline{1}}+ \alpha
e_{\overline{0}1\overline{1}2\overline{2}}+\beta
e_{\overline{1}0\overline{0}2 \overline{2}}+\gamma
e_{\overline{2}0\overline{0}1\overline{1}}
 \eeq
\noindent where $e_{01\overline{1}2\overline{2}}=e_0\wedge
e_{1}\wedge\cdots\wedge e_{\overline{2}}$ etc. Now the
action of ${\partial}$ is just omission of blocks like
$1\overline{1}$ hence \beq \mathcal{R}\equiv
\partial\mathcal{S}=a(e_{01\overline{1}}+e_{02\overline{2}})+
b(e_{10\overline{0}}+e_{12\overline{2}})+\cdots +
\gamma(e_{\overline{2}0\overline{0}}+e_{\overline{2}1\overline{1}})
=\mathcal{R}_{01\overline{1}}e_{01\overline{1}}+\cdots
+R_{\overline{2}1\overline{1}}e_{\overline{2}1\overline{1}}.
\label{Rconstraints}
 \eeq
 \noindent
Now we see that $\partial^2=0$ and that $\mathcal{R}$ contains terms for
which \beq \mathcal{R}_{01\overline{1}}=\mathcal{R}_{02\overline{2}}=a,\quad
\mathcal{R}_{10\overline{0}}=\mathcal{R}_{12\overline{2}}=b,\quad\cdots\qquad
\mathcal{R}_{\overline{2}0\overline{0}}=\mathcal{R}_{\overline{2}1\overline{1}}=\gamma. 
\label{kakukk6}
\eeq
\noindent
As a next step writing out the expansion of $\mathcal{P}\in\wedge^3V$, one can see that precisely the 
same constraints are arising from $\partial\mathcal{P}=0$ for the $12$ components of 
$\mathcal{P}$ having the same index structure as the ones of $\mathcal{R}$ in (\ref{kakukk6}). 
The remaining $8$ components of $\mathcal{P}$ contain no blocks at all and they are of the form
\beq
(\mathcal{P}_{012}, \mathcal{P}_{0\overline{12}},\mathcal{P}_{\overline{0}1\overline{2}},
\mathcal{P}_{\overline{01}2},
\mathcal{P}_{\overline{012}},\mathcal{P}_{\overline{0}12},
\mathcal{P}_{0\overline{1}2},\mathcal{P}_{01\overline{2}}),
\label{vastag}
\eeq
\noindent
where we arranged these $8$ components in a (\ref{familiar}) form also in accordance with the structure of the (\ref{symp}), (\ref{simplasympl}) symplectic forms. 

Now one can write 
\beq
\mathcal{P}=\mathcal{P}_{\psi}+\mathcal{R},\qquad 
\mathcal{P}_{\psi}=\mathcal{P}_{012}e_{012}+\mathcal{P}_{01\overline{2}}e_{01\overline{2}}+\cdots+\mathcal{P}_{\overline{012}}e_{\overline{012}}.
\label{tools}
\eeq
\noindent
Here $\mathcal{P}_{\psi}$ with $8$ terms and with the index structure reminiscent of a three-qubit state $\psi$ 
\beq
(\psi_{000},\psi_{001},\cdots,\psi_{111})\leftrightarrow (\mathcal{P}_{012},\mathcal{P}_{01\overline{2}},\cdots,\mathcal{P}_{\overline{012}})
\label{reminiscent}
\eeq
\noindent
serves as a representative for the coset
$W/\partial(\wedge^{n+2}V)$ with $n=3$.
According to the results of Ref.\cite{Gow} the left hand side of Eq.(\ref{tools}) is also valid for 
arbitrary $n\geq 3$. This time $\psi$ is an $n$-qubit state with $2^n$ components embedded into $\wedge^nV$, $V=V(2n,2)$.
On this $2^n$ dimensional space $Sp(2n,2)$ acts irreducibly. In the literature the corresponding module is called the spin module\cite{Gow,Coss1}.
The embedding trick of $n$-qubits into $n$-vectors of a $2n$-dimensional vector space is frequently used in quantum information\cite{Beny,Djoko1,Djoko2,Oeding,HSL}. 

The $\mathcal{P}=\mathcal{P}_{\psi}+\mathcal{R}$ representation of an element of $W$ where $\mathcal{R}\in\partial({\wedge^{n+2}V})$ is of great value when we connect the $N={2n\choose n}$ component vectors ($N$ is an even number) of Pl\"ucker coordinates to $N/2$-qubit observables. First of all one can prove\cite{Gow} that for $\mathcal{P}_{\psi}+\mathcal{R}$ and $\mathcal{P}^{\prime}_{\psi^{\prime}}+\mathcal{R}^{\prime}$ we have $\langle\mathcal{R},\mathcal{R}^{\prime}\rangle=0$ meaning that $\partial({\wedge^{n+2}V})$ is totally isotropic with respect to our symplectic form.
Since clearly $\langle \mathcal{P}_{\psi},\mathcal{R}^{\prime}\rangle =\langle \mathcal{P}^{\prime}_{\psi^{\prime}},\mathcal{R}\rangle=0$ we get
\beq
\langle\mathcal{P},\mathcal{P}^{\prime}\rangle =\langle \mathcal{P}_{\psi}, \mathcal{P}^{\prime}_{\psi^{\prime}}\rangle.
\label{leegyszer}
\eeq
\noindent
Hence using the dictionary of Eq.(\ref{op}) the commutation properties of the corresponding $N/2$-qubit observables are entirely determined by the $2^{n-1}$-qubit observables encoded in the $2^n$ components of the $\mathcal{P}_{\psi}$ part.
However, according to Eq.(\ref{ujra}) the commutation properties of observables in the bulk are encapsulating the causal structure of $\Re({\rm BULK})$. Then we conclude that the causal structure of $\Re({\rm BULK})$ is determined by the $\mathcal{P}_{\psi}$ part encoding $2^{n-1}$-qubit observables.  

One can define a quadratic form on $\wedge^nV$ by setting $Q(e_{\Lambda})=0$ for all $n$-subsets of $\Omega$ and letting $\langle\cdot,\cdot\rangle$ of (\ref{simplasympl}) be the polarization of $Q$, i.e. a relationship between them of the form of Eq.(\ref{kapcsolat}) should hold.
The explicit form of $Q$ can then be given by the familiar (\ref{kvad}) expression with $N={2n\choose n}$ used instead of $n$ in that formula. This time the $q,p$ pairs showing up in Eq.(\ref{kvad}) should be replaced by pairs like 
$\mathcal{P}_{\Lambda},\mathcal{P}_{\overline{\Lambda}}$ where $\Lambda$ and $\overline{\Lambda}$ are covering all of the $N$ possible $n$-subsets of $\Omega$.
A structure of this kind is exemplified by Eq.(\ref{hquad}).

Then one can show\cite{Gow} that for $n\geq 3$ the quadratic form $Q$ is identically zero on $\partial(\wedge^{n+2}V)$ moreover, 
$\partial(\wedge^{n+2}V)$ is the radical in $W$. As a result of this there is a {\it nondegenerate} quadratic form
$\hat{Q}$
on the spin module $W/\partial(\wedge^{n+2}V)$ of index $2^{n-1}$ invariant under $Sp(2n,2)$.
This means that for a decomposition of the (\ref{tools}) form we have $Q(\mathcal{R})=0$.
Hence
\beq
Q(\mathcal{P})=Q(\mathcal{P}_{\psi}+\mathcal{R})=Q(\mathcal{P}_{\psi})+Q(\mathcal{R})+\langle \mathcal{P}_{\psi},\mathcal{R}\rangle=Q(\mathcal{P}_{\psi})\equiv\hat{Q}(\mathcal{P}_{\psi}).
\label{visszakvad}
\eeq
\noindent

Let us now show that for $\mathcal{P}\in \Re({\rm BULK})$ one has $\hat{Q}(\mathcal{P}_{\psi})=0$.
By virtue of Eq.(\ref{visszakvad}) in order to show this we have to verify that $Q(\mathcal{P})=0$.
Let us first recall that
\beq
{2n\choose n}=\sum_{k=0}^n{n\choose k}{n\choose n-k}.
\label{binom}
\eeq
\noindent
The cardinality of the set of all possible $n$-subsets $\Lambda$ of $\Omega$ is ${2n\choose n}$ and Eq.(\ref{binom}) shows that we have a partition of this set to $n$-subsets containing $k=0,1,\dots n$ overlined elements. 
Let us now represent a point $\mathcal{P}$ of 
$\Re({\rm BULK})$
by an $n-1$-subspace $\mathcal{A}$ with its $n\times 2n$ matrix of the form $({\bf 1}\vert{\bf A})$. Label the first $n$ columns with the numbers $\{0,1,2,\dots,n-1\}$ and the last $n$ ones with their overlined versions.
Consider now the terms of $Q(\mathcal{P})$ of the form ${\mathcal{P}}_{\Lambda}{\mathcal{P}}_{\overline{\Lambda}}$ where the $n$-subset $\Lambda$ is containing $k$ overlined elements from $\Omega$. Let us fix these $k$-numbers. Then there are ${n\choose n-k}$ possibilities to remain for building up an $n$-subset $\Lambda$ of that kind.  Since $\mathcal{P}\in \Re({\rm BULK})$
then $A^t=A$. Taking this into account and using Laplace expansion for the calculation of the determinant of ${\bf A}$ with respect to the fixed $k$-rows one can check that the
the sum of the ${n\choose n-k}$ terms ${\mathcal{P}}_{\Lambda}{\mathcal{P}}_{\overline{\Lambda}}$ coincides with ${\rm Det}(\bf A)$.
Since one can choose the fixed $k$ numbers in ${n\choose k}$ different ways what we get in this manner is 
the sum of ${n\choose k}$ terms all of them calculating ${\rm Det}(\bf A)$. 
Now $Q(\mathcal{P})$ is the sum of ${2n\choose n}/2$ terms of the form ${\mathcal{P}}_{\Lambda}{\mathcal{P}}_{\overline{\Lambda}}$ where $\Lambda$ is running through all $n$-subsets containing $k$ overlined elements of $\Omega$ where $k$ is running from $0,1,\dots \lfloor{n/2}\rfloor$.
Note that for $n$ even in order to avoid double counting one has to take into account the case of $s=n/2$ only ${2s\choose s}/2$ times. 
Note in this respect that for $n=2s$
\beq
\sum_{k=0}^{s-1}{2s\choose k}+\frac{1}{2}{2s\choose s}=2^s.
\label{kiegbinom}
\eeq
\noindent
The upshot of these considerations is that no matter whether $n$ is even or odd  
$Q(\mathcal{P})$ is calculating ${\rm Det}(\bf A)$ in an {\it even} number of times. Since we are over $GF(2)$ this means that $Q(\mathcal{P})=0$ as claimed.

We can conclude that the points of $\Re({\rm BULK})$ are lying inside a hyperbolic quadric  $Q^+(2^n-1,2)$. These points are satisfying the constraint 
$\hat{Q}(\mathcal{P}_{\psi})=0$
where $\mathcal{P}_{\psi}$ is defined by Eq.(\ref{tools}).
Moreover, since the $2^n$ different Pl\"ucker coordinates of $\mathcal{P}_{\psi}$ are alligned in pairs according to Eq.(\ref{familiar}), these points correspond up to sign to symmetric $2^{n-1}$-qubit observables.
In summary we have the situation
\beq
\Re({\rm BULK})\subseteq Q^+(2^n-1,2)\subset{\rm BULK}\subset PG(N-1,2),\qquad n\geq 3.
\label{include}
\eeq
\noindent
where $N={2n\choose n}$ and equality is obtained\cite{LPS} for $n=3$.
Notice that the $n=2$ case is special, since in this case
\beq
\Re({\rm BULK})\subseteq Q^+(5,2)\equiv{\rm BULK}\subset PG(5,2).
\label{spacetimespec}
\eeq
\noindent

Let us summarize. 
Just like in the $n=2$ case even for $n\geq 3$ the real part of the bulk can be embedded into a unique hyperbolic quadric\cite{Coss2}.
As a result of this, commuting $2^n-1$-tuples of $n$-qubit observables (assigned to Lagrangian {\it subspaces} of the boundary) are mapped to symmetric $2^{n-1}$-qubit ones (assigned to the {\it points} of $\Re({\rm BULK})$).
There is an ambiguity in choosing the signs of the observables on both sides of the correspondence.
However, the ambiguity in signs over the boundary can be eliminated. In this case to a member of the fibration one can associate a MUB system in a unique manner\footnote{Provided we follow the rotationally covariant method of Ref.\cite{w3} for building up a quantum net.}. 
Then the remaining sign ambiguity on the  $\Re({\rm BULK})$ side is manifesting itself in the following problem: To a particular basis of the boundary which eigensubspace of the corresponding symmetric $2^{n-1}$-qubit observable one should associate?
A possible answer to this question in the $n=2$ case is encapsulated in Figure 7. However, in order to find a satisfactory solution to the problem of fixing the quantum net also for the bulk for $n\geq 3$ further elaboration is needed.

\subsection{Glueing up the real slice of the bulk from $n$ fibits}

We have seen in Section 4.2 that the process of blowing up the points of the projectivization of the GHW phase space of $n$-qubits yields our boundary.
In arriving at this picture instead of using $2n$ component vectors over $GF(2)$ we were using two component ones over the field extension $GF(2^n)$. For such two component objects we coined the term: {\it fibit}.
Now we show that it is possible to give an elegant characterization of the real slice of the bulk as an $n$-fibit system glued together in a peculiar manner.
This trick also provides a neat way of characterizing our codewords in the bulk as partial ovoids.

\begin{figure}[pth!]
	\centerline{\includegraphics[width=6truecm,clip=]{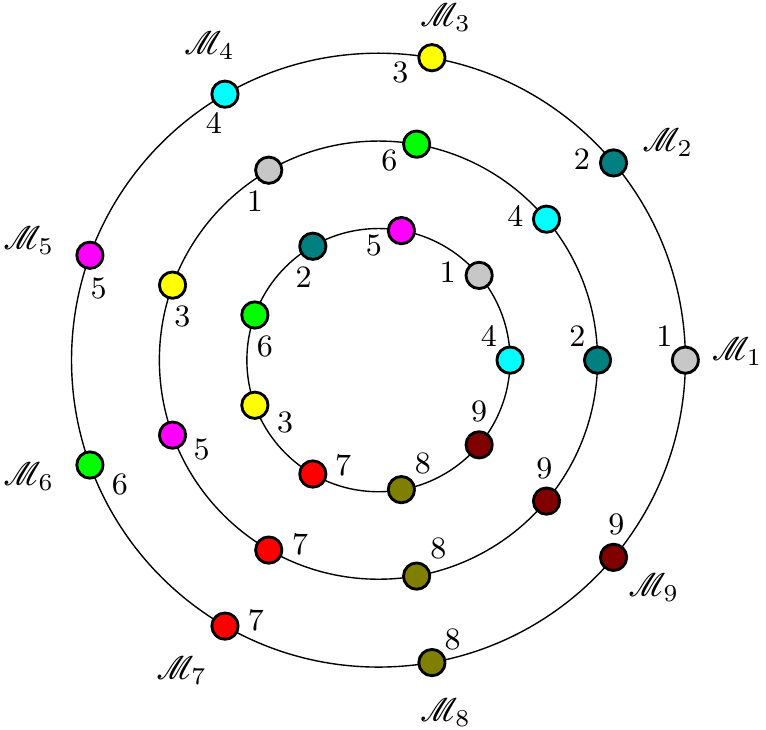}}
\caption{A representation for the state space of $n$ twisted fibits for $n=3$.
The blow up of the projectivization of the  
GHW phase space $PG(1,2^n)$ can be identified with our boundary. This $PG(1,2^n)$ can be represented as a collection of $2^n+1$ points arranged on the outmost circle of the figure. Starting from the state $\vert 00\dots 0\rangle$ associated to ${\mathcal M}_{2^n+1}$, via applying a unitary $U({\mathcal R})$,
to the remaining  points of $PG(1,2^n)$
quantum states can be associated 
in a unique manner.
The $n$ copies of $PG(1,2^n)$s are arranged in an onion like pattern reflecting the twisted tensor product structure of Eq.(\ref{Frobi}). The hyperbolic quadric of Eq.(\ref{include}) containing the real part of our bulk is then created by a glueing up process of this pattern effected by the Pl\"ucker map as explained at the end of Section 4.6.
Combined with Table 4 of Appendix D the colours and the numbers showing up in the figure indicate which state $\vert\varphi_a\rangle$ with $a=1,2,\dots 9$ of Eqs.(\ref{state1})-(\ref{state3}) is to be associated with the coloured points.
}
\end{figure}

The main idea is to take $n$ copies of the GHW phase space ${\bf V}\equiv V(2,2^n)$ equipped with the (\ref{fibitsymp}) symplectic form and the natural tensor product action of $n$ copies of the group $G=Sp(2,2^n)=SL(2,2^n)$. This action is the finite geometric analogue of the action of the SLOCC group\cite{Dur} familiar from quantum information.  
Let us then consider the rank $2^n$ vector space ${\bf V}_n={\bf V}\otimes {\bf V}\otimes\cdots\otimes{\bf V}$ over $GF(2^n)$.
Starting from the $\langle\cdot,\cdot\rangle_0$ symplectic form on ${\bf V}$ of Eq.(\ref{fibitsymp}) one can define a symplectic form $\langle\cdot,\cdot\rangle_n$ on ${\bf V}_n$ by the linear extension of the formula
\beq
\langle x_0\otimes x_1\otimes\cdots\otimes x_{n-1},y_0\otimes y_1\otimes\cdots\otimes y_{n-1}\rangle_n=\prod_{i=0}^{n-1}\langle x_i,y_i\rangle_0.
\label{nsympl}
\eeq
\noindent
One can associate a quadratic form $Q_n:{\bf V}_n\to GF(2^n)$ to this symplectic form via the usual method, as explained in the text showing up in the paragraph after Eq.(\ref{leegyszer}).
Now the state space of $n$ distinguishable fibits is formed by considering
$({\mathbb P}({\bf V}_n),\langle\cdot,\cdot\rangle_n)$ where ${\mathbb P}({\bf V}_n)=PG(2^n-1,2^n)$. 
Inside this state space the space
\beq
\Sigma_{2,2,\dots ,2}\equiv{\mathbb P}({\bf V})\times
{\mathbb P}({\bf V})\times\cdots\times{\mathbb P}({\bf V})=PG(1,2^n)\times PG(1,2^n)\times\cdots\times PG(1,2^n)
\label{nGHW}
\eeq
\noindent
describes the state space of $n$ {\it separable} fibits. The embedding of this space called $\Sigma_{2,2,\dots,2}$ into the full state space is called the Segre embedding\cite{Thas,Pepe}.
On the right hand side of Eq.(\ref{nGHW}) we have $n$ copies of the projectivization of the GHW phase space, i.e. up to the blowing up process $n$ copies of our boundary. 
If we represent an element of ${\bf V}$ as in Eq.(\ref{fibit}) 
then the left hand side of Eq.(\ref{nGHW}) can be represented by elements of the form $x_0\otimes x_1\otimes\cdots\otimes x_{n-1}$, defined up to multiplication by a nonzero element of $GF(2^n)$.

One can introduce correlation into this picture by demanding that the fibits are indistinguishable. Since we are over an extension of $GF(2)$ the only way to do this seems to be to considering "bosonic fibits" via restricting attention to merely those elements of the left hand side of  Eq.(\ref{nGHW}) that are of the form $x\otimes x\otimes\cdots\otimes x$. This amounts to considering as components of a state
the combinations of the form $q^{\alpha}p^{\beta}$ where $\alpha+\beta =n$, i.e. the coordinates are monomials in $q,p$ of degree $n$.
Geometrically this corresponds to using instead of the Segre variety the Veronese variety\cite{Thas,Pepe} inside $PG(2^n-1,2^n)$.

However, for finite fields one also has the possibility of forming twisted tensor products\cite{Sternberg}. This trick introduces a peculiar form of correlation into our system of fibits (see Figure 9.).
The idea is as follows.
Unlike the case of the Veronese embedding where we take merely the diagonal of the Segre embedding let us consider
instead elements of  $PG(2^n-1,2^n)$ of the form
\beq
x\otimes x^2\otimes x^4\otimes\cdots\otimes x^{2^{n-1}},\qquad x^j\equiv q^jE+p^jF,\quad j=1,2,\dots,2^{n-1}.
\label{Frobi}
\eeq
\noindent

Let us elaborate on the rationale for doing this. We would like to obtain an alternative characterization of the structures of the previous section (which are ones over $GF(2)$) as subgeometries of ones over the extension $GF(2^n)$.
In the case of the boundary this approach was fruitful because we managed to relate the boundary to the GHW discrete phase space.
Now we would like to arrive at a similar level of understanding for the codewords embedded in the real part of the bulk.
For a very brief reminder on subgeometries see Appendix E.

	 Over finite fields we have the Frobenius automorphism of the field extension $[GF(2^n):GF(2)]$.
This is simply the map $\phi:\lambda\mapsto \lambda^2$ of order $n$ for an arbitrary element $\lambda\in GF(2^n)$.
For an element of $g\in G$ denote by $g^{\phi^i}$ the $2\times 2$ matrix with its entries raised to the $2^i$th power where $i=0,1,\dots n-1$. Denoting by ${\bf V}^{\phi^i}$ the representation space on which $g^{\phi^i}$ acts
the twisted tensor product is idefined as the tensor product ${\bf V}\otimes{\bf V}^{\phi}\otimes {\bf V}^{\phi^2}\otimes\cdots\otimes
{\bf V}^{\phi^{n-1}}$.
Hence we explicitely have the twisted action $\varrho$ of $G$ as
\beq
\varrho:x_0\otimes x_1\otimes\cdots\otimes x_{n-1}\mapsto gx_0\otimes g^{\phi}x_1\otimes\cdots\otimes g^{\phi^{n-1}}x_{n-1},\qquad g\in G=Sp(2,2^n).
\label{twistagain}
\eeq
\noindent
Let us also define the map\footnote{See Appendix E. and Ref.\cite{Coss1} for some elaboration on the meaning of this map.} $\sigma$ on ${\bf V_n}$ by
\beq
\sigma(x_0\otimes x_1\otimes\cdots\otimes x_{n-1})=x_{n-1}^{\phi}\otimes x_0^{\phi}\otimes\cdots\otimes x_{n-2}^{\phi}.
\label{semilin}
\eeq
\noindent
Then one can show the following set of results\cite{Coss1}. First of all $\varrho$ and $\sigma$ commutes on ${\bf V}_n$ and as a result of this the set
$\mathcal{W}\equiv\{w\in {\bf V}_n: \sigma(w)=w\}$ is a $GF(2)$ subspace of ${\bf V}_n$ stabilized by $\varrho$.
Then any vectors in $\mathcal{W}$ that are linearly independent over $GF(2)$ are linearly independent over $GF(2^n)$. $\mathcal{W}$ has dimension $2^n$ over $GF(2)$ and spans ${\bf V}_n$ over $GF(2^n)$.
Now all of the vectors which are of the form as the one on the left hand side of Eq.(\ref{Frobi}) are fixed by $\sigma$ and hence are in ${\mathcal W}$ and these vectors span $\mathcal{W}$.

These technical results ensure that one can choose $2^n$ basis states of the (\ref{Frobi}) form such that they represent the $\mathcal{P}_{\psi}$ part of Eq.(\ref{tools}) of the previous section\footnote{ Recall that the $\mathcal{P}_{\psi}$ part is residing in the subgeometry $PG(2^n-1,2)$.} in the $PG(2^n-1,2^n)$ language. 
In particular one can show\cite{Coss1} that the $Q_n$ quadratic form is nondegenerate and takes its values in $GF(2)$ on restriction to ${\mathcal W}$. Hence the elements satisfying $Q_n(w)=0$ for $w\in{\mathcal W}$ define a quadric in a $PG(2^n-1,2)$ subgeometry  of $PG(2^n-1,2^n)$. The points of the real part of our bulk are forming a subset of the points of this quadric. This gives an alternative characterization of the result of Eq.(\ref{include}).

\subsection{Characterizing the code variety in terms of fibits}

Let us shed some light on the advantage of the viewpoint provided by the trick of using the extension field $GF(2^n)$.
Let us take a point $x\in{\bf V}$ represented by the pair $(q,p)$. Embed this vector into $V(2n,2^n)$ by forming a vector $v$ as follows
\beq
v\leftrightarrow (q,0,0,\dots,0,p,0,0,\dots, 0),\qquad q,p\in GF(2^n)
\label{peca}
\eeq
\noindent
where 
in the notation of Eq.(\ref{familiar}) $q$ is in the $q_0$th and $p$ is in the $p_0$th slot.
In accord with this in the following we use a labelling convention for the basis vectors of $V(2n,2^n)$ based on the set $\Omega$ of Eq.(\ref{Omegalab}).
Apply now $n-1$ times the map $\sigma$ on $v$ according to the rule\footnote{By an abuse of notation we use the same letter $\sigma$ for the map of Eq.(\ref{semilin}) and a new one introduced here. However, the former is acting on ${\bf V}\otimes\cdots \otimes{\bf V}$ and the latter on ${\bf V}\oplus\cdots \oplus{\bf V}$. Moreover, the basis states of the direct sum are labelled according to the pattern of (\ref{Omegalab}). However, since both of these maps are acting via effecting a cyclic permutation and applying the Frobenius automorphism, we used the same letter to refer to them.}   
\beq
v^{\sigma}=(0,q^2,0,\dots,0,0,p^2,0,\dots,0),
\nonumber
\eeq
\noindent
\beq
v^{\sigma^2}=(0,0,q^4,\dots,0,0,0,p^4,\dots,0),
\nonumber
\eeq
\noindent
\beq
v^{\sigma^{2^{n-1}}}=(0,0,0,\dots,q^{2^{n-1}},0,0,0,\dots,p^{2^{n-1}}).
\nonumber
\eeq
\noindent
Now the vectors fixed by $\sigma$ are precisely of the form $v+v^{\sigma}+\dots +v^{\sigma^{2^{n-1}}}\in V(2n,2^n)$.
Such vectors form a $2n$ dimensional vector space $V^{\prime}\simeq V(2n,2)$ i.e. a one over $GF(2)$.
Of course we are not pretending that $V^{\prime}$ and our original vector space $V$ of Section 1 are the same.
The only point we would like to emphasize here is that this description based on $V^{\prime}$ gives rise to an alternative characterization\cite{Coss2} of spreads of the associated ${\mathbb P}(V^{\prime})$.
Indeed, if we consider the $n-1$ plane
\beq
\Pi(v)\equiv\langle v,v^{\sigma},\dots,v^{\sigma^{2^{n-1}}}\rangle
\eeq
\noindent
then the set
\beq
S=\{\Pi(v): v\in V(2n,2^n)\}
\eeq
\noindent
where $v$ assumes the peculiar (\ref{peca}) form, forms a spread of the ${\mathbb P}(V^{\prime})$ subgeometry\cite{Lunardon,Pepe}.

We can represent the $n-1$ plane $\Pi(v)$ in the (\ref{arrange}) form $(Q\vert P)$ where now $Q$ and $P$ are $n\times n$ diagonal matrices with entries of the form $q^j$ and $p^j$ with $j=1,2,\dots ,2^{n-1}$.
Now when calculating the Grassmannian image of this plane all but $2^n$ of the $n\times n$ submatrices have determinant zero.
These give rise to $2^n$ nonvanishing Pl\"ucker coordinates of our codewords. However, this time we are over $GF(2^n)$.
Hence for  neither $q$ nor $p$ zero the $2^n$ coordinates are of the form $q^{\alpha}p^{\beta}\in GF(2^n)$ where 
$\alpha,\beta =0,1,\dots 2^n-1$ and
$\alpha+\beta =1+2+\dots +2^{n-1}=2^n-1$. In the exceptional cases of the rays of the points $(0,p)$ and $(q,0)$ only one of the Pl\"ucker coordinates is one and the remaining coordinates are zero.
Note that the pair $(q,p)$ of $PG(1,2^n)$ is defined merely up to multiplication with a nonzero $\lambda\in GF(2^n)$. 
However, $(\lambda q)^{\alpha}(\lambda p)^{\beta}=(\lambda)^{2^n-1}q^{\alpha}p^{\beta}=q^{\alpha}p^{\beta}$ hence our Pl\"ucker coordinates are defined merely up to a $GF(2)$ scalar multiple (the only scalar multiple is $1$ over $GF(2)$). We have the same property for the exceptional cases.
This indicates that in spite of working over $GF(2^n)$ our set of $2^n+1$ points with $2^n$ such coordinates is contained in a subgeometry\cite{Lunardon} of $PG(2^n-1,2^n)$ isomorphic to $PG(2^n-1,2)$.

Introduce the notation $\Delta=PG(2^n-1,2)$ for this subgeometry. Then it is known that the set of points in $PG(2^n-1,2^n)$ with coordinates
$q^{\alpha}p^{\beta}$ with $\alpha+\beta =2^n-1$ is the algebraic variety\cite{Lunardon,Pepe} 
\beq
{\mathcal V}_{2,n}\equiv \Delta\cap \Sigma_{2,2,\dots,2}= \Delta\cap{\rm BULK}
\label{nuvariety}
\eeq
\noindent
where $\Sigma_{2,2,\dots,2}$ is defined in (\ref{nGHW}).
This formula should already be familiar from Eq.(\ref{intersect}) in connection with the $n=2$ case, where the Klein quadric $Q^+(5,2)$ has been identified as the ${\rm BULK}$.
There this formula defined the codewords ${\mathcal C}_a$ arising as the Grassmannian image of the message words $\mathcal{M}_a$. In that case the codewords formed an ovoid. 
Based on this observation in the following we will argue that it is plausible to identify ${\mathcal V}_{2,n}$ for $n\geq 3$ as the {\it code variety}.

In order to justify this notice that the $2^n+1$ points arising as the Grassmannian image of a spread consisting of $2^n+1$ $n-1$-planes are lying on a hyperbolic quadric $Q^+(2^n-1,2^n)$, and since they lie in a subgeometry $PG(2^n-1,2)$ they further lie on a quadric  $Q^+(2^n-1,2)$
in that subgeometry\cite{Coss1}.
Indeed this is an alternative construction of the quadric $\mathcal{Q}$ we are already familiar with from Eq.(\ref{visszakvad}) of the previous section. 
Now a partial ovoid of a quadric $\mathcal{Q}$ is a pointset $O$ of the quadric which has at most one point in common with each maximal totally isotropic subspace of $\mathcal{Q}$. In particular an ovoid is arising when each maximal isotropic subspace of the quadric has exactly one point in common with $O$.
For $n=3$ it is known that the Grassmannian image of our spread in the boundary is an ovoid in the bulk\cite{Coss1}.
However, for $n>3$ we have no ovoids on our hyperbolic quadrics. What we have are partial ovoids showing up as pairwise non-orthogonal points. A partial ovoid is said to be complete if it is maximal with respect to set theoretic inclusion.
There is an upper bound for the size of a partial ovoid of a classical polar space\cite{Blokhuis}.
In our special case it turns out that this bound is  precisely $2^n+1$. According to this result for $n\geq 2$ our set of codewords residing on our hyperbolic quadric form a complete partial ovoid attaining this bound\cite{Coss1}.
Interestingly according to Remark 3.7. of Ref.\cite{Coss2} the quadric on which $Sp(2n,2)$ acts via its spin representation
maybe recovered from the knowledge of a complete partial ovoid, i.e. a collection of codewords.
Hence the code variety ${\mathcal V}_{2,n}$ and ${\mathcal Q}$ related to the real part of our bulk are intimately connected.
However, the precise mathematical relationship between these objects needs further clarification.

Let us give the explicit form of the elements of ${\mathcal V}_{2,n}$ in fibit notation.
An $n$-fibit state can be written as
\beq
\psi=\psi_{00\dots 0}E\otimes E\otimes\cdots\otimes E+\psi_{10\dots 0}F\otimes E\otimes\cdots E+\dots +\psi_{11\dots 1}F\otimes F\otimes\cdots\otimes F
\label{nfibitstate}
\eeq
\noindent
see Eq.(\ref{fibit}) for definitions.
It is better to write such expressions in decimal notation using {\it reversed binary labelling}. By this we mean that after introducing the basis vectors $\{\varepsilon_0,\varepsilon_1,\dots,\varepsilon_{2^n-1}\}$ where
\beq
\varepsilon_0=E\otimes E\otimes\cdots\otimes E,\quad
\varepsilon_1 =F\otimes E\otimes\cdots\otimes E,\quad \varepsilon_2 =E\otimes F\otimes\cdots\otimes E,\quad \dots
\eeq
\noindent
for $q\neq 0$ and $p\neq 0$ an element of ${\mathcal V}_{2,n}$ can be written as
\beq
\psi=x\otimes x^2\otimes\cdots\otimes x^{2^{n-1}}=\psi_0\varepsilon_0+\psi_1\varepsilon_1+\dots +\psi_{2^n-1}\varepsilon_{2^n-1}\equiv \psi_{\alpha}\varepsilon_{\alpha},\qquad \psi_{\alpha}=q^{2^n-\alpha-1}p^{\alpha}
\label{decimalstate}
\eeq
\noindent
where summation for repeated indices of $\alpha$ in the range $0,1,\dots 2^n-1$ is understood.
Clearly if neither $q$ nor $p$ is zero then we always have $\psi_0=\psi_{2^n-1}=1$. For elements of the form $(q,0)$ we have $\psi_0=1$ and the remaining components zero. Likewise for elements of the form $(0,p)$ we have $\psi_{2^n-1}=1$ and the remaining ones are zero.
Let us denote these special $n$-fibit states as follows
\beq
\psi^{[2^n]}\equiv \varepsilon_{2^n-1},\qquad \psi^{[2^n-1]}\equiv \varepsilon_0.
\label{lasttwo}
\eeq
\noindent
Now if neither $q$ nor $p$ is zero then for an element $x\leftrightarrow (q,p)$ representing an equivalence class of $PG(1,2^n)$ can be represented in the form $(1,\omega^k)$ for some $k=0,1,\dots 2^n-2$.
Hence in these cases we get further $2^n-2$ states of the form
\beq
\psi^{[k]}\equiv \omega^{\alpha k}\varepsilon_{\alpha},\qquad \alpha=0,1,\dots 2^n-1,\quad k=0,1,\dots 2^n-2,
\label{remaining}
\eeq
\noindent
i.e. $\psi^{[k]}_{\alpha}\equiv \omega^{\alpha k}$ and we have summation for $\alpha$.
Finally what we get is a collection of $2^n+1$ separable $n$-fibit states. Since the $n$-fibit states are separable they are elements of the Segre variety $\Sigma_{2,2,\dots,2}$. Moreover, each fibit has $2^n$ coordinates taken from $GF(2^n)$ with a special form. These give rise to coordinates of $2^n+1$ points in the subgeometry $\Delta=PG(2^n-1,2)$ of $PG(2^n-1,2^n)$. 
This reasoning clarifies the meaning of the first equality of Eq.(\ref{nuvariety}).

We emphasize that although our $n$-fibit states are separable, hence they contain no entanglement however, they are {\it correlated}.
For example had we choosen the constraint $\psi_{\alpha}=q^{n-\alpha}p^{\alpha}$ we would have obtained
a correlated bosonic $n$-fibit state. Correlation in this picture would have come from regarding the fibits indistinguishable.
Likewise in our case the twisted tensor product structure introduces a new type of correlation characteristic merely of fibit systems. This correlation manifests itself in the constraint of choosing amplitudes in the nontrivial cases in the form: $\psi_{\alpha}=q^{2^n-\alpha-1}p^{\alpha}$.

From the paragraph following Eq.(\ref{reminiscent}) we know that the space of separable $n$-fibit states inside ${\bf V}\otimes {\bf V}\otimes\cdots\otimes{\bf V}$ can be embedded into $\wedge^n({\bf V}\oplus{\bf V}\oplus\cdots\oplus{\bf V})$ as the space of separable $n$-vectors. Moreover, this embedding is respecting the SLOCC group action. The space of separable $n$-vectors forms the Grassmannian image of the space of $n-1$-subspaces in $PG(2n-1,2^n)$.
In our special case when we have separable $n$-vectors of the form $x\wedge x^2\wedge\cdots\wedge x^{2^{n-1}}$ it turns out that they are embedded into to the Grassmannian image of the set of $n-1$-subspaces of $PG(2n-1,2)$. This image is precisely our bulk.
This clarifies the meaning of our second equality sign in Eq.(\ref{nuvariety}).

Let us now consider the symplectic form of Eq.(\ref{nsympl}) on ${\bf V}\otimes {\bf V}\otimes\cdots\otimes{\bf V}$.
This induces a corresponding symplectic form on $\wedge^n({\bf V}\oplus{\bf V}\oplus\cdots\oplus{\bf V})$ which is a generalization of the (\ref{ujra}) symplectic form we have used in Eq.(\ref{leegyszer}).
Using this form let us demonstrate that the $2^n+1$ points are pairwise non-collinear hence they form a partial ovoid.
Write $\psi=\psi_{\alpha}\varepsilon_{\alpha}$ and $\varphi=\varphi_{\beta}\varepsilon_{\beta}$.
Then using (\ref{nsympl}) we have
\beq
\langle\psi,\varphi\rangle_n=\sum_{\alpha=0}^{2^{n-1}-1}(\psi_{\alpha}\varphi_{2^n-1-\alpha}+\varphi_{\alpha}\psi_{2^n-1-\alpha}).
\eeq
\noindent
Now we fix two numbers $k>k^{\prime}$ taken from the set $\{0,1,2,\dots,2^n-2\}$. Define $\xi\equiv\omega^{k-k^{\prime}}$
and use $\omega^{2^n-1}=1$ to calculate
\beq
\langle\psi^{[k]},\psi^{[k^{\prime}]}\rangle_n=\sum_{\alpha=0}^{2^n-1}\omega^{k\alpha}\omega^{2^n-1-k^{\prime}\alpha}=
\sum_{\alpha=0}^{2^n-1}\xi^{\alpha}=1+\xi+\cdots +\xi^{2^{n-1}}=\frac{\xi^{2^n}-1}{\xi -1}=1.
\eeq
\noindent
Since this quantity for $k\neq k^{\prime}$ is not zero, then the points represented by the $n$-fibit states $\psi^{[k]}$ and
$\psi^{[k^{\prime}]}$ are not collinear in the bulk. Moreover, we have
\beq
\langle\psi^{[2^n]},\psi^{[2^n-1]}\rangle_n=\langle\psi^{[2^n]},\psi^{[k]}\rangle_n=\langle\psi^{[2^n-1]},\psi^{[k]}\rangle_n=1
\eeq
\noindent
then their bulk representatives are also non collinear. Since the bound of Ref.\cite{Blokhuis,Coss1} is also attained then the $2^n+1$ points corresponding to our codewords are forming a complete partial ovoid.

Let us also demonstrate that our codewords are indeed lying on the hyperbolic quadric defined as the zero locus of $Q_n$.
The explicit form of the quadric related to the symplectic form as in Eq.(\ref{kapcsolat}) is given by the formula
\beq
Q_n(\psi)=\sum_{\alpha=0}^{2^{n-1}-1}\psi_{\alpha}\psi_{2^n-1-\alpha}.
\eeq
\noindent
Clearly for the states of Eq.(\ref{lasttwo}) we have $Q(\psi)=0$ hence the corresponding points are lying on the quadric.
Moreover, the states $\psi^{[k]}$ are also lying on the quadric since $\omega^{\alpha k}\omega^{({2^n-1-\alpha})k}=1$ and we have an even number of terms in the sum. Then we also have $Q_n(\psi^{[k]})=0$ for all  $k=0,1,\dots 2^n-2$.

Let us finally have a look at Figure 9. It summarizes the twisted tensor product structure associated with a collection of specially correlated $n$-fibit separable states for $n=3$. Consider a particular fibit $x$ encoding a message. Then the state 
$\psi=x\otimes x^2\otimes\cdots\otimes x^{2^{n-1}}$ is of that specially correlated kind. Alternatively one can consider ${\mathcal P}_{\psi}=x\wedge x^2\wedge\cdots\wedge x^{2^{n-1}}$ which represents a single point in the bulk: a codeword.
As $x$ is running through all the points in $PG(1,2^n)$ we are exploring the $2^n+1$ points of the outer circle of Figure 9. 
The blow up of these points gives the message words of the boundary. When running through these points the $n$-fibit state $\psi$ changes accordingly.
Then ${\mathcal P}_{\psi}$ is running through $2^n+1$ of our bulk points representing codewords, i.e. through the points of the variety ${\mathcal V}_{2,n}$ inside the bulk.
However, one cannot associate any of these bulk points to a point of the onion-like structure of Figure 9. due to the correlated nature of ${\mathcal P}_{\psi}$. In order to have a pictorial representation of the situation the best we can do is to distribute these points along a radial direction.  
In this picture proceeding deeper in this direction is associated with taking higher and higher powers\footnote{Recall that an alternative presentation for $\psi$ is ${\mathcal P}_{\psi}=x^{\phi^0}\wedge x^{\phi^1}\wedge\cdots\wedge x^{\phi^{n-1}}$.} of the Frobenius automorphism $\phi$. 

On the other hand proceeding along cyclically on the boundary circle ${\mathbb P}({\bf V})=PG(1,2^n)$ amounts to applying the transformation ${\mathcal R}$ of order $2^n+1$
\beq
{\mathcal R}=L^{2^n-1}=\begin{pmatrix}1&1\\b^{-1}&b^{-1}+1\end{pmatrix},\qquad L=\begin{pmatrix}1&b\\1&0\end{pmatrix}
\eeq
\noindent
to a fibit $x$ of ${\bf V}$.
This transformation takes circles to circles in ${\bf V}$ centered at the origin. It is based on the existence of a primitive polynomial for $GF(2^n)$ of the form $z^2+z+b=0$.
Such a transformation\cite{w3} defines circles in ${\bf V}$ of the form $q^2+qp+bp^2=c$.
Notice that for $\lambda\neq 0$ the transformation $(q,p)\mapsto (q^{\prime},p^{\prime})=\lambda(q,p)$ produces a line through the origin, hence a point in $PG(1,2^n)$. For the points on this ray the equation of the circle for the primed coordinates is of the same form except for the radius is scaled as $c\mapsto \lambda^2 c$. This shows that one can partition the nonzero elements of ${\bf V}$ into $2^n-1$ circles centered in the origin. Each of these circles projects to the boundary circle ${\mathbb P}({\bf V})$.
For the $n=3$ case shown in Figure 9. we have the matrix
\beq
{\mathcal R}=\begin{pmatrix}1&1\\\omega^6&\omega^4\end{pmatrix}.
\eeq
\noindent
In this case using Table 4 of Appendix D one can check that the choice $b=\omega$ gives rise to the cyclic permutation of the points in the outer circle in the form $(162783459)$.
Moreover, the reader can identify the precise form of the partitioning of the $63$ observables to seven circles with their radii corresponding to the seven nonzero values of $c$. 
These results give some support for calling the blow up of $PG(1,2^n)$ as some sort of boundary of our bulk.

\subsection{Error correction for $n\geq 3$}

In our error correction scheme we have a set of messages $\mathcal{S}=\{\mathcal{M}_1,\dots,\mathcal{M}_{2^n+1}\}$ which is a set of totally isotropic $n-1$ subspaces forming a partition of the boundary. 
$\mathcal{S}$ is encoding a collection of maximal sets of mutually commuting $n$-qubit observables, alternatively a set of states taken from a particular MUB.
See for example  Eqs.(\ref{stab1})-(\ref{state3}) of Appendix D.
The Grassmannian image of this collection of messages is a set $\mathcal{O}=\{\mathcal{C}_1,\dots,\mathcal{C}_{2^n+1}\}$ of 
codewords which is a set of points in the real part of our bulk. $\mathcal{O}$ is encoding a collection of $2^{n-1}$-qubit observables defined up to sign. For an example of such observables see the bold faced parts of Eqs.(\ref{ovi1})-(\ref{ovi3}) of Appendix D.

The set of codewords $\mathcal{O}$ forms an algebraic variety in the bulk. The code variety is given in terms of Eq.(\ref{nuvariety}) as the complete intersection of the bulk with a projective space $\Delta=PG(2^n-1,2)$. Hence an algebraic characterization of $\Delta$ fixes our set of codewords. A characterization of this kind is encapsulated in Eqs.(\ref{decimalstate})-(\ref{remaining}), where the $2^n$ coordinates of our $2^n+1$ codewords are given in terms of the amplitudes of a separable but correlated $n$-fibit state.

Let us now generalize the results of Section 3.5. By error correction we mean the following process. Suppose that we are intending to send a message $\mathcal{M}$, but instead a corrupted one $\mathcal{E}$ is received which is a subspace such that ${\rm dim}({\mathcal E})\leq n-1$. We suppose that this corrupted subspace satisfies
$d({\mathcal M},\mathcal{E})<n$  where for the definition of our metric see Eq.(\ref{dist}).
Let $\Omega(\mathcal{E})$, with elements denoted by $\chi$, be the space of subspaces of dimension $n-1$ with nonzero intersection with $\mathcal{E}$, i.e. ${\rm dim}(\chi\cap \mathcal{E})\geq 0$.
$\Omega(\mathcal{E})$ is called a Schubert variety\cite{Hodge}. It is known\cite{Hodge} that its Grassmannian image is given by a linear section of the bulk with a projective subspace $\Sigma$.  
Then by calculating ${{2n}\choose n}-2^n$ linear equations with coefficients taken from the vectors spanning $\mathcal{E}$ we calculate $\Sigma\cap\Delta$. Only one of the vectors contained in $\Sigma\cap\Delta$ will be separable i.e. an element of our bulk. This will be our codeword $\mathcal{C}$. The corresponding message word of the boundary then will be ${\mathcal{M}}$.
Since our message words are located in the middle of the lattice of subspaces of our boundary this error correction procedure dualizes nicely. The result of this is that we can use the same algorithm for recovery also from errors with ${\rm dim}({\mathcal E})> n-1$\cite{Stokes}.

Let us clarify these abstract results via recapitulating some features of the $n=2$ case. The message ${\mathcal M}$ to be 
sent is a projective line with dimension one, corresponding to a vector subspace of rank two.  
For point errors we have the corrupted message  ${\mathcal{E}}$ which is projectively a point of dimension zero, corresponding to a vector subspace of rank one. Hence we have ${\rm dim}(\mathcal{E})=0<1=n-1$, and $d(\mathcal{M},\mathcal{E})=2+1-2\cdot 1=1\leq n$. 
Now $\Omega(\mathcal{E})$ consists of all those lines which go through the point $\mathcal{E}$. As we know there are seven such lines with their Grassmannian images as seven points in the bulk forming an $\alpha$-plane. Hence one of the $\alpha$-planes on the right hand side of Figure 5. is the Grassmannian image of the Schubert variety $\Omega(\mathcal{E})$. Physically it is the finite geometric analogue of a complexified light ray.
Notice that the set of lines going through the point $\mathcal{E}$ located in the boundary is represented by the set of bulk points $\mathcal{P}^{(\mathcal{E},\chi)}$, where $\chi$ with changing coordinates is representing the other point of the line.
$\mathcal{P}^{(\mathcal{E},\chi)}$ with the coordinates of $\chi$ running, span the linear subspace $\Sigma$.
Notice that the $n=2$ case is very special since $\Sigma$ is contained in the bulk, which is not true for $n$ arbitrary.
Let us now recall the first two expressions on the left hand side of Eq.(\ref{important}).
These give rise to two constraints $\mathcal{P}^{(\mathcal{E},\chi)}$ should satisfy. These give rise to the ${4\choose 2}-2^2$ 
linear equations for finding the unique point $\mathcal{C}$ in $\Sigma\cap\Delta$.
As we demonstrated in Section 3.5 the explicit solution of these equations in the form of Eq.(\ref{elegans}) then determines $\mathcal{M}$
completing the error correction process.

Notice that had we choosen in the boundary merely the three isotropic lines going through the point $\mathcal{E}$ (see Figure 4.), we would have obtained three points in the bulk with light like separation, i.e. an ordinary light ray.
Generally we could have defined $\Omega_0(\mathcal{E})$ as the space of totally isotropic subspaces of dimension $n-1$ with nonzero intersection with $\mathcal{E}$ as a generalized Schubert variety. Hence for point errors, in the $n=2$ case, the Grassmannian image of this object is a light ray. In Section 3.5. for the $n=2$ case we have demonstrated that the error correction process works no matter whether it is based on lines or isotropic lines, i.e. whether for the representation of errors we use $\Omega(\mathcal{E})$ or $\Omega_0(\mathcal{E})$. According to the results of Ref.\cite{Stokes} the error correcting process based on the use of the image of $\Omega(\mathcal{E})$ generalizes for $n>2$. We conjecture that it also works in the case of errors represented by $\Omega_0(\mathcal{E})$.
An argument in favour of our conjecture is supported by the fact that the explicit error correction algorithms of Ref.\cite{Stokes} are designed to decode any subspace code whose Pl\"ucker coordinates give an algebraic variety of the Pl\"ucker embedding of the Grassmannian.
Since according to Eq.(\ref{zaldivar}) the Lagrangian Grassmannian of totally isotropic $n-1$ planes is mapped to the real part of the bulk which is an algebraic variety\cite{LagCode}, this generalization should work. However, further elaboration on this point is still needed.

It would be interesting to understand better the causal structure of points in the bulk representing all types of errors in the boundary.
For this one should explore the properties of Grassmannian images of the Schubert varieties $\Omega_0(\mathcal{E})$ for
all types of errors $\mathcal{E}$. Due to projective duality it would be enough to consider merely errors with 
${\rm dim}({\mathcal E})\leq n-1$.
In order to highlight the new stuctures present in the general case
in the following we elaborate on the $n=3$ case.

For $n=3$ we have partitions of the boundary by totally isotropic (Lagrangian) planes. The set of such planes is of cardinality $135$.
It is also known\cite{Edge} that we have $960$ such partitions of the boundary. One of such partitions is the one given by
Eqs.(\ref{stab1})-(\ref{state3}) with each partition containing $9$ message words. 
The Grassmannian image of the set of messages comprises the set of codewords. They are points of the quadric $\mathcal{Q}$ which is a hyperbolic one $Q^+(7,2)$. This object for $n=3$ coincides with the real part of the bulk. Since the number of points of this quadric is $135$, one can map bijectively the set of messages to the set of points of the real part of the bulk\cite{LPS}.
Now in our case we have two types of nontrivial errors: point errors and line errors.

Let us first have a point error.
At the end of Appendix D. the reader can find a detailed investigation for the point error $\mathcal{E}$ correponding to the observable $YXZ$.
This point is featuring the message plane $\mathcal{M}_4$. Since $Sp(6,2)$ acts transitively on the set of Lagrange planes this example can be used to show that for an arbitrary $\mathcal{M}$ and its point error $\mathcal{E}$ there are altogether $15$ Lagrangian planes to be considered for $\Omega(\mathcal{E})$. Such planes are intersecting in either the point $\mathcal{E}$, or in a line containing this point. In order to understand the finite geometric structures associated with  $\Omega(\mathcal{E})$ one can proceed as follows.

The image of these $15$ planes gives $15$ points in the real part of the bulk. All of these points are light-like separated, forming the light cone of $\mathcal{M}$. However, this light cone also has an interesting substructure.
Indeed, it turns out that this set of $15$ points can be given the incidence structure of a doily. Collinearity for this doily is defined when three commuting four-qubit observables produce the identity up to a sign. This situation is the Grassmannian image of the case when the corresponding boundary planes are intersecting in a line.

It is easy to identify why this happens by noticing that
the fifteen planes comprising $\Omega(\mathcal{E})$ are also forming the incidence structure of a doily.
This time incidence is defined when triples of planes are intersecting in special lines. We have $15$ such triples producing $15$ special lines. The speciality of such lines is that they have a common point: $\mathcal{E}$. A special case for the complementary situation occurs when a set of {\it five} planes is intersecting just in the common error point. In the language of incidence in the doily this case corresponds to having five disjoint lines forming a partition (a spread) of the doily. It turns out that the Grassmannian image of this spread forms an ovoid of the {\it bulk} doily.
Hence suprisingly for the $n=3$ case we have found at the level of a single point error of $\mathcal{M}$ the same structure as we have found for the $n=2$ case for the full collection of messages (see Figure 3.). See Appendix D. for more details.

Let us then have a line error such that this line is fully contained in one of the messages say $\mathcal{M}$.
For a line error $\mathcal{E}$ of this kind in the boundary we have three isotropic planes containing it one of them being $\mathcal{M}$ itself. The image of 
$\Omega_0(\mathcal{E})$ in the bulk is a line which is just the usual light ray of $\mathcal{C}$ consisting of the three points. 
When we consider $\Omega(\mathcal{E})$ instead, it turns out that its image in the bulk will be a $3$-dimensional subspace in the bulk. This is the analogue of a complexified light ray containing our ordinary one\footnote{Generally, the set of $n-1$ dimensional projective subspaces of $PG(2n-1,2)$ intersecting in a fixed $n-2$ dimensional projective subspace, corresponds to the points of an $n$-dimensional projective subspace of the bulk. For a proof see Lemma 6. of Ref.\cite{Stokes}.}. Then unique decoding is clearly possible. 
However, one can also have the situation when the error line is intersecting with a message plane $\mathcal{M}$ merely in a point, a situation which can also happen for point errors.
It is easy to see (Appendix D.) that due to this intricacy there is no possibility for a unique decoding for the case ${\rm dim}(\mathcal{M}\cap\mathcal{E})=0$. Indeed, the general statement for $n$ odd is\cite{Gorla}: that a unique decoding is only possible for ${\rm dim}(\mathcal{M}\cap\mathcal{E})\geq (n-1)/2$.

One can try to evade this subtlety by restricting the set of possible errors on physical grounds to ones that are living entirely inside a message word. For example in the $n=3$ case this would mean that we should only allow such point an line errors that are lying within a message plane. Since a message is associated with a complete set of $2^n-1$ nontrivial commuting observables, then an error of this restricted type would encapsulate the notion of transmitting incomplete {\it subsets} of a particular complete set of observables to the receiver. 
In this picture the physical representative of the lack of unique decoding of the previous paragraph is our inability to reconstruct the complete set of observables in a unique manner from the data provided by the transmitted commuting subset. The reason for this is clear: we also have observables in this set anticommuting with some elements of the message set. 
In the example of Appendix D. the transmitted incomplete commuting set is $\{YXZ, XIY, ZXX\}$ with each of its three observables
is featuring merely one from the three messages $\mathcal{M}_{2,4,5}$. Hence the remaining two ones are acting like error operations transforming the ray of the stabilizer states $\vert\varphi_{2,4,5}\rangle$  to different subspaces of the Hilbert space.

Note however, that having anticommuting error operators is not tantamount to failure of error correction.
Indeed, in Section 3.8. we have seen that the effect of dualization of errors, is to have larger dimensional subspaces than the message ones.
This happens for instance in the $n=3$ case when for a plane-message we have a space-error which is containing the message. In this case we necessarily have new observables that are anticommuting with our maximal set.
However, being dual to correctable line errors, these space errors can also be corrected\cite{Stokes}.

Clearly apart form the exploration of the bulk causal structure of errors there are many more interesting issues amenable for elaboration. For example: can we find a nice algebraic generalization of the explicit error correcting algorithm culminating in the appearance of Eq.(\ref{nice})? 
How to characterize algebraically  the space of possible messages (totally isotropic spreads) of the boundary? For $n=2,3$ these spaces are of cardinality $6$ and $960$. Although an explicit construction is given for them in Section 3.6 and in Ref.\cite{Edge}, but a unified treatment for $n>3$ is not known to us. We must bear in mind that this is a central question, since according to the basic philosophy we adopted here and also in AdS/CFT space-time is a {\it collection} of error correcting codes. Hence one should consider not merely one particular geometric subspace code, but a collection of such codes. 
This is an unusual way of looking at error correction via subspace codes.

Finally, Section 3.8 we initiated a study of error correction as a topic related to properties of quantum nets. An interesting project would be the exploration of the physical meaning and the possible generalizations the ideas culminating in Eq.(\ref{meaning}) imply.
Though in our conclusions we would like to share with the reader some speculations on such issues, at this point we stop and postpone the exploration of many interesting pathways for a future work.

\section{Conclusions}

\subsection{Summary of results}

According to the idea of holography, bulk space-time structure is encoded into degrees of freedom residing in the boundary.
In the AdS/CFT literature it has been suggested that this encoding is accomplished by the bulk-boundary correspondence functioning as a collection of error correcting codes. 
In this paper we investigated a finite geometric toy model illustrating how this might be realized.

In a minimalist representation (see Figure 9.) our finite geometric model of the boundary is the projectivization of the GHW phase space for $n$-qubits. Equivalently we can regard this space as the space of states of a single fibit, a two component quantity with elements taken from the field extension $GF(2^n)$ of $GF(2)$. 
The space of states of a fibit is equipped with a symplectic form inherited from the symplectic structure of the GHW phase space.
 Taking $n$-copies of this space and glueing them together in a specially correlated fashion creates a minimalist representation for our bulk.
This coarse grained representation of the bulk is that of a state space of $n$-indistinguishable fibits based on the twisted tensor product structure of Eq.(\ref{twistagain}). 

Doing field reduction however, reveals an intricate fine-structure of Figure 9.
This blowing up process generates from the space of states for a fibit , $PG(1,2^n)$, our boundary which is $PG(2n-1,2)$.
Under field reduction the the set of points of $PG(1,2^n)$ is mapped to a set of Lagrangian $n-1$-subspaces of $PG(2n-1,2)$ partitioning the point set of the boundary.
Such a partitioning or fibration of the boundary is called a spread of $n-1$-subspaces.
We regarded this set of $n-1$-subspaces of cardinality $2^n+1$ as a set of message words for a subspace code, a spread code\cite{Gorla}.

The next item in the elaboration of the fine structure of Figure 9 is the identification of the bulk and the correspondence map relating the bulk to the boundary.
In this paper we have choosen as our correspondence map the (\ref{Pluckermap}) Pl\"ucker map. The bulk then is simply the image of the set of $n-1$-subspaces of the boundary under this map.
We called the bulk image of the set of message words of the boundary, codewords.
The codewords form a special set of $2^n+1$ points (a partial ovoid) in the bulk.
We called this collection of special points as the code variety with its explicit algebraic description given by Eq.(\ref{nuvariety}).
The code variety has an elegant physical interpretation in terms of a collection of $2^n+1$ separable but correlated $n$-fibit states with explicit form given by Eqs.(\ref{decimalstate})-(\ref{remaining}).

The $n=2$ case of our correspondence is just a finite geometric analog of the twistor correspondence.
In this case the bulk is the $GF(2)$ version of complexified and compactified Minkowski space-time, the boundary $PG(3,2)$ playing the role of the twistor space.
The fibration of the boundary with messages here, corresponds to the so-called twistor fibration of ${\mathbb C}P^3$
to complex lines,  ${\mathbb C}P^1$s, Riemann spheres. 
In the $n\geq 3$ cases the boundary $PG(2n-1,2)$ provides a finite geometric version of hypertwistor space. Now the bulk corresponds to the Brody-Hughston quantum space-time structure\footnote{Note that the adjective "quantum" in this context could be misleading. For the original meaning see Refs.\cite{Brody1,Brody2}.}
equipped with the chronometric form of Eq.(\ref{ujra}).

Using the simple map of Eq.(\ref{op}) one can associate observables to the points of both the boundary and the bulk.
However, these associations are plagued by sign ambiguities. We pointed out that these ambiguities are related to the existence of Mermin square-like configurations\cite{Mermin1,Mermin2,HS,LHSmagic}.
As a result of this, there is no way of having a unique assignment of observables to the {\it points} of our spaces featuring the correspondence.
However, there is a unique association of observables to their {\it isotropic spreads}. In particular there is a unique association of observables to the message 
$n-1$-subspaces of our boundary.
We have shown that the possibility of carrying out this unique assignment of observables is ensured by the existence of a unique rotationally covariant construction of the GHW quantum net\cite{w3}.
Physically this process of creating the quantum net is equivalent to constructing mutually unbiased basis systems (MUBs) starting from a distinguished basis. This construction is effected by conjugating the mutually commuting set of observables related to this basis by a suitable unitary transformation of order $2^n+1$.
As a result, to a particular collection of message $n-1$-subspaces one can associate a collection of stabilizers,  and a collection of (generally entangled) stabilized states in a unique manner. For an illustration for $n=3$ of such a collection see Eqs.(\ref{stab1})-(\ref{state3}). 
In the minimalist representation of our bulk the states assigned to the messages, are also associated to the points of the concentric circles representing the bulk in Figure 9. The precise form of this association is controlled by the Frobenius automorphism of the field extension.

The error correction process relating the bulk and the boundary is based on a {\it geometric subspace code}\cite{Stokes}.
This term means that unlike in conventional discussions concerning subspace codes\cite{Gorla}, in order to facilitate efficient decoding the Grassmannian image of the codewords is also considered.
In this paper we have made a somewhat unusual twist to the usual nomenclature of subspace codes. Indeed in our interpretation the message words belonging to a spread are residing in the boundary and the codewords are residing in the bulk\footnote{In Ref.\cite{Gorla} the {\it the codewords} are belonging to a spread.}.
This is in accord with the conception of regarding the bulk as a higher dimensional code space encoding  messages located in the lower dimensional boundary. For example in the $n=2$ case
the message words are rank two {\it subspaces} of $V(4,2)$ and the codewords are rank one {\it subspaces} of $V(6,2)$. 
The reader should compare this with the situation in e.g. the classical Hamming code\cite{Nielsen} where the message words are {\it vectors} 
of $V(4,2)$ and the codewords are {\it vectors} of $V(7,2)$.

In the language of subspace codes the ${\mathcal E}$ errors are subspaces of the boundary of different dimension with nontrivial intersection with one of the message subspaces ${\mathcal M}_1,\dots ,{\mathcal M}_{2^n+1}$. 
The aim is to ensure a unique recovery, meaning to find conditions and an explicit decoding algorithm for identifying ${\mathcal M}$ 
from the above list for a given ${\mathcal E}$.
In order to complete this task one has to identify the bulk image of the space of $n-1$-subspaces with nonzero intersection with
${\mathcal E}$. In the case of succesful recovery this image is connected to the light-cone (causal) structure of a unique codeword ${\mathcal M}$.
In the $n=2$ case we have given a detailed analysis of the error structure corresponding to point errors and plane errors.
We studied the bulk image of different types of errors. These images indeed turned out to be connected to the light cone (causal) structure of the codewords representing the messages subjected to certain errors (see Figures 5 and 6).
For the $n=3$ case we revealed a fine structure of point errors, and some subtleties obstructing unique error correction.
In this case a unique decoding is only possible for ${\mathcal M}$ and ${\mathcal E}$ intersecting at least in a line.
The condition for unique decoding for $n$ odd is\cite{Gorla} 
${\rm dim}(\mathcal{M}\cap\mathcal{E})\geq (n-1)/2$.

Based on our detailed study of Section 3.4-3.5
in Section 4.7 we sketched the $n\geq 3$ error correction process. 
We conjectured that a straightforward version of the decoding algorithm applied to spread codes of Ref.\cite{Stokes} should be valid for arbitrary $n$, when adapted to our situation where the messages are isotropic spreads.
In this case the challenge is to find the $n$-qubit generalization of the explicit algebraic decoding formula of Eq.(\ref{nice}) we have found for $n=2$.
Notice that this formula of the form $\mathcal{F}=JR\mathcal{E}$, with $\mathcal{M}=\{\mathcal{E},\mathcal{F},\mathcal{E}+\mathcal{F}\}$ being the reconstructed message line, is universal in two respects. First, the codeword structure is hidden inside the structure of the $R$ recovery matrix for a whole {\it collection of codes}. These are connected to the six possible ovoids of the bulk representing the six possible isotropic spreads of the boundary.
The set of codewords in all of the six possible cases and their corresponding $R$ matrices are explicitely given by Eqs.(\ref{recliff})-(\ref{codewordscliff}).
Second, this formula can be used for recovery from both point and plane errors, i.e. for all types of errors.
Notice also that the quantities like $\mathcal{M}$, $\mathcal{E}$, $\mathcal{F}$ are boundary and the ones $J,R$ are bulk related. Interestingly, according to Eq.(\ref{recliff}) the data which uniquely specifies the recovery matrix $R$ is given in terms of observables forming a six dimensional Clifford algebra. However, these observables are {\it off} the Klein quadric, hence they are not related to any of the boundary observables in an explicit way.
We mention that according to Section 4.6 the codewords for arbitrary $n$ are forming a partial ovoid of the bulk. This means
that the observables corresponding to the $2^n+1$ partial ovoid points are comprising $2^n+1$ pairwise anticommuting observables.This indicates that a characterization of our codewords in terms of observables off the quadric $\hat{\mathcal Q}$ showing up at the end of Section 4.4, also forming a Clifford algebra might be possible.
A result of that kind could pave the way for an elegant algebraic characterization of the decoding algorithm for arbitrary $n$.

We emphasize that the error correction code featuring our investigations is a {\it classical} subspace code.
However, the existence of the quantum net structure on the boundary side indicates that some sort of quantum generalization of our considerations might be feasible.
In Section  3.8. we already have given few hints on this possibility. 
The key observation in this respect is encapsulated in Eq.(\ref{meaning}).
Namely, if we restrict the possibility of errors to subspace errors that are contained entirely inside a particular message, then one can consider a {\it flag} which is a nested sequence of subspaces with {\it decreasing dimension}. For a fixed quantum net structure when choosing a fixed message plane one can consider its associated stabilizer state. Then for a flag for a sequence of errors what we would get is set of subspaces with {\it increasing dimension} of the $n$-qubit Hilbert space, i.e. a nested set of stabilizer subspaces, containing the stabilizer state.
It would be interesting to study the Grassmannian image of the associated bulk quantum net. However, here we are bogged down due to our sign ambiguities. Indeed, the Pl\"ucker map by itself not telling us anything on how to associate stabilizer states to spreads in the bulk given a quantum net in the boundary.
In any case Eq.(\ref{meaning}) indicates that a "quantum version of the Pl\"ucker" map seem to be reversing the order of the embedding of the corresponding subspaces in the bulk Hilbert space. Hence in the boundary we are having a sequence Hilbert subspaces of increasing and in the bulk of decreasing dimension.
Further exploration on this point would be desirable.

\subsection{Comments and speculations}

The final point we would like to briefly discuss is concerning the issue of entanglement. Indeed, in the AdS/CFT dictionary for bulk reconstruction, boundary entanglement serves as a glue.
In our finite geometric considerations boundary entanglement seems to be not playing any role at all.
Nevertheless, boundary entanglement is hidden in the structure of the stabilizer states that we associate to the fibration of the boundary. In fact such multipartite correlations in mutually unbiased bases have already been investigated\cite{Kraus}.
The point where boundary entanglement should connect to error correction rests on the fact that our boundary can be regarded as a {\it collection} of messages, giving rise to a collection of bulk codewords.

In order to see this let us associate states to the set of messages in the following manner. Choose the system of MUBs
based on the state $\vert 00\dots 0\rangle$.
This state is belonging to the set of eigenstates of the complete set of $n$-qubit observables containing only combinations of the $Z$ and $I$ Pauli operators. Then the rotationally covariant generation of MUBs is effected with the help of our unitary operator of order $2^n+1$ generalizing the one of the $n=2$ case we presented in Eq.(\ref{unitary9}). Acting with different powers of this operator on the initially fully separable state $\vert 00\dots 0\rangle$ we obtain the $2^n+1$ states of our quantum net. Clearly the entanglement properties of such $2^n+1$ states are hidden in the explicit structure of the different powers of this operator.
This unitary can be regarded as some sort of discretized time evolution operator relating the different slices of the boundary. At the level of Figure 9 these time evolution steps are the ones which cyclically rotate the states associated to the $2^n+1$ points of the boundary circle.
Now in this picture changing the sets of messages amounts to switching from one spread to another one. This can be effected by repeating our construction based on some other state different from $\vert 00\dots 0\rangle$. According to the illustrative calculations of Section 3.9 these changes in the spreads are accompanied by changes in the stabilizer states with the assistance of unitaries acting as quantum gates. These are in turn again changing the entanglement types, of the states assigned to the points of the boundary circle of Figure 9.
Unfortunately it is not at all obvious how this boundary entanglement of the MUBs should manifest itself in the bulk codewords. 
In order to have some progress on this issue one should be able to define a quantum net, similar to the GHW one, in a consistent manner for the full bulk.

In our approach the issue of what do we mean by subsystems is also unclear. Since the boundary is associated with an $n$-qubit system then the possible subsystems should arise naturally from taking different partitions of the set $\{1,2,\dots ,n\}$ of qubits.
This is the natural approach of the previous paragraph, and also the one followed by Ref.\cite{Kraus}.
In this spirit one can define measures of multipartite quantum correlations based on classical corelations in our MUBs.
However, according to the spirit of AdS/CFT an approach where subsystems correspond to subregions of the boundary would be more desirable.
In order to have an approach of this kind one can regard our boundary as the space ${\mathbb P}({\bf V})$ showing up in Eq.(\ref{nGHW}). One can then consider as subregions of the boundary certain subsets of the $2^n+1$ points of the outer circle of Figure 9. Let us suppose that we have already fixed the quantum net structure. This means that we have a fixed set of messages i.e. a fixed set of MUBs assigned to these points. Denote the basis vectors of these MUBs as $\vert\varphi_a^{(j)}\rangle$ with $a=1,2,\dots,2^n+1$ and $j=1,2\dots ,2^n$. Then for an arbitrary $n$-qubit state $\psi$  one can calculate the probabilities $p_{aj}=\vert\langle\psi\vert\varphi_a^{(j)}\rangle\vert^2$. 
Consider now the R\'enyi entropies $S^{(a)}\equiv -\log_2(\sum_jp_{aj}^2)$ quantifying our inability to predict the outcome of the $a$th measurement.
For a collection of points forming a subregion $I$ where $I\subset\{1,2,\dots,2^n+1\}$ one can consider $\langle S\rangle_I$ i.e. the average of the $S^{(a)}$ for $a\in I$. Obviously one can also regard $I\subset {\mathbb P}(\bf V)$. It is known that for $I$ covering the whole boundary we have\cite{Werner}  
\beq
\langle S\rangle_{{\mathbb P}(\bf V)}=\frac{1}{2^n+1}\sum_a S^{(a)}\geq\log_2(2^n+1)-1.
\label{egyenlotlenseg}
\eeq
Recall now that for rotationally invariant states\cite{w3} the $S^{(a)}$ are equal for all $a$. As a result of this such states minimize the average R\'enyi entropy, hence in this sense they are of minimum uncertainty. In this sense such states are like coherent states approximating classical ones. Such states are the eigenstates of the unitary operator\cite{w3} of order $2^n+1$ generating cyclic evolution on the boundary.

We note that relations similar to (\ref{egyenlotlenseg}) have also been shown featuring the Shannon entropy\cite{Winter}.
These results show that for a given quantum state one can associate entropic quantities to subregions of ${\mathbb P}(\bf V)$.
We emphasize that the inequality of (\ref{egyenlotlenseg}}) gives an example of an entropic uncertainty relation. It is known that such relations provide an alternative means for quantifying multipartite correlations of $\psi$. Indeed, quantum correlated states exhibit strong classical correlations in the measurement outcomes of local complementary observables. This is the basic idea how MUBs can be used to define new measures of quantum entanglement\cite{Kraus}.
This also shows that the two notions based on entanglement between particles (qubits) and entanglement between subregions (modes) are interlinked in a nontrivial manner.
The situation we are given here is reminiscent of the one when in holography apart from the usual spacial splittings of the boundary region into two regions one is also forced to consider splittings of more general kind.
This is the situation one is faced with in field theories with holographic duals having (internal) gauge symmetries.
Note that the symmetries that show up for systems of $n$ indistinguishable systems can be regarded of that kind\cite{Sarosi}. Recall in this respect the $n$-fold twisted tensor product structure we have found in Section 4.6.
In holography these subtleties boil down to the presence of correlations of unusual kind culminating in the phenomenon called entwinement\cite{entwine}, which can be regarded as some sort of generalization of the traditional notion of entanglement. In order to embark in an 
exploration of the viablility of this analogy further insight is needed.

In this paper we subscribed to the view of regarding our bulk as some sort of finite geometric version of a space-time structure.
This need not be the case. In the following paragraphs we comment on the possibility of an alternative interpretation of the bulk.

Recall that the space-time interpretation was supported by the occurrence of the chronometric form on the bulk side of our correspondence. In the very special $n=2$ case this form boiled down to the Minkowski line element.
The representation of correctable errors as the causal structure of the closest codewords was motivated by having light-like or non light-like separation for bulk points with respect to this chronometric form.
However, we also pointed out that the errors have an intricate fine structure manifesting itself in the fine structure of the light cone.

This fine structure has made its appearance due to the fact that locally the $n\times n$ matrix underlying the chronometric form can have different ranks, see Eq.(\ref{ujra}) and Figure 8 for illustrations.
In fact, Eq.(\ref{ujra}) can be regarded as a basic formula relating a metric stucture of the boundary and the bulk.
On the right hand side we have information coming from intersection properties of $n-1$-planes ${\mathcal M}$ and ${\mathcal N}$ represented by $n\times 2n$ matrices.
This is translated to the information on the special relationship between points  ${\mathcal{P}}^{(\mathcal M)}$ and ${\mathcal{P}}^{(\mathcal N)}$ in the bulk. Hence for the special case of two $n-1$-subspaces the ${\mathcal{P}}^{(\mathcal M)}\wedge {\mathcal{P}}^{(\mathcal N)}=0$ constraint means that the corresponding subspaces have nontrivial intersection. The properties of the intersection are classified by the rank structure of a $2n\times 2n$ matrix coming from the two $n\times 2n$ ones\cite{Gorla}.
For arbitrary subspaces on the boundary side this information is to be combined with the one provided by the (\ref{dist}) metric $d({\mathcal M},{\mathcal N})$.

Now it is known that this metric represents the distance of a geodesic\cite{cheche} between ${\mathcal M}$ and ${\mathcal N}$ in the undirected Hasse graph representing the lattice of subspaces of $V(2n,2)$ partially oredered by inclusion. This notion means that ${\mathcal M}\preceq {\mathcal N}$ if and only if ${\mathcal M}$ is a subspace of ${\mathcal N}$. In this graph the vertices correspond to the elements of our boundary $PG(2n-1,2)$ and an edge joins a subspace ${\mathcal M}$ with a subspace ${\mathcal N}$ if and only if $\vert{\rm dim}({\mathcal M})-{\rm dim}({\mathcal N})\vert =1$ and either $\mathcal{M}\subset {\mathcal N}$ or ${\mathcal N}\subset {\mathcal M}$.
Clearly this structure is mapped into a dual metric structure in the bulk expressed in terms of ${\mathcal{P}}^{(\mathcal M)}$ and  ${\mathcal{P}}^{(\mathcal N)}$.
On the boundary side we have basically a metric structure describing which subspaces contain which other subspaces related to geodesics with respect to the metric of Eq.(\ref{dist}). Hence how the ${\mathcal{P}}^{(\mathcal M)}$ is related to the one
${\mathcal{P}}^{(\mathcal N)}$ causally, translates to how the boundary subspaces are organised with respect to one another. 

This observation resonates with the one of Ref.\cite{Bartek} observed in the context of integral geometry and $AdS_3/CFT_2$. 
In their investigation the authors introduce the concept of kinematic space which acts as an intermediary translator between the languages of the boundary and the bulk.
The kinematic space organizes the data about subsets of the Hilbert space associated with the boundary quantum field theory. At the classical level this data of Hilbert space subsets is related to which interval in the boundary contains which other intervals. 
In this approach the points in the kinematic space are associated with intervals in the boundary. Since the intervals are having a partial ordering, this renders the kinematic space partially ordered too. 
In some sense the causal structure of kinematic space encodes the containment relation for boundary intervals.
In the special case of when considering the static slice of $AdS_3$ the kinematic space is a de Sitter space\cite{Bartek}.

Notice now that our finite geometric setting resonates with these findings. 
Using this observation one also has a possibility of interpreting the image of the boundary under the Pl\"ucker map as some sort of finite geometric version of kinematic space. 
Amusingly according to Eq.(\ref{nuvariety}) the code variety is obtained as a slice of this version of kinematic space precisely in the same way as de Sitter space is obtained in twistor theory\footnote{See the discussion at the end of Section 3.6. 
on this point.}.
Of course adopting this interpretation the question left to be answered is: What kind of a space should then play the role of space-time in our finite geometric toy model?

\section{Acknowledgement}

The authors would like to express their gratitude to Zsolt Szab\'o for his work with the figures of this paper.
This work was supported by the Franche-Comt\'e Regional Research Council, 'Mobilit\'e internationale des chercheurs'.
It was also supported by the French 'Investissements d'Avenir' program, project ISITE-BFC (contract ANR-15-IDEX-03) and the National Research, Development and Innovation Fund of Hungary within the Quantum Technology National Excellence Program (Project Nr. 2017-1.2.1-NKP-2017-00001).

\section{Appendices}
\subsection{Appendix A: Details of the Klein Correspondence}

In $PG(3,2)$ we have $15$ points, $35$ lines and $15$ planes. Since in our setting the underlying vector space $V(4,2)$ of $PG(3,2)$ is equipped with a symplectic form, the $35$ lines can be split into $15$ isotropic and $20$ non-isotropic
ones. 
In the following lists one can easily check that through a particular point there are $7$ lines, $3$ of them are isotropic and $4$ are non-isotropic.
Dually a particular plane consisting of $7$ points and $7$ lines, is containing $3$ isotropic and $4$ non-isotropic lines. 

The $15$ isotropic lines taken together with the $15$ points of $PG(3,2)$ form a point-line incidence structure of a generalized quadrangle $GQ(2,2)$, the doily. The structure of the doily with its isotropic lines and a sketch of the structure of non isotropic ones, in two-qubit observable labelling, can be seen in Figure 1.

The Klein Correspondence relates the geometric objects of $PG(3,2)$ and certain geometric objects of a hyperbolic (Klein) quadric $Q^+(5,2)$ in $PG(5,2)$. For the definition of a hyperbolic quadric recall the form of Eq.(\ref{kvad}). 
There are $35$ points and $30$ planes lying entirely in the Klein quadric. It turns out that the planes can be partitioned into two classes, with cardinalities $15$-$15$ each. One class of planes is called $\alpha$-planes the other $\beta$-planes. 

Now, the objects and their cardinalities featuring the Klein Correspondence are as follows (see Table 2):
\beq
15\quad{\rm points}\leftrightarrow 15\quad{\alpha-\rm planes}
\nonumber
\eeq
\beq
35\quad{\rm lines}\leftrightarrow 35\quad{\rm points}
\nonumber
\eeq
\beq
15\quad{\rm planes}\leftrightarrow 15\quad{\beta-\rm planes}.
\nonumber
\eeq
\noindent
In terms of $2$ and $3$-qubit Pauli operators the explicit form of the correspondence for the $20$ non-isotropic lines is
\beq
{ IX,IY,IZ}\leftrightarrow IXI,\qquad
{ XX,XY,IZ}\leftrightarrow IXX,\qquad
{ ZZ,ZY,IX}\leftrightarrow IXZ,
\nonumber
\eeq
\noindent
\beq
{ XZ,XY,IX}\leftrightarrow XXI,\qquad
{ XX,XZ,IY}\leftrightarrow XXX,\qquad
{ YY,YZ,IX}\leftrightarrow XXZ,
\nonumber
\eeq
\noindent
\beq
{ ZX,ZY,IZ}\leftrightarrow ZXI,\qquad
{ YY,YX,IZ}\leftrightarrow ZXX,\qquad
{ ZZ,ZX,IY}\leftrightarrow ZXZ,
\nonumber
\eeq
\noindent
\beq
{ XI,YI,ZI}\leftrightarrow IZI,\qquad
{ ZZ,YZ,XI}\leftrightarrow IZX,\qquad
{ XX,YX,ZI}\leftrightarrow IZZ,
\nonumber
\eeq
\noindent
\beq
{ ZX,YX,XI}\leftrightarrow XZI,\qquad
{ YY,ZY,XI}\leftrightarrow XZX,\qquad
{ XX,ZX,YI}\leftrightarrow XZZ,
\nonumber
\eeq
\noindent
\beq
{ XZ,YZ,ZI}\leftrightarrow ZZI,\qquad
{ ZZ,XZ,YI}\leftrightarrow ZZX,\qquad
{ YY,XY,ZI}\leftrightarrow ZZZ,
\nonumber
\eeq
\noindent
\beq
{ YZ,YX,IY}\leftrightarrow YXY,\qquad
{ ZY,XY,YI}\leftrightarrow YZY.
\nonumber
\eeq
\noindent
For the $15$ isotropic lines  we have
\beq
{\bf IY,XY,XI}\leftrightarrow XIX,\qquad
{\bf IY,YI,YY}\leftrightarrow YIY,\qquad
{\bf IY,ZY,ZI}\leftrightarrow ZIZ,
\nonumber
\eeq
\noindent
\beq
{\bf XI,IZ,XZ}\leftrightarrow IIX,\qquad
{\bf XI,IX,XX}\leftrightarrow XII,\qquad
{\bf ZI,IZ,ZZ}\leftrightarrow ZII,
\nonumber
\eeq
\noindent
\beq
{\bf ZX,ZI,IX}\leftrightarrow IIZ,\qquad
{\bf YY,ZZ,XX}\leftrightarrow IYY,\qquad
{\bf YY,XZ,ZX}\leftrightarrow YYI,
\nonumber
\eeq
\noindent
\beq
{\bf YX,YI,IX}\leftrightarrow XIZ,\qquad
{\bf YI,IZ,YZ}\leftrightarrow ZIX,\qquad
{\bf ZY,XX,YZ}\leftrightarrow XYY,
\nonumber
\eeq
\noindent
\beq
{\bf XZ,YX,ZY}\leftrightarrow YYX,\qquad
{\bf XY,YX,ZZ}\leftrightarrow ZYY,\qquad
{\bf XY,ZX,YZ}\leftrightarrow YYZ.
\nonumber
\eeq
\noindent

In $PG(3,2)$ we have $7$ lines through each point and $7$ lines contained in each plane. (Lines through a point, and lines on a plane.) Under the Klein Correspondence the $7$ lines of the former type are mapped to $7$ points of the Klein Quadric. These points are forming an $\alpha$-plane. The lines of the latter type are mapped to $7$ points of the Klein Quadric comprising the $\beta$-planes.
In terms of $2$ and $3$-qubit Pauli operators, this observation gives rise to the explicit correspondence
\beq
XX\leftrightarrow {IZZ,XZZ,IXX,XXX,{\bf IYY,XYY,XII}}
\nonumber
\eeq
\noindent
\beq
YY\leftrightarrow {XXZ,XZX,ZXX,ZZZ,{\bf IYY,YYI,YIY}}
\nonumber
\eeq
\noindent
\beq
ZZ\leftrightarrow {ZXZ,ZZX,IZX,IXZ,{\bf IYY,ZYY,ZII}}
\nonumber
\eeq
\noindent
\beq
IX\leftrightarrow {XXZ,IXZ,XXI,IXI,{\bf IIZ,XIZ,XII}}
\nonumber
\eeq
\noindent
\beq
XI\leftrightarrow {XZX,IZX,IZI,XZI,{\bf XII,XIX,IIX}}
\nonumber
\eeq
\noindent
\beq
IZ\leftrightarrow {IXX,ZXX,ZXI,IXI,{\bf ZIX,ZII,IIX}}
\nonumber
\eeq
\noindent
\beq
ZI\leftrightarrow {IZZ,ZZZ,ZZI,IZI,{\bf ZIZ,IIZ,ZII}}
\nonumber
\eeq
\noindent
\beq
IY\leftrightarrow {XXX,ZXZ,YXY,IXI,{\bf YIY,ZIZ,XIX}}
\nonumber
\eeq
\noindent
\beq
YI\leftrightarrow {XZZ,ZZX,YZY,IZI,{\bf XIZ,YIY,ZIX}}
\nonumber
\eeq
\noindent
\beq
XZ\leftrightarrow {XXX,ZZX,XXI,ZZI,{\bf YYX,YYI,IIX}}
\nonumber
\eeq
\noindent
\beq
ZX\leftrightarrow {XZZ,ZXZ,XZI,ZXI,{\bf YYI,IIZ,YYZ}}
\nonumber
\eeq
\noindent
\beq
XY\leftrightarrow {IXX,ZZZ,YZY,XXI,{\bf ZYY,XIX,YYZ}}
\nonumber
\eeq
\noindent
\beq
YX\leftrightarrow {IZZ,ZXX,XZI,YXY,{\bf XIZ,YYX,ZYY}}
\nonumber
\eeq
\noindent
\beq
ZY\leftrightarrow {XZX,IXZ,ZXI,YZY,{\bf XYY,YYX,ZIZ}}
\nonumber
\eeq
\noindent
\beq
YZ\leftrightarrow {IZX,XXZ,ZZI,YXY,{\bf XYY,ZIX,YYZ}}
\nonumber
\eeq
\noindent
where the boldface triples represent isotropic lines. 

The explicit form of the dual correspondence relating the $15$ planes with the $\beta$ planes is
\beq
{\bf ZZ},XX,YY,XY,YX,ZI,IZ\leftrightarrow {IXX,ZXX,IZZ,ZZZ,{\bf IYY,ZYY,ZII}}
\nonumber
\eeq
\noindent
\beq
{\bf YY},XX,ZZ,IY,YI,ZX,XZ\longleftrightarrow {ZZX,ZXZ,XZZ,XXX,{\bf IYY,YYI,YIY}}
\nonumber
\eeq
\noindent
\beq
{\bf XX},YY,ZZ,IX,XI, ZY,YZ\longleftrightarrow {XZX,XXZ,IXZ,IZX,{\bf IYY,XYY,XII}}
\nonumber
\eeq
\noindent
\beq
{\bf IZ},ZI,ZZ,XI,YI,XZ,YZ\longleftrightarrow {ZZX,IZX,ZZI,IZI,{\bf IIX,ZIX,ZII}}
\nonumber
\eeq
\noindent
\beq
{\bf ZI},IZ,ZZ,IX,IY,ZX,ZY\longleftrightarrow {ZXZ,IXZ,IXI,ZXI,{\bf ZII,ZIZ,IIZ}}
\nonumber
\eeq
\noindent
\beq
{\bf IX},XI,XX,ZI,YI,ZX,YX\leftrightarrow {IZZ,XZZ,XZI,IXI,{\bf XIZ,XII,IIZ}}
\nonumber
\eeq
\noindent
\beq
{\bf XI},IX,XX,IZ,IY,XZ,XY\leftrightarrow {IXX,XXX,XXI,IXI,{\bf XIX,IIX,XII}}
\nonumber
\eeq
\noindent
\beq
{\bf IY},YI,YY,XI,ZI,XY,ZY\leftrightarrow {ZZZ,XZX,YZY,IZI,{\bf YIY,XIX,ZIZ}}
\nonumber
\eeq
\noindent
\beq
{\bf YI},IY,YY,IX,IZ,YX,YZ\leftrightarrow {ZXX,XXZ,YXY,IXI,{\bf ZIX,YIY,XIZ}}
\nonumber
\eeq
\noindent
\beq
{\bf ZX},XZ, YY,XY,YZ,IX,ZI\leftrightarrow {ZZZ,XXZ,ZZI,XXI,{\bf YYZ,YYI,IIZ}}
\nonumber
\eeq
\noindent
\beq
{\bf XZ},ZX,YY,YX,ZY,XI,IZ\leftrightarrow {ZXX,XZX,ZXI,XZI,{\bf YYI,IIX,YYX}}
\nonumber
\eeq
\noindent
\beq
{\bf ZY},YX,XX,XZ,YZ,IY,ZI\leftrightarrow {IZZ,XXX,YXY,ZZI,{\bf XYY,ZIZ,YYX}}
\nonumber
\eeq
\noindent
\beq
{\bf YZ},ZY,XX,ZX,ZY,YI,IZ\leftrightarrow {IXX,XZZ,ZXI,YZY,{\bf ZIX,YYZ,XYY}}
\nonumber
\eeq
\noindent
\beq
{\bf XY},YX,ZZ,ZX,YZ,XI,IY\leftrightarrow {ZXZ,IZX,XZI,YXY,{\bf ZYY,YYZ,XIX}}
\nonumber
\eeq
\noindent
\beq
{\bf YX},XY,ZZ,XZ,ZY,IX,YI\leftrightarrow {IXZ,ZZX,XXI,YZY,{\bf ZYY, XIZ, YYX}}.
\nonumber
\eeq
\noindent
Notice that in each row of this correspondence the boldfaced observables on the left hand side are commuting with the remaining six accompanying ones.
On the right the observables are mutually commuting.
This illustrates the fact that the planes in the left are isotropic, and in the right are totally isotropic.

In our $3$-qubit Pauli operator labelling of the $35$ points of the Klein quadric, the middle qubit plays a special role.
Indeed, there are
the $15$ bold faced $3$-qubit operators which are featuring either $I$ or $Y$ in the middle position, while the remaining $20$ ones are featuring  $X$ or $Z$. 
According to Eqs.(\ref{alap}) and (\ref{pluck})
this can be traced back to our considerations about isotropic lines resulting in the constraint of Eq.(\ref{laggrass}).
The $15$ bold-faced $3$-qubit Pauli operators are special. They form a new copy of a doily. This time it is living inside the Klein quadric (see Figure 3).
The $15$ triples showing up as lines inside {\it both} $\alpha$ and $\beta$-planes (see Figure 4) are giving rise to a new set of $15$ isotropic lines of this new doily.
One can easily characterize this $35=15+20$ split in an algebraic manner.

For this notice that the $15$ observables from this split are commuting, and the $20$ ones are anticommuting with the special one
$\Gamma\equiv IYI$ coming from the vector $w$ with components $(010010)$. The vector $w$ generates a transvection $T_w$
acting on an arbitrary $v\in V(4,2)$ as $v\mapsto v+\langle v,w\rangle w$.
One can lift this transvection to a unitary action by conjugation on our observables as follows\cite{Cherchiai}
\beq
\mathcal{O}_v\mapsto \mathcal{U}(T_w)\mathcal{O}_v\mathcal{U}^{\dagger}(T_w),\qquad \mathcal{U}(T_w)=
I\otimes\frac{1}{\sqrt{2}}(I+iY)\otimes I.
\eeq
\noindent 
This unitary operator is leaving invariant the $15$
bold face observables and changing the $X$ in the middle slot ($10$ observables) to a $Z$ ($10$ observables) and $Z$ to $-X$.
Hence the squared action of this operation is {\it minus} the identity.
Then $\mathcal{U}(T_w)$ is acting like a ${\it conjugation}$ by leaving invariant the $15$
doily observables and exchanging a $10$ element set of points with its conjugate $10$ element set. 
Hence this copy of the doily living inside the Klein quadric can be regarded as a finite geometric analogue of the real part of the complexified and compactified Minkowski space-time familiar from twistor theory\cite{Penrose1}.

\subsection{Appendix B: Linking transvections to quantum gates}

As was explained in Section 2 the symplectic group $Sp(2n,2)$ is generated by transvections. These transvections can be lifted to unitary operators acting on the observables of the $n$-qubit Hilbert space via conjugation. 
To a vector $v\in V(2n,2)$ one can associate a transvection $T_v$, and an observable $\mathcal{O}_v$ up to sign.
Furthermore, according to the prescription of Ref.\cite{Cherchiai} to a transvection one can associate an unitary $\mathcal{U}(T_v)\equiv \mathcal{U}(\mathcal{O}_v)$ with the explicit form\footnote{We have slightly changed the definition of Ref.\cite{Cherchiai} which employs the phase factor $e^{-i\pi/4}$. For convenience we rather use the phase factor $e^{i\pi/4}$.}
\beq
\mathcal{U}(\mathcal{O}_v)\equiv \eta\frac{1}{\sqrt{2}}({\bf 1}+i\mathcal{O}_v),\qquad \eta=\frac{1}{\sqrt{2}}(1+i)=e^{i\pi/4}.
\label{liftoper}
\eeq
\noindent
The action of this unitary via conjugation is as follows\cite{Cherchiai}
\beq
\mathcal{U}(\mathcal{O}_v)\mathcal{O}^{\prime}_w\mathcal{U}(\mathcal{O}_v)^{\dagger}=\begin{cases}\mathcal{O}^{\prime}_w,\qquad\quad{\rm if}\quad [\mathcal{O}_v,\mathcal{O}^{\prime}_w]=0,\\
i\mathcal{O}_v\mathcal{O}^{\prime}_w,
\quad{\rm if}\quad \{\mathcal{O}_v,\mathcal{O}^{\prime}_w\}=0,\end{cases}
\label{liftingexplicit}
\eeq
\noindent
where in the second case $\{\cdot,\cdot\}$ is the anticommutator.
Notice that in this definition based on observables there is a sign ambiguity since the mapping $v\mapsto\mathcal{O}_v$ is merely up to sign.
However, for a fixed spread of totally isotropic $n-1$-planes one can always find a set of signs such that the collection of such $n-1$-planes is positive. 
Fixing the signs in this way means that the 
corresponding set of $2^n-1$ commuting observables forms a stabilizer for a unique state. 

In the special case of $n=2$ we have $16$ observables up to sign, and $Sp(4,2)\simeq S_6$, where $S_6$ is the symmetric group on six letters. Now it is well-known that $W(3,2)$ (doily) can be labelled by duads. The isotropic lines in this picture are formed by triples of duads like $(12,34,56)$. For a labelling of the doily in terms of duads we use the conventions of Figure 1 of Ref.\cite{LHSmagic}.
Let us now adopt the sign convention of Table 1. meaning we use the sign convention of our canonical spread of lines. Hence all of the observables should be taken with a positive sign except for the ones $-XZ$ and $-ZX$ taken with a negative sign.
Then all of the five lines of our spread are positive ones.
Now we explicitely have 
\beq
IX\leftrightarrow 14,\quad XI\leftrightarrow 23,\quad XX\leftrightarrow 56,\qquad
IZ\leftrightarrow 16,\quad ZI\leftrightarrow 35,\quad ZZ\leftrightarrow 24
\label{d1}
\eeq
\noindent
\beq
IY\leftrightarrow 46,\quad YI\leftrightarrow 25,\quad YY\leftrightarrow 13,\qquad
XY\leftrightarrow 15,\quad YZ\leftrightarrow 34,\quad -ZX\leftrightarrow 26
\label{d2}
\eeq
\noindent
\beq
YX\leftrightarrow 36,\quad ZY\leftrightarrow 12,\quad -XZ\leftrightarrow 45.
\label{d3}
\eeq
\noindent
Now it is easy to check that transvection $T_v$ corresponding to the observable $\mathcal{O}_v$ generates a transposition of the form $(jk)$. For example $T_{(1101)}=T_{ZY}=T_{(12)}$. In this language the isomorphism $Sp(4,2)\simeq S_6$ translates
to the fact that any permutation can be represented as a sequence of transpositions. For more details on this point see Ref.\cite{Cherchiai}. 

Now according to our (\ref{d1})-(\ref{d3}) dictionary and the (\ref{liftoper}) prescription one can prove that the permutations $(14)(26)(35)$ and $(16)(23)(45)$ can be lifted to the unitaries
\beq
\mathcal{U}(ZI)\mathcal{U}(IX)\mathcal{U}(-ZX)=-C_{12},\qquad
\mathcal{U}(IZ)\mathcal{U}(XI)\mathcal{U}(-XZ)=-C_{21},
\label{cnottrans}
\eeq
\noindent
i.e. up to a sign they are the controlled CNOT gates.
A similar calculation for the permutation $(13)(24)(56)$ shows that
\beq
\mathcal{B}\equiv\mathcal{U}(XX)\mathcal{U}(YY)\mathcal{U}(ZZ)=-S_{12},
\label{swaptrans}
\eeq
\noindent
which up to a sign is the SWAP gate.
Hence the unitary operator $\mathcal{D}^{-1}=C_{21}C_{12}=C_{12}S_{12}$ of Eq.(\ref{atteres}) cyclically permuting the entries of the message words $\mathcal{M}_k$, $k=0,1,2$ is arising from the lifts of a sequence of transvections based on the permutation:
$(13)(24)(56)(14)(26)(35)=(125)(364)$.
From this form it is easy to find the explicit element of $Sp(4,2)$ corresponding to the inverse of this permutation $(152)(346)$. It is
\beq
D\equiv\begin{pmatrix}1&1&0&0\\1&0&0&0\\0&0&0&1\\0&0&1&1\end{pmatrix}\in Sp(4,2), \qquad \mathcal{D}\equiv\mathcal{U}(D),
\eeq
\noindent
where the second equation shows that the lift of $D$ is just the unitary operator $\mathcal{D}=S_{12}C_{12}$.
Similarly we have
\beq
B\equiv\begin{pmatrix}0&1&0&0\\1&0&0&0\\0&0&0&1\\0&0&1&0\end{pmatrix}\in Sp(4,2), \qquad \mathcal{B}\equiv\mathcal{U}(B).
\eeq
\noindent
One can check that the pair $(B,D)$ generates a copy of the symmetric group on three letters i.e. we have the presentation
\beq
S_3\equiv\langle B,D\vert D^3=B^2=(BD)^2\rangle \subset S_6\simeq Sp(4,2).
\eeq
\noindent
The operators $(\mathcal{B},\mathcal{D})=(-S_{12},S_{12}C_{12})$ serving as quantum gates are forming  a representation of this $S_3$.

Having a representation of the symplectic group on $V\equiv V(4,2)$ on the boundary defines a corresponding action of the same group on the bulk.
Indeed, for $n=2$ we have the Pl\"ucker map ${\rm span}\{v,u\}\mapsto v\wedge u$.
Hence having an action $v\mapsto Mv$ for $v\in V$ and $M\in Sp(4,2)$ we have a corresponding $Sp(4,2)$ action on $\wedge^2V$ via calculating $Mv\wedge Mu$ and writing it in terms of the six Pl\"ucker coordinates. 
For example for $M=D^{-1}$ under $v\wedge u\mapsto D^{-1}v\wedge D^{-1}u$ we get the map
\beq
(P_{12},P_{13},P_{14},P_{34},P_{24},P_{23})\mapsto (P_{12},P_{23}+P_{24},P_{23},P_{34},P_{23}+P_{13},P_{13}+P_{24}+P_{23}+P_{14}).
\eeq
\noindent
Under this map the constraint (\ref{laggrass}) is left invariant hence this is a transformation leaving the doily inside the Klein quadric invariant (see Figure 3).
Moreover, using the convention of Eq.(\ref{elsokonv}) we see that under this transformation the first qubit is left invariant.
The second and third qubits are transformed by the matrix
\beq
M=\begin{pmatrix}0&0&1&1\\0&0&0&1\\1&0&0&1\\1&1&1&1\end{pmatrix}=
\begin{pmatrix}1&0&0&0\\0&0&0&1\\0&0&1&0\\0&1&0&0\end{pmatrix}
\begin{pmatrix}0&0&1&1\\1&1&1&1\\1&0&0&1\\0&0&0&1\end{pmatrix}.
\eeq
\noindent
The latter two matrices are the ones of the transvections $T_{IY}$ and $T_{YZ}$.
Lifting these transvections gives
\beq
\mathcal{U}(M)=\mathcal{U}(IY)\mathcal{U}(-YZ)=i(H\otimes I)C_{12}(I\otimes H)C_{21}(X\otimes X).
\eeq
\noindent
After defining
\beq
{\bf D}\equiv I\otimes \mathcal{U}^{\dagger}(M)
\eeq
\noindent
we get back to the transformation rule of Eq.(\ref{kakukk}).

Let us consider now the following lift of the permutation (another automorphism of the doily) $(56421)=(56)(45)(25)(15)$.
Using again the (\ref{liftoper}) definition and the (\ref{d1})-(\ref{d3}) dictionary a lift of this permutation is the unitary 
\beq
U\equiv \mathcal{U}(-XX)\mathcal{U}(XZ)\mathcal{U}(YI)\mathcal{U}(-XY)=\frac{1}{2}\begin{pmatrix}i&1&-i&1\\-1&-i&-1&i\\
1&i&-1&i\\i&1&i&-1\end{pmatrix},\qquad U^5={\bf 1}.
\label{marhajo}
\eeq
\noindent
One can then show that conjugate action of this unitary is permuting the message words of the boundary in the following way
\beq
(\{ZZ,IZ,ZI\},\{IY,-YI,-YY\},\{XX,IX,XI\},\{XY,-ZX,YZ\},\{ZY,-XZ,YX\}).
\label{orbitexplicit}
\eeq
\noindent
It is understood that the cycle notation $(\cdot,\cdot,\cdot,\cdot,\cdot)$ means that the first, second and third entries of the corresponding $\{\cdot,\cdot,\cdot\}$s are permuted by the permutation of order five. 
Hence for example $ZZ\mapsto IY\mapsto XX\mapsto\dots$ etc.
The reader must notice that this set is not precisely the same message set we used in Table 1. Indeed, instead of the message word $\{IY,YI,YY\}$ the one with the sign combination $\{IY,-YI,-YY\}$ shows up. 
As a result the corresponding stabilizer state
of Table 1 has to be changed. As explained at the end of Section 3.9 $U$ generates all of the message words by conjugation starting from the one: $\{ZZ,IZ,ZI\}$. Alternatively repeated action of $U$ on the initial stabilizer state $\vert 00\rangle$ generates all of the corresponding stabilizers of the message words. 

Let us now define
\beq
W\equiv C_{21}P_{12}(X\otimes I)(H\otimes H),\qquad P_{12}=(I\otimes H)C_{12}(I\otimes H),
\label{w}
\eeq
\noindent
where $P_{12}$ the controlled phase gate $P_{12}={\rm diag}\{1,1,1,-1\}$.
Then we have 
\beq
\mathcal{U}(-XX)\mathcal{U}(XZ)=(P\otimes I)(X\otimes X)W(P\otimes I)(X\otimes X),\qquad
\mathcal{U}(YI)\mathcal{U}(-XY)=S_{12}WS_{12}
\label{resz}
\eeq
\noindent
where the SWAP gate can be expressed in terms of CNOT gates if desired as $S_{12}=C_{12}C_{21}C_{12}$. 
These results give the (\ref{unitary9}) explicit form of Section 3.9 for our unitary $U$ in terms of the basic gates: $X,P,H$ and $C_{ij}$.

\subsection{Appendix C: The field $GF(2^n)$}

Elements of the field extension $GF(2^n)$ of $GF(2)$ can be thought of as degree $n-1$ polynomials with binary ($GF(2)$) coefficients.
Multiplying two such polynomials of $GF(2^n)$ is defined modulo a {\it primitive} polynomial $\pi_n(x)$ of order $n$ which we write in the form
\beq
\pi_n(x)=a_0+a_1x+a_2x^2+\dots +a_{n-1}x^{n-1}+x^n.
\label{primike}
\eeq
\noindent
$\pi_n(x)$ is an irreducible polynomial, i.e one which cannot be factored in $GF(2)$. For example the polynomial $x^2+x+1$ cannot be factored over $GF(2)$.
Some primitive polynomials of $GF(2^n)$ for $n=2,3,4$ are
\beq
\pi_2(x)=x^2+x+1,\qquad \pi_3(x)=x^3+x^2+1,\qquad \pi_4(x)=x^4+x+1.
\eeq
\noindent

The extension to $GF(2^n)$ is effected by adjoining to the two elements of $GF(2)$, i.e. $0$ and $1$, a new element $\omega$ which by definition satisfies the primitive polynomial $\pi_n(\omega)=0$. It then turns out that the product structure in $GF(2^n)$ is such that after excluding the zero element the powers of $\omega$ form a cyclic group of order $2^n-1$. Hence a list of all the elements of $GF(2^n)$ can be written as
\beq
\{0,1,\omega,\omega^2,\dots,\omega^{2^n-2}\},\qquad \omega^{2^n-1}=1,\qquad \pi_n(\omega)=0.
\label{elemek}
\eeq
\noindent
Clearly the requirement of closedness also under addition forces other elements like $\omega+1$ into existence. The primitive polynomial then can be used to relate these extra elements to the powers of $\omega$.
For example using $\pi_3$, for $n=3$ the (\ref{elemek}) list of $8$ elements can be written in the form
\beq
\{0,1,\omega,\omega^2,\omega^3,\omega^4,\omega^5,\omega^6\}=\{0,1,\omega,\omega^2,1+\omega^2,1+\omega+\omega^2,1+\omega,\omega+\omega^2\}
\eeq
\noindent
i.e. all of them are polynomials with degree at most $n-1=2$.

The upshot of these considerations is that an arbitrary element of $GF(2^n)$ can be written in the form
\beq
x=\sum_{k=0}^{n-1}x_ke_k
\eeq
\noindent
where we say that $\{e_1,e_2,\dots,e_{n-1}\}$, as a collection of $GF(2^n)$ elements, forms a {\it field basis}. Here the expansion coefficients $x_k$ are elements of $GF(2)$.
Hence $GF(2^n)$ can be regarded as a rank $n$ vector space over $GF(2)$.

For any element of $x$ of $GF(2^n)$ one can introduce its {\it trace} by the formula
\beq
{\rm Tr}(x)=x+x^2+x^{2^2}+x^{2^3}+\dots +x^{2^{n-1}}.
\label{tracefield}
\eeq
\noindent
The trace operation is linear, and takes values in $GF(2)$.
For example for $n=3$ we have
\beq
{\rm Tr}(\omega^4)=\omega^4+\omega^8+\omega^{16}=\omega^4+\omega+\omega^2=(1+\omega+\omega^2)+\omega+\omega^2=1
\nonumber
\eeq
\noindent
Using the trace for a field basis $\{e_0,e_1,e_2,\dots,e_{n-1}\}$ one can define its dual
$\{\tilde{e}_0,\tilde{e}_1,\tilde{e}_2,\dots,\tilde{e}_{n-1}\}$
by the formula
${\rm Tr}(\tilde{e}_je_k)=\delta_{jk}$.
For example in the $n=3$ case for the basis
$\{e_0,e_1,e_2\}=\{1,\omega,\omega^2\}$ its dual is
$\{\tilde{e}_0,\tilde{e}_1,\tilde{e}_2\}=\{\omega^4,\omega^3,\omega^5\}$ since for example
${\rm Tr}(\tilde{e}_1e_1)={\rm Tr}(\omega^4)=1$ etc.

The points of the GHW discrete phase space are parametrized by pairs of numbers $(q,p)$ where $q,p\in GF(2^n)$.
However, for reasons explained in the text $q$ is expanded with respect to the dual field basis, and $p$ is expanded in the field basis. Hence for $n=3$ we have the expansions
\beq
q=\sum_{j=0}^2q_j\tilde{e}_j=q_0\cdot \omega^4+q_1\cdot\omega^3+q_2\cdot\omega^5,\qquad
p=\sum_{j=0}^2p_j{e}_j=p_0\cdot 1+p_1\cdot\omega+p_2\cdot\omega^2.
\label{haromexpansion}
\eeq
\noindent
According to the method of field reduction any such pair can be regarded as a $2n$ component vector over $GF(2)$.
As a result of this, for example for the pair $(\omega, \omega^4)$ we have
\beq
(q,p)=(\omega,\omega^4)=(\omega^3+\omega^4,1+\omega+\omega^2)\leftrightarrow (110,111)\leftrightarrow \pm YYX
\label{peldared}
\eeq
\noindent
where the last of the arrows illustrates the ambiguous association of observables to phase space points.

\subsection{Appendix D: The case of three-qubits}

In order to describe the a set of message words for three qubit systems (a totally isotropic plane-spread) we use the method of Section 4.1 for partitioning the $63$ points of our boundary $PG(5,2)$ into $9$ planes, with $7$ points each.
This is equivalent to finding a MUB for three-qubits.
A plane in $PG(5,2)$ is arising from a rank $3$ subspace of $V\equiv V(6,2)$. Such subspaces are generated by taking the linear span of three linearly independent vectors: $\langle u,v,w\rangle$. Following the pattern of Eq.(\ref{arrange}) of the $n=2$ case we write the three linearly independent vectors as a $3\times 6$ matrix in the form $(Q\vert P)$. In the coordinate patch where ${\rm Det}Q\neq 0$ one can alternatively write a representative as: $({\bf 1}\vert {\bf A})$ where ${\bf 1}$ is the $3\times 3$ unit matrix.
Now we would like to present our spread corresponding to our choice of message words in the form
\beq
({\bf 1}\vert {\bf A}_k),\quad ({\bf 0}\vert {\bf 1}),\quad ({\bf 1}\vert {\bf 0}), \quad k =1,2,\dots ,7
\eeq
\noindent
where the seven $3\times 3$ matrices ${\bf A}_{k}$ are symmetric, invertible and also satisfying the constraint
of Eq.(\ref{jaj}).

Let us choose
\beq
{\bf A}_1=\begin{pmatrix}1&0&1\\0&0&1\\1&1&1\end{pmatrix},\quad
{\bf A}_2=\begin{pmatrix}0&0&1\\0&1&0\\1&0&1\end{pmatrix},\quad
{\bf A}_3=\begin{pmatrix}0&1&0\\1&0&0\\0&0&1\end{pmatrix},\quad
\label{elso1}
\eeq
\noindent
\beq
{\bf A}_4=\begin{pmatrix}1&0&0\\0&1&1\\0&1&0\end{pmatrix},\quad
{\bf A}_5=\begin{pmatrix}0&1&1\\1&1&0\\1&0&0\end{pmatrix},\quad
{\bf A}_6=\begin{pmatrix}1&1&0\\1&1&1\\0&1&1\end{pmatrix},\quad
{\bf A}_7=\begin{pmatrix}1&1&1\\1&0&1\\1&1&0\end{pmatrix}.
\label{masodik2}
\eeq
\noindent
Using the pattern $({\bf 1}\vert {\bf A}_k)$ for converting this code to observables and choosing the signs appropriately a set of 
$S_a, a=1,2,\dots 9$ stabilizers is
\beq
S_1=\langle YIX,IZX,XXY\rangle,\quad
S_2=\langle -ZIX,-IYI,XIY\rangle,\quad
S_3=\langle ZXI,XZI,IIY\rangle,\quad
\label{stab1}
\eeq
\noindent
\beq
S_4=\langle -YII,IYX,-IXZ\rangle,\quad
S_5=\langle ZXX,XYI,XIZ\rangle,\quad
S_6=\langle YXI,XYX,-IXY\rangle,\quad
\label{stab2}
\eeq
\noindent
\beq
S_7=\langle YXX,-XZX,XXZ\rangle,\quad
S_8=\langle XII,IXI,IIX\rangle,\quad
S_9=\langle ZII,IZI,IIZ\rangle,\quad
\label{stab3}
\eeq
\noindent
where the last two stabilizers correspond to the planes $({\bf 0}\vert {\bf 1})$ and $({\bf 1}\vert {\bf 0})$.

Using the notation of Eqs.(\ref{underline})-(\ref{tilde}) one can check that
 the states stabilized by the corresponding stabilizers are
\beq
\vert\varphi_7\rangle=\frac{1}{\sqrt{2}}(\vert 0\tilde{0}\tilde{1}\rangle +i\vert 1\tilde{1}\tilde{0}\rangle),\quad
\vert\varphi_1\rangle=\frac{1}{\sqrt{2}}(\vert \tilde{0}0\overline{0}\rangle +\vert\tilde{1}1\overline{1}\rangle),\quad
\vert\varphi_2\rangle=\frac{1}{\sqrt{2}}(\vert 0\tilde{1}\overline{1}\rangle +i\vert 1\tilde{1}\overline{0}\rangle),
\label{state1}
\eeq
\noindent
\beq
\vert\varphi_3\rangle=\frac{1}{\sqrt{2}}(\vert 0\overline{0}\tilde{0}\rangle +\vert 1\overline{1}\tilde{0}\rangle),\quad
\vert\varphi_4\rangle=
\frac{1}{\sqrt{2}}(\tilde{1}0\tilde{0}\rangle)-\vert\tilde{1}1\tilde{1}\rangle),\quad
\vert\varphi_5\rangle=\frac{1}{2}(\vert 00\tilde{0}\rangle+\vert 10\tilde{1}\rangle +i\vert 01\tilde{1}\rangle
+i\vert 11\tilde{0}\rangle),
\label{state2}
\eeq
\noindent
\beq
\vert\varphi_6\rangle=\frac{1}{2}(\vert 0\tilde{0}0\rangle +\vert\tilde{1}0\tilde{1}\rangle+
\vert 0\tilde{1}1\rangle-\vert 1\tilde{1}0\rangle),\qquad
\vert\varphi_8\rangle=
\vert \overline{0}\overline{0}\overline{0}\rangle,\qquad \vert\varphi_9\rangle=\vert 000\rangle.
\label{state3}
\eeq
\noindent
Notice that $\vert\varphi_2\rangle$, $\vert\varphi_3\rangle$ and $\vert\varphi_4\rangle$ are biseparable of type $2(13)$, $(12)3$ and $1(23)$, $\vert\varphi_8\rangle$ and $\vert\varphi_9\rangle$ are separable.
The remaining four states belong to the GHZ-class.

In our error correcting scheme the $\{S_a\}, a=1,\dots 9$ correspond to a collection of messages $\{\mathcal{M}_a\}$.
For example $\mathcal{M}_7=\{YZZ,ZYI,-ZIY,-XZX,YXX,XXZ,-IYY\}$ gives rise to a {\it positive}
totally isotropic plane in $PG(5,2)$.
This means that if we multiply together all of the commuting observables of $\mathcal{M}_7$ one gets $+III$.
This is the generalization of the notion of positive lines familiar from studies concerning Mermin square-like configurations\cite{Mermin1,Mermin2,HS}.
Moreover, all the isotropic lines contained in our plane are in turn {\it positive}.
Indeed, several positive lines of that type are: $\{XXZ,-IYY,-XZX\}$, $\{-IYY,YXX,YZZ\}$ etc.
Hence in this case if $\mathcal{M}_7$ is the message plane, a corrupted message can be the first of these lines, or merely a point e.g. $XXZ$ on this line. Alternatively one can say that starting from the corresponding stabilizer state $\vert \varphi_7\rangle$ in the first case one gets a two dimensional, in the second case a four dimensional corrupted subspace of the Hilbert space.

Let us now use the results of Appendix C. to build up the GHW phase space. The points of this space are parametrized by pairs $(q,p)$ where $q,p\in GF(8)$. We adopt the convention that the coordinates are expanded according to the dual field basis, and the momenta according to the field basis.
Hence for $n=3$ we have the expansions of Eq.(\ref{haromexpansion}).
According to the pattern of Eq.(\ref{peldared}) using field reduction each pair $(q,p)$ can be converted to an observable up to sign.
In this way to each of the $64$ points of the GHW phase space we can assign an observable up to sign.
There are different possibilities for fixing the signs of these observables.
A possible choice of signs compatible with our choice of stabilizers of Eqs.(\ref{stab1})-(\ref{stab3}) can be seen in Table 4.
Notice that there is a unitary transformation $D$ of order seven acting by conjugation that can be used to get all of the seven stabilizers $S_k$ starting from $S_1$.

This unitary is of the form as Eq.(18) of Ref.\cite{LSV}:
\beq
\mathcal{D}=C_{12}C_{21}C_{12}C_{31}C_{23}C_{12}C_{31}
\label{Wootterstraf}
\eeq
\noindent
with its action on the messages
\beq
\mathcal{M}_k=\mathcal{D}^{k-1}\mathcal{M}_1\mathcal{D}^{1-k},\qquad k=1,2,\dots 7.
\eeq
\noindent
In terms of the GHW phase space coordinates the seven orbits are described as $qp=\omega^{k-1}$.
Notice also that $\mathcal{D}$ is an unitary representation on the $3$-qubit Hilbert space, of the $SL(2,8)$ transformation
acting on the points of the GHW phase space as
\beq
\begin{pmatrix}q\\p\end{pmatrix}\mapsto \begin{pmatrix}\omega^4&0\\0&\omega^3\end{pmatrix}\begin{pmatrix}q\\p\end{pmatrix}
\eeq
\noindent
alternatively instead of $\mathcal{D}$ one can choose the unitary $C_{13}S_{13}S_{12}$ calculated in Ref.\cite{W2}.
This unitary fits into the general scheme valid for arbitrary $n$ as can be seen from Eqs. (\ref{Lambda})-(\ref{Wootlift}).

\begin{table}[t]
\begin{center}
\caption{An association of observables to the points of the GHW discrete phase space for three-qubits.
This association corresponds to our choice of stabilizers of Eqs.(\ref{stab1})-(\ref{stab3}). The elements of the stabilizers $S_8$ and $S_9$ correspond to the horizontal and vertical lines.
The seven nontrivial elements of the remaining stabilizers correspond to the points belonging to lines passing through the origin satisfying an equation of the form $p=\omega^kq$ where $k\in\{1,2,3,4,5,6,7\}$ is fixed.
This choice of stabilizers corresponds to a quantum net, i.e. an association of states to the lines of the GHW phase space.
In particular to the lines through the origin we can associate the states of Eq.(\ref{state1})-(\ref{state3}).
}

\vspace*{0.3cm}
\begin{tabular}{||ll|llllllll||}
\hline \hline
&&&&&&&&&\\[-.2cm]
q/p & & $0$ & $1$ & $\omega$ & $\omega^2$ & $\omega^3$ & $\omega^4$ & $\omega^5$ & $\omega^6$ \\[1.mm]
 & & $000$ & $100$ & $010$ & $001$ & $101$ & $111$ & $110$ & $011$ \\[1.mm]
\hline
&&&&&&&&&\\[-.2cm]
$0$ & $000$ & $III$ & \col ${\bf XII}$ & \col ${\bf IXI}$ & \col ${\bf IIX}$ & \col ${\bf XIX}$ & \col ${\bf XXX}$ & \col ${\bf XXI}$ & \col ${\bf IXX}$ \\
$1$ & $111$ & \csl ${\bf ZZZ}$ & \crd $YZZ$ & $ \cgr ZYZ$ & \cpn $ZZY$ & \clb $-YZY$ & \cyl $YYY$ & \ctl $-YYZ$ & \cg $ZYY$ \\
$\omega$ & $110$ & \csl ${\bf ZZI}$ & \cg $YZI$ & \crd $ZYI$ & \cgr $ZZX$ & \cpn $-YZX$ & \clb $-YYX$ & \cyl $-YYI$ & \ctl $ZYX$ \\
$\omega^2$ & $101$ & \csl ${\bf ZIZ}$ & \ctl $YIZ$ & \cg $ZXZ$ & \crd $-ZIY$ & \cgr $-YIY$ & \cpn $YXY$ & \clb $YXZ$ & \cyl $ZXY$ \\
$\omega^3$ & $010$ & \csl ${\bf IZI}$ & \cyl $XZI$ & \ctl $-IYI$ & \cg $IZX$ & \crd $-XZX$ & \cgr $XYX$ & \cpn $XYI$ & \clb $IYX$ \\
$\omega^4$ & $100$ & \csl ${\bf ZII}$ & \clb $-YII$ & \cyl $ZXI$ & \ctl $-ZIX$ & \cg $YIX$ & \crd $YXX$ & \cgr $YXI$ & \cpn $ZXX$ \\
$\omega^5$ & $001$ & \csl ${\bf IIZ}$ & \cpn $XIZ$ & \clb $-IXZ$ & \cyl $IIY$ & \ctl $XIY$ & \cg $XXY$ & \crd $XXZ$ & \cgr $-IXY$ \\
$\omega^6$ & $011$ & \csl ${\bf IZZ}$ & \cgr $-XZZ$ & \cpn $IYZ$ & \clb $IZY$ & \cyl $XZY$ & \ctl $-XYY$ & \cg $-XYZ$ & \crd $-IYY$ \\[1.mm]
\hline \hline
\end{tabular}
\end{center}
\end{table}

Now we look at the Grassmannian image of the Lagrangian planes arising from the stabilizers ${S}_{a}$ where $a=1,2,\dots 9$.
We have already seen that they encapsulate the messages $\mathcal{M}_a$ providing a fibration (spread) of the boundary.
In the level of the GHW discrete phase space they provide striations\cite{W2} of this space.
For three qubits we have $N={6\choose 3}=20$ Pl\"ucker coordinates. We arrange them according to the pattern of Eq.(\ref{familiar}). Then the Grassmannian image of the messages provides $9$ points in the real part of the bulk. These will be the codewords $\mathcal{C}_{a}$. They form an ovoid in $\Re({\rm BULK})=Q^+(7,2)$ (see Eq.(\ref{include})). Following the patterns suggested by Eqs.(\ref{vastag})-(\ref{tools}) we have:
\beq
S_1\mapsto \mathcal{C}_1={\bf YYIX}YXZYXZ,\quad
S_2\mapsto \mathcal{C}_2={\bf YZYX}IYIIYI,\quad
S_3\mapsto \mathcal{C}_3={\bf YIIY}IIYIIY
\label{ovi1}
\eeq
\noindent
\beq
S_4\mapsto \mathcal{C}_4={\bf YYXZ}YIIYII,\quad
S_5\mapsto \mathcal{C}_5={\bf YIYZ}ZYXZYX,\quad
S_6\mapsto \mathcal{C}_6={\bf YXYX}YZYYZY
\label{ovi2}
\eeq
\noindent
\beq
S_7\mapsto \mathcal{C}_7={\bf YYZZ}XYYXYY,\quad
S_8\mapsto \mathcal{C}_8={\bf XIII}IIIIII,\quad
S_9\mapsto \mathcal{C}_9={\bf ZIII}IIIIII
\label{ovi3}
\eeq
\noindent
This correspondence can be visualized by associating to the differently coloured striations (message words) of the boundary showing up in 
Table 4 differently coloured points (code words) in the bulk.
The bold faced part of the codewords describes the $\mathcal{P}_{\psi}$ part (the $4$-qubit part) and the remaining six qubits refer to the $\mathcal{R}$ part. Notice that the doubled pattern of the latter part is in accord with Eqs.(\ref{kakukk6}). 
The bold faced parts and the remaining six-qubit part are both symmetric observables. This fact is in accordance with the results
$\hat{Q}(\mathcal{P}_{\psi})=Q(\mathcal{R})=0$.
Hence the $\mathcal{P}_{\psi}$ parts of the $9$ points are lying on the hyperbolic quadric $Q^+(7,2)$ which is the zero locus of the quadratic form $\hat{Q}$ as it has to be. 
Notice that the bold faced ($4$-qubit) parts and the full set $\mathcal{C}_{a}$ forms a nine dimensional Clifford algebra, i.e. we have $\{\mathcal{C}_{a},\mathcal{C}_{b}\}=2\delta_{ab}$. The pairwise anticommuting nature is in accord with the ovoid property, which says that no three points from this $9$ points are collinear.
However the ovoid property also says that every maximal totally isotropic $3$-subspace contains precisely one point from this set of $9$ points. Indeed, it is known that the quadric $Q^+(7,2)$ is doubly ruled by two set of generators i.e. it is of the form $PG(3,2)\times PG(3,2)$.
Notice also that due to the fact $\Re({\rm BULK})=Q^+(7,2)$ this case is very special. Indeed in this case the number of Lagrangian subspaces of the boundary coincides with the number of points of $Q^+(7,2)$ which is $135$. For the explicit form of the bijection between the points of the quadric and the 135 Lagrangian planes see Ref.\cite{LPS}.

Let us finally give a simple description of the set of Lagrangian planes in terms of generators of a seven dimensional Clifford algebra\cite{LPS}.
Define
\beq
\{\Gamma_1,\Gamma_2,\Gamma_3,\Gamma_4,\Gamma_5,\Gamma_6,\Gamma_7\}=\{IIY,ZYX,YIX,YZZ,XYX,IYZ,YXZ\}.
\label{Cliffordlabel}
\eeq
\noindent
Introduce the notation
\beq
\{k\}\equiv \Gamma_k\qquad \{km\}\equiv i\Gamma_k\Gamma_m,\qquad \{kml\}\equiv i\Gamma_k\Gamma_m\Gamma_l,\qquad 1\leq k<m<l\leq 7.
\nonumber
\eeq
\noindent
Then for example up to sign the observables of $\mathcal{M}_4$ corresponding to the stabilizer $S_4$ can be written as 
$\mathcal{M}_4=\{7,16,25,34,167,257,347\}$. Since the structure of this arrangement is determined by the first four entries we use the abbreviation as: $\mathcal{M}_4=\{7,16,25,34\}$.
Then the full set of message words in this notation is
\beq
\mathcal{M}_1=\{3,24,15,67\},\quad \mathcal{M}_2=\{2,13,47,56\},\quad\mathcal{M}_3=\{1,27,36,45\},\quad
\mathcal{M}_4=\{7,16,25,34\}
\nonumber
\eeq
\noindent
\beq
\mathcal{M}_5=\{6,57,14,23\},\quad \mathcal{M}_6=\{5,46,37,12\},\quad\mathcal{M}_7=\{4,35,26,17\},
\label{emmek1}
\eeq
\noindent
\beq
\mathcal{M}_8=\{124,235,346,457,156,267,137\},\quad \mathcal{M}_9=\{126,237,134,245,356,467,157\}.
\nonumber
\eeq
\noindent
Notice that the last two messages are simply generated by a cyclic shift of the combination $124$ and $126$ followed by normal ordering. 
Moreover, one can get all of the remaining seven message words from a reversed cyclic shift of the first.
Now it is easy to generate $105$ of the $135$ Lagrangian planes\cite{LPS}.

Indeed, from this notation it is clear that one can generate altogether $15$ planes intersecting at least in a point.
For example choosing this point to be the one represented by $\{7\}$ we get a list of such planes where the first entry is fixed to $7$ and the remaining three ones are consisting of all possible triples of duads of the form: $\{12,34,56\}$, $\{13,26,45\},\dots$ etc. These are the ones that make a partition of the full set $\{1,2,3,4,5,6\}$ to triples of duads. Notice that the structure of $\mathcal{M}_4$ is of this form. Hence if $\mathcal{E}=\{7\}$ is representing a point error then the Schubert variety $\Omega(\mathcal{E})$ is consisting of $15$ planes where the message word $\mathcal{M}_4$ is one of them.
Notice now that the incidence structure of such triples of duads is just the one of the doily, which is shown on the left hand side of Figure 1.
The $15$ duads correspond to its points and the $15$ triples to its lines. Since the doily is self dual one can also label its points with triples of duads and its lines with the duads.
When we take the Grassmannian image of these $15$ planes to the real part of the bulk what we get is a collection of $15$ points, with their corresponding $10$-qubit observables mutually commuting. However, for labelling what we really need is merely their four-qubit part. (For an example see the bold faced part of (\ref{ovi1})-(\ref{ovi3}).) 

Now this set of $15$ points in the real part of the bulk labelled by $4$-qubit observables can be equipped with an incidence structure according to whether the corresponding points are collinear or not.
At the level of observables collinearity is defined when the product of triples of our commuting observables give the identity up to sign.
It is easy to check that this incidence structure is again that of the doily.
Collinear points in the bulk represent planes intersecting in a line in the boundary, and non collinear ones represent ones
intersecting merely in a point.
Since all of these $15$ points are light-like separated in the bulk, this extra incidence structure gives rise to the fine structure of the "light cone" referred to in Section 4.7.
Notice that taking a spread in the duad labelling of the doily in this set of $15$ boundary planes , gives rise to an ovoid in 
the bulk doily. For example, taking $\mathcal{M}_4$ as one of the planes we have a triple of duads of the form $\{16,25,34\}$.
Choosing the remaining four triples as $\{12,35,48\}$, $\{13,26,45\}$, $\{15,24,36\}$, and $\{14,23,56\}$ we have a spread of the doily taken from five planes. Clearly these planes are intersecting merely in the error point $\{7\}$.
The image of these planes in the bulk is five non collinear points such that every isotropic line of this bulk doily contains one such point. This means that these points form an ovoid. 

On the other hand if any of our triples of duads contains a common duad, then it means that the corresponding three planes are intersecting in a line, hence their image in the bulk doily is just three collinear points.
An example for this is: $\{16,25,34\}$, $\{16,23,45\}$. In this case the corresponding planes are intersecting in the boundary line labelled by the triple $(7,16,167)$.

Take now the error line $\mathcal{E}=\{56,567,7\}$. It is easy to check that this line is intersecting with three message planes namely: $\mathcal{M}_2$,
$\mathcal{M}_4$, $\mathcal{M}_5$ in a point. Hence a unique recovery from this type of error is not possible.
However, if we are restricting the range of possible errors to point and line errors living entirely inside of some message word, then recovery from point errors is also possible. As advertised in the text these types of errors are compatible with the quantum net structure related to our association of states to planes and subspaces of ever increasing dimension to lines and planes living {\it inside} our messages.

We remark in closing, that now it is easy to generate $105$ of the $135$ Lagrangian planes.
Just take the $15$ planes intersecting the $\mathcal{M}_4$ one in $\{7\}$ Then rotate cyclically the labels of the corresponding planes. The result is $7\times 15=105$ planes. For the cyclic generation of the missing $28$ planes from four seed ones see Ref.\cite{LPS}. Clearly these $14+14$ planes are the ones which are intersecting with the remaining message planes $\mathcal{M}_8$ and $\mathcal{M}_9$ at least in a point.
The latter $15+15$ special planes, featuring symmetric three-qubit observables, are of course the $\alpha$ and $\beta$-planes living on the Klein quadric.

\subsection{Appendix E: Subgeometries}

Let us consider $\Sigma=PG(2^n-1,2)$ and $\Sigma^{\ast}=PG(2^n-1,2^n)$ as the geometries corresponding to their lattices of their subspaces. Since $GF(2)$ is a subfield of $GF(2^n)$ we say that the former is the {\it canonical subgeometry} of the latter. 
For each subspace $S^{\ast}$ of $\Sigma^{\ast}$ the set $S=S^{\ast}\cap \Sigma$ is a subspace of $\Sigma$ whose rank is at most equal to the rank of $S^{\ast}$. We say that a subspace $S^{\ast}$ of $\Sigma^{\ast}$ is a subspace of $\Sigma$ whenever $S$ and $S^{\ast}$ have the same rank\cite{Lunardon}.

In our special case used in the text we use further definitions. A linear map $\sigma$ between two vector 
spaces over $GF(2^n)$ is called a {\it semilinear map} if $\sigma (\lambda v)
=\lambda^2 \sigma(v)$. This means that our map is linear up to a twist generated by a field automorphism, which is in our case the Frobenius automorphism. 
By a {\it collineation} we mean a bijective map between projective spaces such that the images of collinear points (i.e.the ones lying on the same line) are
themselves collinear.

There is a fundamental lemma which is proved in Ref.\cite{Lunardon} which is implicitly used in our considerations of Section 4.

LEMMA: If $\sigma$ is a semilinear collineation of $\Sigma^{\ast}$ having as fixed points exactly the ones of $\Sigma$, then $S^{\ast}$ is a subspace of $\Sigma$ if and only if $S^{\ast}$ is fixed by $\sigma$.

An explicit example for a map of that kind is the one used in the text
satisfying $\sigma(\lambda v_1\otimes v_2\otimes\cdots \otimes v_n)=\lambda^2\sigma(v_n\otimes v_1\otimes\cdots \otimes v_{n-1})$, whenever each of $v_j$ with $j=1,2,\dots n$ is taken from the set of basis vectors $\{E,F\}$ of the vector space 
${\bf V}$ and $\lambda\in GF(2^n)$ of Eq.(\ref{fibit}).
If for an $x\in{\bf V}$ written in the form of Eq.(\ref{fibit}) we write $x^{\phi}$ as in Eq.(\ref{Frobi}) with $j=2$ then from the definition of $\sigma$ it follows that $\sigma(x_1\otimes x_2\otimes\cdots\otimes x_n)
=x^{\phi}_n\otimes x^{\phi}_1\otimes\cdots\otimes x^{\phi}_{n-1}$.

\end{document}